\documentclass[journal]{IEEEtran}
\usepackage[utf8]{inputenc}

\usepackage{xcolor,soul,framed} 

\colorlet{shadecolor}{yellow}
\usepackage[pdftex]{graphicx}
\graphicspath{{../pdf/}{../jpeg/}}
\DeclareGraphicsExtensions{.pdf,.jpeg,.png}
\usepackage{cite}
\usepackage[cmex10]{amsmath}
\usepackage{array}
\usepackage{mdwmath}
\usepackage{mdwtab}
\usepackage{eqparbox}
\usepackage{url}
\usepackage{multirow}
\usepackage{amssymb}
\usepackage{booktabs}

\usepackage[tikz]{bclogo}
\usepackage[framemethod=tikz]{mdframed}
\usepackage{tikz}
\usepackage[edges]{forest}
\usepackage{multicol}
\usetikzlibrary{trees,positioning,mindmap,backgrounds}
\usetikzlibrary{calc,positioning,intersections}
\usepackage{listings}
\usetikzlibrary{trees,positioning,shapes,shadows,arrows.meta}
\definecolor{hidden-draw}{RGB}{195,82,66}
\definecolor{hidden-blue}{RGB}{194,232,247}
\definecolor{hidden-orange}{RGB}{195,82,66}
\definecolor{hidden-yellow}{RGB}{242,244,193}
\definecolor{tree-level-1}{RGB}{195,82,66}
\definecolor{tree-level-2}{RGB}{195,82,66}
\definecolor{tree-level-3}{RGB}{195,82,66}
\definecolor{tree-leaf}{RGB}{195,82,66}

\begin{document}
\bstctlcite{IEEEexample:BSTcontrol}
    \title{Resource-Oriented Optimization of Electric Vehicle Systems: A Data-Driven Survey on Charging Infrastructure, Scheduling, and Fleet Management}
    
\author{
\IEEEauthorblockN{
Hai Wang\IEEEauthorrefmark{1},
Baoshen Guo\IEEEauthorrefmark{2},
Xiaolei Zhou\IEEEauthorrefmark{1},
Shuai Wang\IEEEauthorrefmark{2},
Zhiqing Hong\IEEEauthorrefmark{3},
Tian He\IEEEauthorrefmark{4}
}
\\
\IEEEauthorblockA{
\IEEEauthorrefmark{1}The Sixty-third Research Institute, National University of Defense Technology,
\IEEEauthorrefmark{2}Southeast University,
\IEEEauthorrefmark{3}Hong Kong University of Science and Technology (Guangzhou),
\IEEEauthorrefmark{4}JD Logistics
}
\\
\{haiwang, zhouxiaolei\}@nudt.edu.cn, 
\\
\{guobaoshen, shuaiwang\}@seu.edu.cn,
zhiqinghong@hkust-gz.edu.cn,
dr.tianhe2@gmail.com
}

\maketitle

\begin{abstract}
Driven by growing concerns over air quality and energy security, 
electric vehicles (EVs) has experienced rapid development and are reshaping global transportation systems and lifestyle patterns.
Compared to traditional gasoline-powered vehicles, EVs offer significant advantages in terms of lower energy consumption, reduced emissions, and decreased operating costs. However, there are still some core challenges to be addressed:
(i) Charging station congestion and operational inefficiencies during peak hours,
(ii) High charging cost under dynamic electricity pricing schemes, and
(iii) Conflicts between charging needs and passenger service requirements.
Hence, in this paper, 
we present a comprehensive review of data-driven models and approaches proposed in the literature to address the above challenges.
These studies cover the entire lifecycle of EV systems, including charging station deployment, charging scheduling strategies, and large-scale fleet management. 
Moreover, we discuss the broader implications of EV integration across multiple domains, such as human mobility, smart grid infrastructure, and environmental sustainability, and identify key opportunities and directions for future research.

\end{abstract}


\IEEEpeerreviewmaketitle

\maketitle

\section{Introduction}\label{Sec1-Introduction}

With advances in energy technologies, electric vehicles (EVs) are finding increasingly widespread applications \cite{li2015growing}. Compared with conventional gasoline-powered vehicles, EVs offer several notable advantages, including (i) lower operating costs, (ii) independence from fossil fuels, (iii) rapid acceleration, and (iv) zero tailpipe emissions. Reflecting these benefits, the EV market has experienced exponential growth, with global sales surpassing 17 million units in 2024 and projected to exceed 20 million in 2025, accounting for more than one-quarter of total automobile sales worldwide \cite{GlobalEV}. At the industrial level, leading corporations have emerged, most prominently Tesla in the United States \cite{Tesla} and BYD in China \cite{BYD}.

The key difference between EVs and gasoline-powered vehicles is the energy replenishment mechanism, i.e., refueling vs. charging. Compared to the refueling pattern, the charging pattern has two key characteristics, which generate significant influence on the operational paradigm of EVs, and make the traditional management approach of gasoline-powered vehicles can not directly deploy on EVs \cite{wang2021data}. The first characteristic is the long charging duration, typically ranging from half an hour to two hours \cite{ullah2022prediction}. Different from the refueling duration of around two minutes, charging space allocation accompanied by congestion problems become a major issue for EV management \cite{wang2018bcharge}. To handle it, it is essential to rethink the whole process from deploying the charging stations to fleet management for EVs by adding the consideration of getting full availability of the charging points. The second characteristic is the dynamic charging price. Affected by the variations in electricity consumption, the charging price varies in the different periods of the day, and drivers are more prone to charging during off-peak electricity price periods.
The inclusion of the time factor and human tendency adds complexity to EV management in this scenario.

Fortunately, the research community produces several papers that shed light on the application and promotion of EVs from various angles. Some scholars focus on policy design as a means to tackle related issues \cite{ji2018plug,lamonaca2022state,harrison2017exploratory}, while others employ social research and empirical analysis \cite{akinlabi2020configuration,berkeley2018analysing,debnath2021political,lopez2021societal,singh2020review}. Over the past decade, there have been significant advancements in hardware devices and the deep learning algorithms that underpin them. Consequently, the daily functioning of society generates vast quantities of data. Statistical data suggest that, as of 2022, there are approximately 25.9 million EVs being used worldwide \cite{Number}. Within the legal boundaries, specific commercially operated EVs have their trajectory data, charging data, and operation data recorded \cite{wang2020understanding}. Therefore, mining valuable insights from historical data emerges as a promising avenue to develop effective EV management strategies. The availability of expansive datasets presents numerous possibilities for obtaining useful information, potentially leading to the discovery of patterns that can significantly enhance the operational efficiency of EVs.

\begin{figure*}[htbp]
\centering
    \includegraphics[width = 0.99\linewidth]{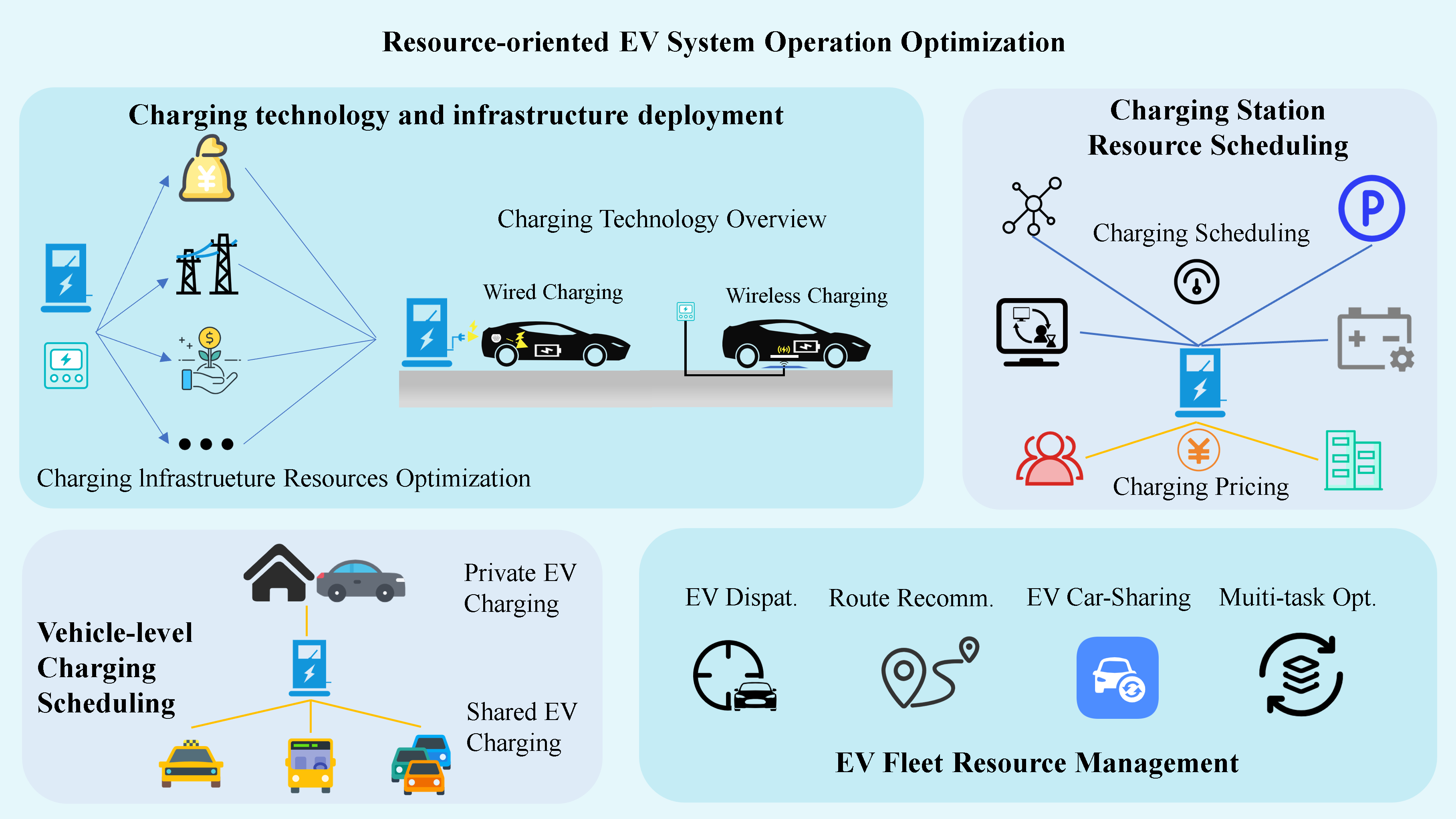}
    \caption{Resource-oriented EV System Operation Optimization}
    \label{Fig.resource-oriented_EV_system_operation_optimization}   
\end{figure*}

In our paper, we delve into the data-driven aspects of EVs, a subset within the field of intelligent transportation. Fig.~\ref{Fig.resource-oriented_EV_system_operation_optimization} illustrates the framework for resource-oriented EV system operation optimization. The primary objective of managing EVs is to enhance the availability of charging points and optimize operational efficiency for commercial EVs. Achieving this goal necessitates addressing various subtasks from multiple perspectives: (1) Charging infrastructure deployment. In this task, researchers must extract spatio-temporal information to discern the charging patterns and demands across the entire city. This insight can then be utilized to design an improved charging infrastructure layout. (2) EVs charging schedule. Given the limited resources available, it is imperative to develop an advanced scheduling methodology that ensures efficient utilization of related resources, including electric energy, charging prices, and available charging sites. (3) EVs fleet management. For businesses utilizing EVs as operational vehicles, establishing a robust fleet management model becomes crucial, as it guarantees profitability and sustainable development. Fleet management tasks comprise the optimal allocation of EV locations, route recommendations for EVs, and car-sharing patterns among EVs. Within our paper, we comprehensively explore the application of data-driven techniques in addressing the aforementioned three tasks, providing a wealth of insightful content.

\tikzstyle{my-box}= [
    rectangle,
    draw=hidden-draw,
    rounded corners,
    text opacity=1,
    minimum height=1.5em,
    minimum width=5em,
    inner sep=2pt,
    align=center,
    fill opacity=.5,
]
\tikzstyle{leaf}=[
my-box, 
minimum height=1.5em,
fill=yellow!27, 
text=black,
align=left,
font=\scriptsize,
inner xsep=5pt,
inner ysep=4pt,
align=left,
text width=45em,
]
\tikzstyle{leaf2}=[
my-box, 
minimum height=1.5em,
fill=purple!22, 
text=black,
align=left,
font=\scriptsize,
inner xsep=5pt,
inner ysep=4pt,
]
\tikzstyle{leaf3}=[
my-box, 
minimum height=1.5em,
fill=hidden-blue!57, 
text=black,
align=left,
font=\scriptsize,
inner xsep=5pt,
inner ysep=4pt,
]
\tikzstyle{leaf4}=[
my-box, 
minimum height=1.5em,
fill=green!17, 
text=black,
align=left,
font=\scriptsize,
inner xsep=5pt,
inner ysep=4pt,
]
\begin{figure*}[!htbp]
\centering
\scalebox{0.9}{
  \begin{forest}
    forked edges,
    for tree={
      grow=east,
      reversed=true,
      anchor=base west,
      parent anchor=east,
      child anchor=west,
      base=left,
      font=\small,
      rectangle,
      draw=hidden-orange,
      rounded corners,
      align=left,
      minimum width=4em,
      edge+={darkgray, line width=1pt},
      s sep=3pt,
      inner xsep=2pt,
      inner ysep=3pt,
      ver/.style={rotate=90, child anchor=north, parent anchor=south, anchor=center, align=center, my-box},
    },
    where level=1{text width=10em,font=\scriptsize,}{},
    where level=2{text width=8em,font=\scriptsize,}{},
    where level=3{text width=6.4em,font=\scriptsize,}{},
    where level=4{text width=6.4em,font=\scriptsize,}{},
[
                \textbf{Resource-oriented EV System Operation Optimization}, ver
                 [  
                    \textbf{Charging technology and} \\ \textbf{infrastructure deployment}, ver
                    [
                            \textbf{Charging Technology} \\ \textbf{Overview}
                            [
                            \textbf{Wired Charging} 
                                [
                                    \textbf{Fast charging}: Hu et al.~\cite{hu2014new}{,} 
                                    MSbordone et al.~\cite{sbordone2015ev}{,}
                                    \textbf{Ultra fast charging}:
                                    Chen et al. \cite{chen2020enabling}, leaf, text width=32em
                                ]
                            ]
                            [
                            \textbf{Wireless Charging} 
                                [
                                    \textbf{Static wireless charging}: Mohamed et al. \cite{mohammed2021comprehensive,mohamed2022comprehensive}{,} 
                                    \textbf{Dynamic wireless charging}:
                                    Machura et al. \cite{machura2019critical}, leaf, text width=32em
                                ]
                            ] 
                    ]
                    [ 
                        \textbf{Charging Infrastructure} \\ \textbf{Resources Optimization}
                        [
                            \textbf{Wired charging}  \\ \textbf{station deployment}
                            [
                                \textbf{Profit-oriented}
                                [
                                Albert et al.~\cite{lam2014electric}{,} 
                                Bae et al.~\cite{bae2020game}{,}
                                Zhao et al.~\cite{zhao2020deployment}, leaf, text width=24em
                                ]
                            ]
                            [
                                \textbf{Power-grid aware}
                                [
                                Mirzaei et al.~\cite{mirzaei2015probabilistic}{,} 
                                Zhang et al.~\cite{zhang2016pev,zhang2017second}{,}
                                Wang et al.~\cite{wang2018coordinated}{,}
                                Ahmad et al.~\cite{ahmad2023placement}, leaf, text width=24em
                                ]
                            ]
                            [
                                \textbf{Social Benefits aware}
                                [
                                Li et al.~\cite{li2015growing}{,} 
                                He et al.~\cite{he2016incorporating}{,}
                                Liu et al.~\cite{liu2016optimal,liu2017locating}{,}
                                Guo et al.~\cite{guo2024approach}{,}
                                Zheng et al.~\cite{zheng2024effects}, leaf, text width=24em
                                ]
                            ]
                            [
                                \textbf{Learning-based}
                                [
                                Liu et al.~\cite{liu2023placement}{,} 
                                Liu et al.~\cite{liu2024random}{,}
                                Tang et al.~\cite{tang2025stochastic}{,}
                                Zhao et al.~\cite{zhao2023optimal}, leaf, text width=24em
                                ]
                            ]
                        ]
                        [
                           \textbf{Wireless charging} \\ \textbf{station deployment}
                            [
                                Yan et al.~\cite{yan2018employing}{,} 
                                Riemann et al.~\cite{riemann2015optimal}{,}
                                Yan et al.~\cite{yan2017catcharger}{,}
                                Li et al.~\cite{li2019wireless}{,}
                                Elmeligy et al.~\cite{elmeligy2024optimal}{,}
                                Bai et al.~\cite{bai2022robust}, leaf, text width=32em
                            ]
                        ] 
                    ]
                ]
                [  
                    \textbf{Charging Station} \\ \textbf{Resource Scheduling}, ver
                    [
                        \textbf{Charging Scheduling}
                        [
                            \textbf{Network-level} \\ \textbf{Optimization}
                            [
                                Yan et al. \cite{yan2022mobicharger}{,} 
                                Gusrialdi et al. \cite{gusrialdi2014scheduling}, leaf2, text width=32em
                            ]
                        ]
                        [
                        \textbf{Parking-based} 
                            [
                                Kong et al. \cite{kong2016line}{,} 
                                Iversen et al. \cite{iversen2014optimal}, leaf2, text width=32em
                            ]
                        ]
                        [
                        \textbf{Demand response} \\ \textbf{mechanisms}
                            [
                                Zhao et al. \cite{zhao2013peak}{,} 
                                Shaaban et al. \cite{shaaban2014real}{,}
                                Duan et al.~\cite{duan2025study}{,}
                                Bokopane et al.~\cite{bokopane2024optimal}, leaf2, text width=32em
                            ]
                        ]
                        [
                        \textbf{Real-time} \\ \textbf{Charging Scheduling}
                            [
                                Du et al. \cite{du2018demand}{,} 
                                Wang et al. \cite{wang2019tcharge, wang2019sharedcharging}{,}
                                Dong et al. \cite{dong2017rec}{,}
                                Du et al. \cite{du2018demand}{,}
                                Wang et al. \cite{wang2019tcharge,wang2019sharedcharging}, leaf2, text width=32em
                            ]
                        ]
                        [
                        \textbf{Battery management}
                            [
                                Kim et al. \cite{kim2014real, kim2015modeling}{,} 
                                Li et al. \cite{li2019data}{,}
                                Yang et al. \cite{yang2017charging}{,}
                                Cao et al. \cite{cao2021online}, leaf2, text width=32em
                            ]
                        ]
                    ]
                    [
                    \textbf{Charging Pricing} 
                    [
                        \textbf{for Customers}
                            [
                            Monhsenian et al. \cite{mohsenian2010optimal}{,} 
                            Shi et al. \cite{shi2011real}{,}
                            Ma et al. \cite{ma2016efficient}, leaf2, text width=32em
                            ]
                    ]
                    [
                        \textbf{for Companies} 
                            [
                            Zou et al. \cite{zou2014cost}{,} 
                            Fan et al. \cite{fan2020enabling}{,}
                            Zhang et al. \cite{zhang2018optimal}{,}
                            Sarker et al. \cite{sarker2017opportunistic}{,}
                            Wang et al. \cite{wang2020pricing}, leaf2, text width=32em
                            ]
                    ]
                    ]
                ]
                [
                \textbf{Vehicle-level}\\ \textbf{Charging Scheduling}, ver
                    [
                        \textbf{Private EV Charging} 
                        [
                        \textbf{User-Centric} 
                            [
                            \textbf{Model-based}:
                            Gusrialdi et al. \cite{gusrialdi2017distributed}{,} 
                            Lin et al. \cite{lin2021toward}{,} \\
                            \textbf{RL-based}:
                            CoordiQ \cite{blum_coordiq_2021}{,}
                            Li et al. \cite{li_coupling_2022}{,} 
                            Liu et al. \cite{liu2021reservation}{,}
                            Valogianni et al. \cite{valogianni_effective_2014}, leaf3, text width=32em
                            ]
                        ]
                        [
                        \textbf{System-Level} 
                            [
                            \textbf{Multi-agent}:
                            Zhang et al. \cite{zhang_intelligent_2021}{,}
                            Suanpang et al. \cite{suanpang2022intelligent}{,}
                            Zhang et al.~\cite{zhang2022rlcharge}{,}
                            \textbf{Graph RL}:
                            Xing et al. \cite{xing2022graph}{,} \\
                            \textbf{Game-based}:
                            Wei et al. \cite{wei2015charging}{,}
                            \textbf{Online recommendation}:
                            Xu et al. \cite{xu2022real}{,}
                            Sun et al. \cite{sun_orc_2020}{,}\\
                            \textbf{Multi objective}:
                            Klein et al. \cite{klein2023electric}{,}
                            Tan et al. \cite{tan2023fair}{,}
                            Li et al. \cite{li2025charging}, leaf3, text width=32em
                            ]
                        ]
                    ]
                    [
                        \textbf{Shared EV Charging} 
                        [
                            \textbf{Taxi Charging} 
                            [
                            Tian et al. \cite{tian_real-time_2016}{,} 
                            REC \cite{dong_rec_2017}{,}
                            Wang et al. \cite{wang_tcharge_2019,wang_foretaxi_2023}{,}
                            R2E \cite{wang_r2e_2018}{,}
                            Wang et al. \cite{wang_joint_2022}{,}
                            Yuan et al.~\cite{yuan2025stochastic}{,}
                            FairCharge \cite{wang_faircharge_2020}, leaf3, text width=32em
                            ]
                        ]
                        [
                            \textbf{Bus Charging} 
                            [
                            bCharge \cite{wang_bcharge_2018}{,} 
                            Ma et al. \cite{ma2021optimal}{,}
                            Hu et al. \cite{hu2022joint}{,}
                            Houbbady et al. \cite{houbbadi2019optimal}{,}
                            Jahic et al. \cite{jahic2019charging}{,} 
                            Rodrigues et al. \cite{rodrigues_optimized_2020}{,} \\
                            Duan et al. \cite{duan2022bidding}{,}
                            Kang et al. \cite{kang2015centralized}{,}
                            Li et al. \cite{li2020joint}{,}
                            Wang et al. \cite{wang2020pricing}{,}
                            Qi et al.~\cite{qi2025optimizing}{,}
                            Zhou et al.~\cite{zhou2024electric}, leaf3, text width=32em
                            ]
                        ]
                        [
                            \textbf{Heterogeneous} \\ \textbf{EV Charging} 
                            [
                            Wang et al. \cite{wang2019sharedcharging}, leaf3, text width=32em
                            ]
                        ]
                    ]                
                ]
                [  
                    \textbf{EV Fleet Resource Management}, ver
                    [
                        \textbf{EV Dispatching} 
                        [   
                            \textbf{EV Repositioning}
                            [
                                Tu et al.~\cite{tu2024deep}{,} 
                                Wang et al. \cite{wang2021data}{,}
                                Kullman et al. \cite{kullman2022dynamic}{,}
                                Hu et al.~\cite{hu2021effective}{,}
                                Ma et al.~\cite{ma2025crowdsourced}{,}
                                Yi et al. \cite{yi2021framework}, leaf4, text width=32em
                            ]
                        ]
                        [
                            \textbf{Order Dispatching}
                            [
                                Yan et al. \cite{yan2023online}{,} 
                                Chen et al. \cite{chen2023electric}{,}
                                Guo et al.~\cite{guo2021concurrent}{,}
                                Wang et al.~\cite{wang2023gcrl}{,}
                                Shi et al. \cite{shi2022memory}, leaf4, text width=32em
                            ]
                        ]
                    ]
                    [
                        \textbf{Route Recommendation} 
                        [   
                            \textbf{Heuristic}:
                            Nolz et al. \cite{nolz2022consistent}{,}
                            Li et al.~\cite{li2023route}{,}
                            Erdelic et al. \cite{erdelic2022goods}{,}
                            \textbf{Optimization-based}:
                            Stodola et al.~\cite{stodola2020hybrid}{,}
                            Wu et al. \cite{wu2019brainstorming}{,}\\
                            \textbf{Learning-based}:
                            Basso et al. \cite{basso2021electric}{,}
                            Basso et al. \cite{basso2022dynamic}{,}
                            Lin et al. \cite{lin2021deep}, leaf4, text width=40em
                        ]
                    ]
                    [
                        \textbf{EV Car-Sharing} 
                        [
                            Folkestad et al. \cite{folkestad2020optimal}{,} 
                            Gambella et al. \cite{gambella2018optimizing}{,}
                            Wang et al. \cite{wang2021record}{,}
                            Luo et al. \cite{luo2021rebalancing}, leaf4, text width=40em
                        ]
                    ]
                    [
                        \textbf{Multi-task Optimization}
                        [
                            Li et al.~\cite{li2025multi}{,} Huang et al.~\cite{huang2024prediction}{,} Che et al.~\cite{che2023battery}{,} Bao et al.~\cite{bao2024dual}, leaf4, text width=40em
                        ]
                    ]                
                ]
]
\end{forest}
}
\caption{Summary of Resource-oriented EV System Operation Optimization}
\label{fig:har_settings}
\end{figure*}
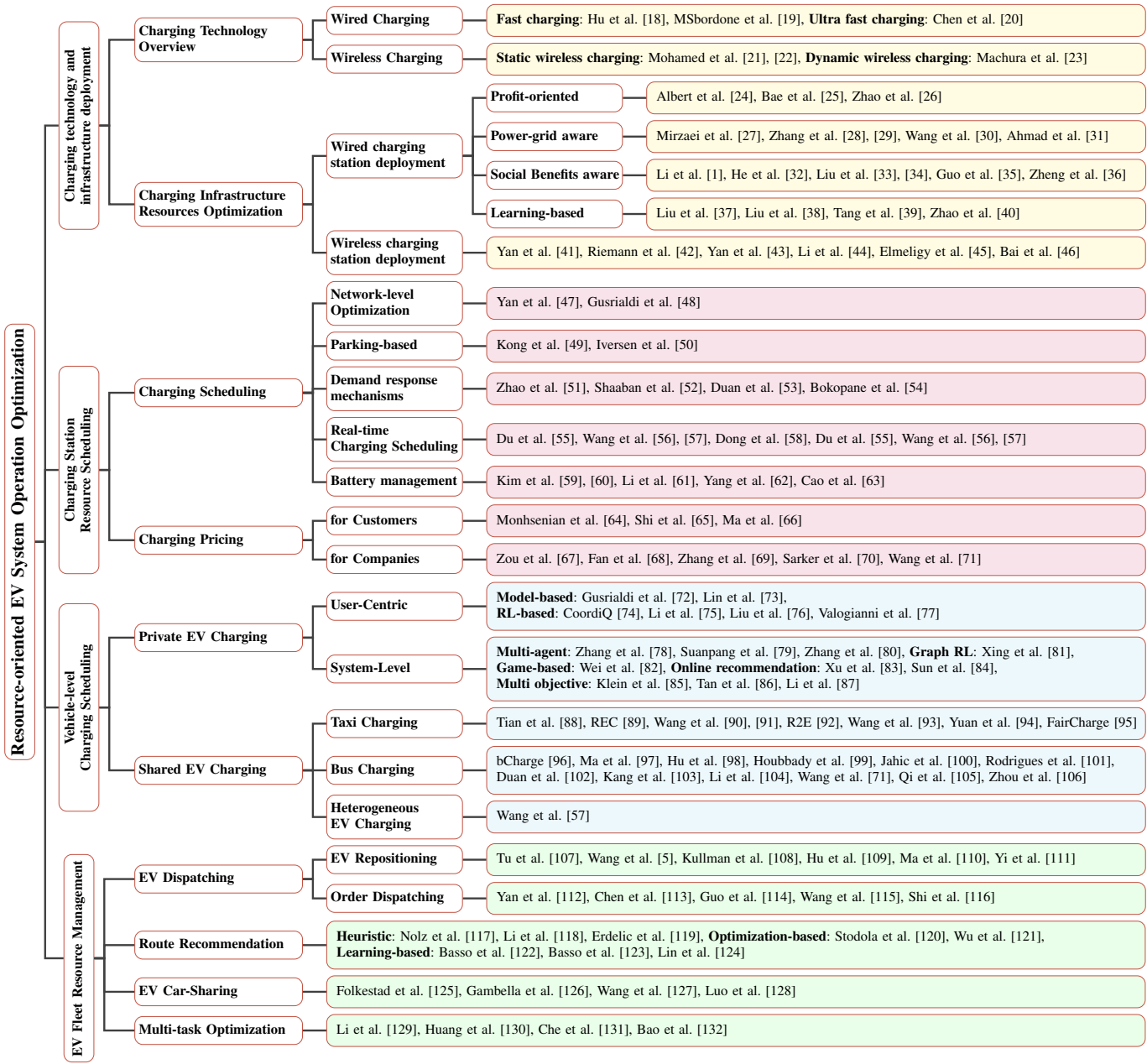

\noindent\textbf{Organization:}
The subsequent sections of this survey are structured as follows. Section II offers a meta-review of recent studies surveying electric vehicle research. Section III reviews the state-of-the-art charging technologies for EVs and discusses algorithmic strategies used in optimizing charging station deployment. Section IV focuses on the existing literature regarding charging schedules for EVs, encompassing energy scheduling, price scheduling, and charging space scheduling. Section V provides a synthesis of these studies in relation to fleet management for EVs, with specific emphasis on commercial operations for EV-based companies. Section VI discusses the societal implications arising from the development of EVs. Lastly, the concluding section offers a summary of our review and presents potential avenues for future research.

\section{Related Surveys}\label{Sec2-EV Survey}

The growing adoption of EVs has sparked extensive research into their integration into transportation and energy systems. This survey covers key areas including charging methodologies, station deployment, scheduling strategies, fleet management, and the broader impacts of EVs on infrastructure and the environment. Together, these topics reflect the current state and future directions of EV research.

\subsubsection{Survey of EV Charging Methodologies and Charging Station Deployment}
The evolution of EV technology has led to significant research efforts aimed at improving charging methodologies, which serve as a cornerstone for enhancing EV usability, efficiency, and integration into the power grid. 
Madaram et al.~\cite{madaram2024advancement} offer a broad overview of the latest EV technologies and charging methods, comparing electric and hybrid vehicles while analyzing conductive, wireless, and battery swap charging solutions. 

Prior to this, Singh et al.~\cite{singh2023comprehensive} focus specifically on the reliability assessment of EVs and the broader implications of different charging methodologies. They delve into the role of grid integration, converter technologies, and charging infrastructure, particularly examining the bidirectional impacts of vehicle-to-grid (V2G) and grid-to-vehicle (G2V) systems on distribution network stability.
Expanding further into the technical aspects of charging systems, Saraswathi et al.~\cite{saraswathi2024comprehensive} classify and analyze various EV charging technologies, including on-board and off-board chargers, as well as unidirectional and bidirectional power flow systems. 

While the above works focus on the technical and infrastructural dimensions of EV charging, Garofalaki et al.~\cite{garofalaki2022electric} shift the focus to cybersecurity considerations in smart charging systems. They investigate vulnerabilities in the Open Charge Point Protocol (OCPP), particularly in its latest version 2.0, and identify key entities involved in smart charging ecosystems.

Some survey papers that examine key challenges and advancements in the deployment and optimization of EV charging stations, with a focus on location planning, technological innovation, and operational strategies.
Suhail et al.~\cite{suhail2023objective} emphasize the importance of smart charging station design and placement, noting that the success of electrified transportation systems hinges on well-planned infrastructure. 
Building on the foundation of strategic infrastructure planning, Ren et al.~\cite{ren2024understanding} explore the emerging field of Wireless Charger Networks (WCNs), which offer a promising alternative to traditional plug-in stations.

\subsubsection{Survey of EV Charging scheduling}
This section presents a series of recent survey papers that systematically review key challenges and advances in EV charging scheduling, covering algorithmic approaches, optimization techniques, and system-level integration strategies.

Zhao et al.~\cite{zhao2024reinforcement} provide a systematic review of reinforcement learning (RL) approaches for EV charging scheduling, highlighting their potential to simultaneously improve user satisfaction, charging station profitability, and grid reliability. 
Building on this foundation, Salam et al.~\cite{salam2024charge} emphasize the need to optimize EV charging schedules in the face of unpredictable vehicle arrival patterns and energy demands. 

In a broader system-level perspective, Al-Alwash et al.~\cite{al2024optimization} examine both centralized and decentralized charging strategies, focusing on how optimized scheduling can prevent grid overload and charging station congestion. 
Elghanam et al.~\cite{elghanam2024optimization} further categorize and analyze deterministic optimization techniques for EV charging coordination, distinguishing among temporal scheduling, spatial assignment, and spatio-temporal strategies. 

Expanding the scope to include renewable energy integration and bidirectional energy flow, Alaee et al.~\cite{alaee2023review} explore coordinated EV charging strategies that incorporate vehicle-to-grid (V2G) technologies and renewable energy sources. 
Zhang et al.~\cite{zhang2023vehicle} focus on the specific application of electric buses, reviewing recent advances in battery electric bus scheduling and identifying key research gaps such as robustness, flexibility in charging modes, and integrated planning across routes and charging facilities. 

In a more integrated system design perspective, Kalakanti et al.~\cite{kalakanti2023computational} present a broad overview of computational approaches used in EV systems, including routing, charging scheduling, station placement, and energy management. 
Finally, Dahiwale et al.~\cite{dahiwale2024comprehensive} offer a comprehensive review of smart charging strategies from the perspective of control topology and methodology.

\subsubsection{Survey of EV Fleet Management}
Recent surveys have explored various aspects of EV fleet management, ranging from routing and scheduling to optimization and machine learning-based decision-making.

Wen et al.~\cite{wen2024survey} review recent advances in machine learning applications for ride-hailing systems, focusing on strategies for matching riders with vehicles and optimizing vehicle repositioning. The authors introduce a taxonomy to classify existing approaches, discuss widely used datasets and simulation tools, and identify promising research directions for enhancing the efficiency and responsiveness of ride-hailing platforms through intelligent planning.
Complementing this, Teusch et al.~\cite{teusch2023systematic} present a broader systematic review of machine learning techniques in shared mobility systems, including car-sharing, ride-sharing, and micromobility services. 
Bruglieri et al.~\cite{bruglieri2023survey} provide a comprehensive overview of optimization challenges in car-sharing systems, covering both conventional and emerging models such as electric, autonomous, and multi-modal fleets. 
Finally, Soto et al.~\cite{soto2025vehicle} focus on satellite-based vehicle routing problems in urban logistics, examining how such models can be applied to optimize EV fleet operations in last-mile delivery.

\subsubsection{Survey of EV Impacts}
This section reviews a set of comprehensive survey papers that examine the technical, operational, and environmental impacts of electric vehicles (EVs), particularly in relation to power systems, charging infrastructure, and energy network integration.
Wang et al.~\cite{wang2021grid} offer a foundational review of fast charging technologies for EVs, examining their technical characteristics, including high-charging-rate batteries, charging infrastructure, and associated grid impacts. 
Franzese et al.~\cite{franzese2023fast} build on this by focusing on ultrafast charging stations and their implications for distribution grid stability. 
Expanding the perspective to the role of EVs in future energy systems, Inci et al.~\cite{inci2024power} explore how EVs can actively support smart grids through bidirectional technologies such as vehicle-to-grid (V2G), grid-to-vehicle (G2V), and vehicle-to-vehicle (V2V) systems. 
Ismail et al.~\cite{ismail2023impact} further emphasize the transformative potential of V2G in enabling a more sustainable and flexible energy ecosystem. 
Rashid et al.~\cite{rashid2024comprehensive} examine the integration of solar photovoltaic (PV) systems with EV charging infrastructure, discussing the design challenges and benefits of solar-powered mobility.

\begin{figure*}[h]
\centering
    \includegraphics[width = 0.75\linewidth]{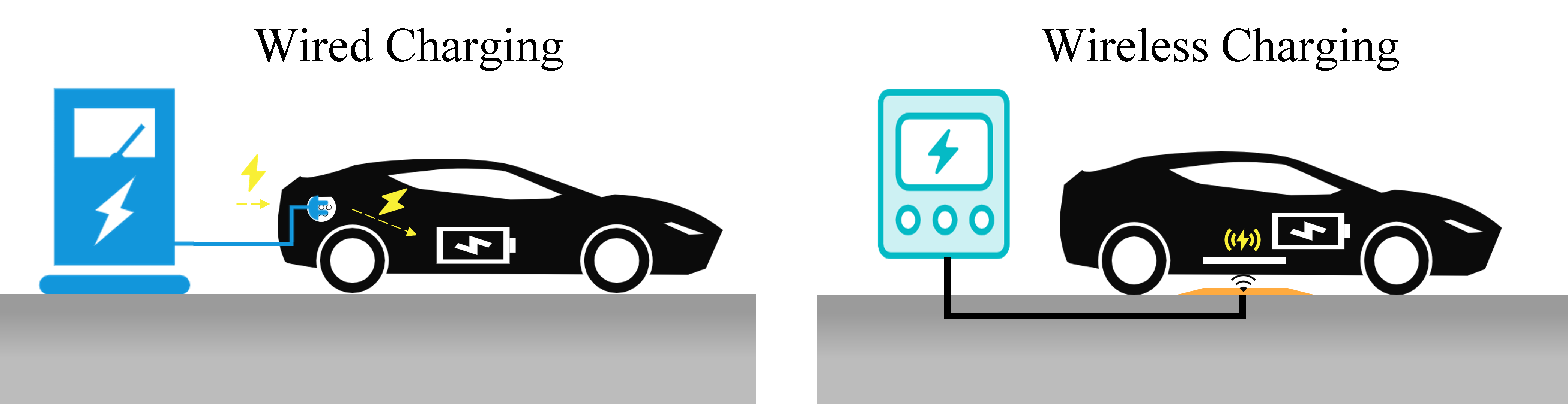}
    \caption{Illustrative schematic diagram delineating both wired and wireless charging methodologies for electric vehicles}
    \label{Fig.stationary-dynamic-charging}   
\end{figure*}

\section{EV Charging Technologies and Infrastructure Deployment}

In this section, we first provide an overview of the charging technologies employed in EVs, followed by a comprehensive review of charging infrastructure deployment, which can be regarded as the spatial optimization of static EV infrastructure resources.

\subsection{Overview of EV Charging Technology}\label{Sec2-1-Charging Technology}
As shown in Fig.~\ref{Fig.stationary-dynamic-charging}, Charging technology can be broadly categorized into wired charging and wireless charging \cite{bai2022charging}. Wired charging involves a crucial aspect wherein electrical power is directly transmitted to the EV battery by means of a connected cable \cite{hemavathi2022study,chen2020review,EVStationTrend}. 
Conversely, wireless charging represents a nascent technology that offers the convenience of transmitting energy without the reliance on traditional physical cables \cite{sathik2022comprehensive}.

\subsubsection{\textbf{Overview of Wired Charging Technologies}}

Wired charging represents the standard modality for EV energy replenishment~\cite{hemavathi2022study}. 
In particular, fast charging (power > 50 kW) enables rapid charging to 80\% capacity within 20–30 minutes, significantly improving user convenience~\cite{rahman2016review,kettles2015electric} of EV services. 
Hu et al. \cite{hu2014new} introduce a novel three-phase switch reluctance motor drive with integrated charging capability, designed specifically for plug-in hybrid electric vehicles (PHEVs), which not only enables rapid charging but also offers advantages such as cost-effectiveness, robustness, and reliability. 
Sbordone et al. \cite{sbordone2015ev} discuss two main fast charging modes for EVs: Mode 3 and Mode 4. Mode 3 uses AC charging via on-board chargers and the Mennekes protocol, with charging time varying by conditions. Mode 4 adopts DC charging with off-board chargers and the CHAdeMO protocol, typically delivering an 80\% charge in 20–30 minutes, though actual duration depends on current (125A) and voltage (500V) limits of the connector.
Moreover, Chen et al. \cite{chen2020enabling} introduce the ultra-fast charging technology as an innovative solution to mitigate the time required for recharging EVs relative to refueling conventional gasoline vehicles. 
Overall, advancements in wired charging technologies play a pivotal role in enhancing the practicality, scalability, and mass adoption of EVs by addressing the energy replenishment issues.

\subsubsection{\textbf{Overview of Wireless Charging Technologies}}

Wireless charging technology for EVs refers to the charging technology that transfers electricity from a charger to an EV through electromagnetic induction or radiation \cite{wu2011review}. Unlike traditional wired charging methods, wireless charging technology eliminates physical interfaces such as connectors and plugs, making it more convenient to charge EVs. 
Wireless charging technology typically consists of two parts: a transmitter (charger) and a receiver (device being charged). The transmitter transmits energy by generating a high-frequency alternating magnetic field or high-frequency radiation field, while the receiver obtains and converts energy by sensing or receiving the radiation field.

The advancement of wireless charging technology categorizes it into two types: static wireless charging and dynamic wireless charging. In the case of static wireless charging, a transmitter coil is affixed to the parking pad, while a receiver coil is installed on the EV. Consequently, the vehicle can be conveniently charged during stationary periods. On the other hand, dynamic wireless charging remains relatively nascent \cite{hemavathi2022study}, with trials solely conducted in select countries and regions. In this scenario, a transmitter is typically positioned on the road and comprises an arrangement of underground coils operated by high-frequency AC power to generate a magnetic field that varies with time.

Mohamed et al. \cite{mohammed2021comprehensive,mohamed2022comprehensive} present a comprehensive mathematical model for a wireless charging system with far-field charging as a wireless charging that caters to both dynamic and static modes of operation for EVs. 
Machura et al. \cite{machura2019critical} outline several challenges that persist within the domain of wireless charging technology, encompassing low transmission efficiency, complicated coil design, foreign object detection, safety concerns, and limited charging power.  
Currently, wireless charging technology is increasingly employed in specific scenarios such as car charging in parking lots, gas stations, and public roads, ultimately furnishing convenient charging services to EV owners. However, despite these advancements, the adoption of wireless charging technology still encounters challenges in high-power and long-distance applications due to factors like low power transmission efficiency, distance constraints, and interference issues.

\begin{figure}[htbp]
\centering
    \begin{minipage}[t]{0.99\linewidth}      
        \includegraphics[width = \linewidth]{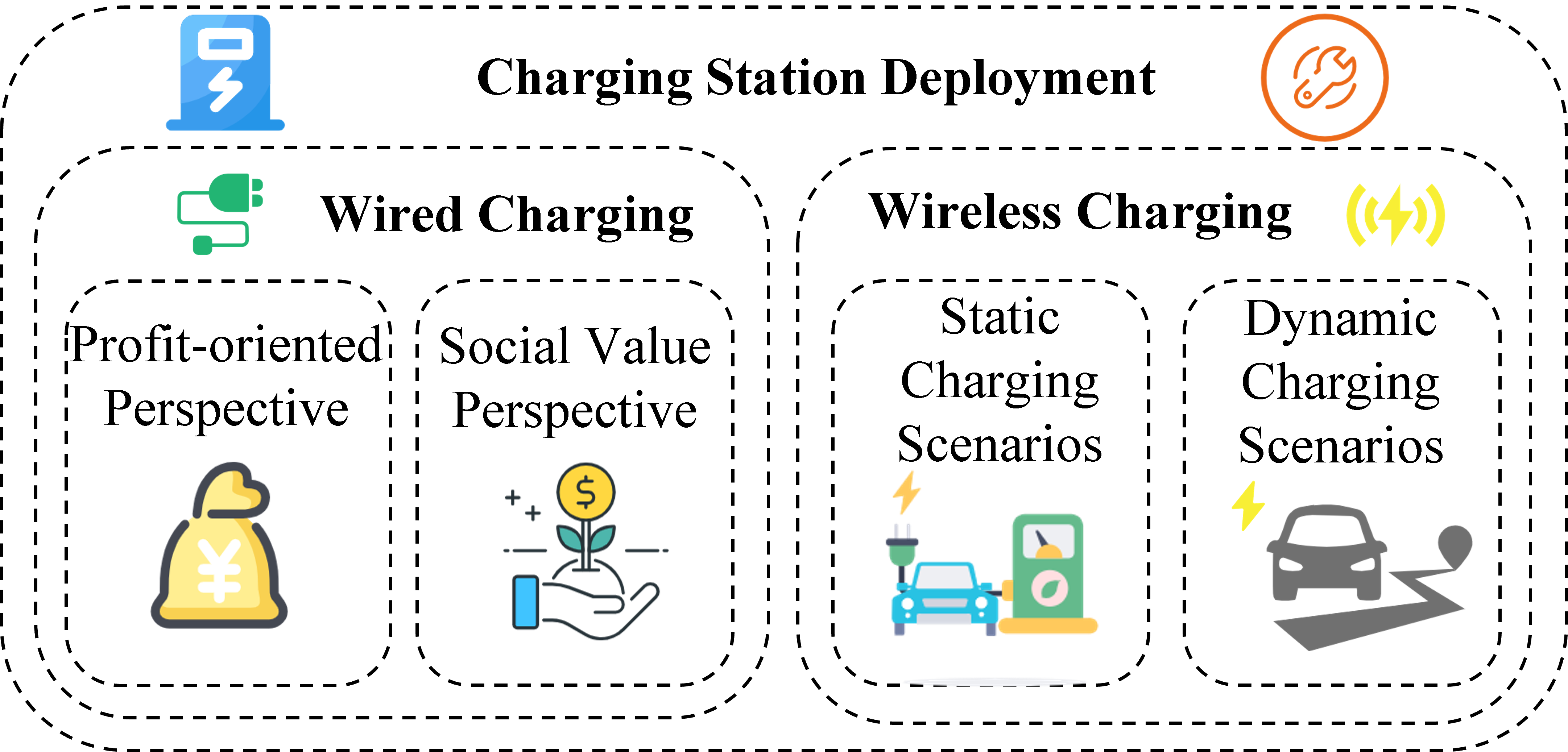}
        \caption{Classification of charging station deployment}
        \label{Fig.dep-charging}
    \end{minipage}
\end{figure}

\subsection{Wired Charging Station Deployment} 
\label{Sec3-Wired Charging Station Deployment}   
The deployment and optimization of wired EV charging stations have become a focal point in urban infrastructure planning, particularly as fast-charging technologies continue to mature and gain widespread adoption. 
As shown in Table~\ref{tb: Wired Charging Station Deployment for EV}, a significant body of literature has emerged focusing on various aspects of charging station deployment, including cost modeling, power network integration, environmental considerations, and user behavior. 
These studies can be broadly categorized into (i) profit-oriented strategies, (ii) power-grid-aware models, (iii) socially beneficial models, and (iv) data-driven learning-based approaches. 

\subsubsection{\textbf{Profit-oriented Strategies}} 
One of the key research directions in charging station deployment concerns economic viability and return on investment. 
Albert et al.~\cite{lam2014electric} formulate the EV charging station placement problem as a cost-minimization optimization model that accounts for both coverage range and driver convenience. 
They prove its NP-completeness and propose several solution methods, including mixed-integer linear programming, greedy algorithms, and chemical reaction optimization. 
Similarly, Bae et al.~\cite{bae2020game} develop a bi-level optimization framework from the perspective of investor profits in competitive markets, incorporating game theory and k-means clustering to balance station deployment and retail pricing while also considering user preferences and congestion levels. Zhao et al.~\cite{zhao2020deployment} further explore this profit-driven approach by formulating fast-charging station deployment as a nonlinear integer programming problem, solved via a heuristic genetic algorithm. 

\subsubsection{\textbf{Power-grid Aware Models}} 
Another important line of existing works focuses on the technical challenges associated with integrating EV charging stations into existing power distribution networks. 
Mirzaei et al.~\cite{mirzaei2015probabilistic} use probabilistic models and point estimation techniques to simulate EV driving patterns and optimize the capacity and location of parking lots within the grid. Arias et al.~\cite{arias2017robust} address the stochastic nature of EV demand by proposing a robust mixed-integer linear programming model that ensures service availability within a given confidence level. 
Zhang et al.~\cite{zhang2016pev,zhang2017second} and Wang et al.~\cite{wang2018coordinated} extend this work by jointly optimizing the transportation and power networks, enabling coordinated planning of charging stations, transportation lanes, and distribution lines. 
Ahmad et al.~\cite{ahmad2023placement} further incorporate renewable energy integration, battery storage systems, and vehicle-to-grid technology into their model.

\subsubsection{\textbf{Social Benefits-aware Models}} 
In addition to economic and technical considerations, some studies emphasize the public service function of EV charging stations and aim to maximize societal benefits. Li et al.~\cite{li2015growing} analyze large-scale taxi trajectory data to minimize the distance and waiting time for EV charging. He et al.~\cite{he2016incorporating} take an environmental perspective, integrating factors such as population density, land use, and spatial constraints into their site selection process. Liu et al.~\cite{liu2016optimal,liu2017locating} propose multi-layer optimization models that consider the perspectives of investors, drivers, government regulators, and road users, aiming to reduce social costs and enhance service quality. Guo et al.~\cite{guo2024approach} build upon this by introducing a personalized driver anxiety model and generating realistic charging demand scenarios through Monte Carlo simulation. 
Zheng et al.~\cite{zheng2024effects} explore the broader economic impact on nearby businesses of EV charging station deployment, highlights EVCS as drivers of local economic growth and stresses the economic benefits of multi-host EVCS setups.

\subsubsection{\textbf{Learning-based Models}} 
More recently, researchers have increasingly adopted data-driven and intelligent optimization techniques to better capture real-world uncertainties and dynamic behaviors. Liu et al.~\cite{liu2023placement} introduce a charging demand force model based on attraction and repulsion forces and apply a Markov Decision Process to dynamically optimize mobile charging station deployment, significantly improving charging success rates and reducing delays. Liu et al.~\cite{liu2024random} combine point and flow demand models with reinforcement learning-based path planning and an improved multi-population genetic algorithm to address driver behavior uncertainty across multiple planning stages. Tang et al.~\cite{tang2025stochastic} propose a stochastic Markov (dis)charging behavior model that incorporates personal features, state of charge, electricity price, and bidirectional charging station locations. Using a hybrid multi-objective optimization algorithm, they demonstrate the financial and operational benefits of energy trading, especially when ultracapacitors are integrated. Zhao et al.~\cite{zhao2023optimal} propose a reinforcement learning framework using a recurrent neural network with an attention mechanism to optimize the deployment of EV fast charging stations, aiming to maximize quality of service while addressing computational challenges in high-dimensional optimization.

\begin{table*}[h]
\label{private}
\caption{Wired Charging Station Deployment for EVs}
\fontsize{6}{10}
\selectfont
\centering
\label{tb: Wired Charging Station Deployment for EV}
\resizebox{\textwidth}{!}{
\begin{tabular}{cccccc}
\hline
\multirow{2}{*}{Paper} & \multirow{2}{*}{Year} & \multirow{2}{*}{Model} & \multicolumn{3}{c}{Optimization Objective}                                               \\ \cline{4-6} 
    & & & deployment costs & revenues & service quality \\ \hline
Lam et al.~\cite{lam2014electric} & 2014 & MILP and heuristic algorithm  &\checkmark &  & \\ \hline
Mirzaei et al.~\cite{mirzaei2015probabilistic}  & 2015 & probabilistic approach  &\checkmark &  & \\ \hline
Arias et al.~\cite{arias2017robust}  & 2017 & genetic algorithm &\checkmark &  & \\ \hline
Zhang et al.~\cite{zhang2016pev} & 2016 & mixed-integer linear programming &\checkmark &  & \\ \hline
Zhang et al.~\cite{zhang2017second}  & 2017 & mixed-integer linear programming &\checkmark &   & \\ \hline
Gan  et al.~\cite{gan2020fast}  & 2020 & heuristic algorithm & & \checkmark & \\ \hline
Zhao et al.~\cite{zhao2020deployment}  & 2020 & mixed-integer linear programming & & \checkmark & \\ \hline
Bae et al.~\cite{bae2020game} & 2020 &game theoretical and k-means clustering &  & \checkmark & \\ \hline
Liu et al.~\cite{liu2016optimal}  & 2016 & a bilevel optimization model &  & \checkmark & \checkmark \\ \hline
Liu et al.~\cite{liu2017locating}  & 2017 & tri-level programming & & \checkmark & \checkmark\\ \hline
Du et al.~\cite{du2018demand} & 2018 & greedy algorithm &  &  & \checkmark \\\hline
Sun et al.~\cite{sun2021data}  & 2021 & greedy algorithm &  &  & \checkmark \\\hline
Vazifeh et al.~\cite{vazifeh2019optimizing} & 2019 & greedy search and ant colony algorithms &  & \checkmark & \\ \hline
Liu et al.~\cite{liu2019social}  & 2019 & heuristic algorithm & & \checkmark & \\ \hline
Von et al.~\cite{von2022reinforcement}  & 2022 &  reinforcement learning &\checkmark &  & \checkmark \\ \hline
Liu et al.~\cite{liu2023placement}  & 2023 &  Markov Decision Process &  & \checkmark & \checkmark \\ \hline
Zhao et al.~\cite{zhao2023optimal}  & 2023 &  reinforcement learning &  &  & \checkmark \\ \hline
Ahmad et al.~\cite{ahmad2023placement}  & 2023 &  optimization model & \checkmark &  &  \checkmark \\ \hline
Guo et al.~\cite{guo2024approach}  & 2024 &  simulated annealing algorithm & \checkmark &  & \checkmark \\ \hline
Liu et al.~\cite{liu2024random}  & 2024 &  multipopulation genetic algorithm &  &  & \checkmark \\ \hline
Tang et al.~\cite{tang2025stochastic}  & 2025 &  hill climbing and particle swarm optimization & \checkmark &  & \checkmark \\ \hline
\end{tabular}
}
\end{table*}

\subsection{Wireless Charging Infrastructure Deployment}
Wireless power transmission offers a convenient and flexible alternative to traditional wired charging for EVs, with the added advantage of enabling charging while vehicles are in motion~\cite{sanguesa2021review}. However, its widespread adoption is currently hindered by high deployment costs and the fact that the technology remains in its early stages of development~\cite{li2018electric}. 
As shown in Fig.~\ref{Fig.dep-charging}, this section focuses on the deployment strategies of wireless charging stations, considering both static charging scenarios and dynamic charging scenarios (summarized in Table~\ref{tb: Wireless Charging Station Deployment for EV}). 

Static wireless charging serves as a complementary solution to wired charging by extending the effective range of EVs. For example, Yan et al.~\cite{yan2018employing} propose a station-based wireless charging system specifically designed for taxis. By strategically placing wireless charging pads at locations where taxis frequently stop—such as taxi stands or passenger pickup zones—the system aims to reduce idle cruising time and the time drivers spend searching for available chargers.
On the other hand, dynamic wireless charging introduces new challenges and opportunities, particularly in determining optimal placement of charging infrastructure within road networks. 
Riemann et al.~\cite{riemann2015optimal} introduce the Flow Refueling Location Model (FRLM) with the objective of maximizing the volume of captured traffic flow that can be effectively served by deployed charging infrastructure. 
Yan et al.~\cite{yan2017catcharger} employ entropy minimization clustering to identify candidate locations that exhibit characteristics such as low vehicle speed and high access frequency—ideal conditions for efficient energy transfer.
Li et al.~\cite{li2019wireless} present a bi-objective optimization model that simultaneously maximizes the energy gain obtained by a vehicle passing through a landmark while minimizing the associated traffic delays.

In a related effort focused on public transportation, 
Elmeligy et al.~\cite{elmeligy2024optimal} present a comprehensive methodology for optimizing the deployment of dynamic wireless charging (DWC) lanes alongside distributed generation resources within urban road networks. 
Bai et al.~\cite{bai2022robust} propose a wireless power transfer (WPT) system tailored for electric buses. They develop a mixed-integer programming model combined with a robust optimization approach to jointly optimize WPT locations, charging lane lengths, bus allocation, and battery size. This ensures reliable operation even under uncertain energy consumption and charging conditions.

\begin{table*}[htbp]
\fontsize{6}{9}
\selectfont
\caption{Wireless Charging Station Deployment for EVs}
\centering
\label{tb: Wireless Charging Station Deployment for EV}
\resizebox{\textwidth}{!}{
\begin{tabular}{cccccc}
\hline
\multirow{2}{*}{Paper} & \multirow{2}{*}{Year} & \multirow{2}{*}{Model} & \multicolumn{3}{c}{Optimization Objective}                                               \\ \cline{4-6} 
    & & & deployment costs & revenues & service quality \\ \hline
Riemann et al.~\cite{riemann2015optimal} & 2015 & mixed-integer nonlinear program & &\checkmark  & \\ \hline
Yan et al.~\cite{yan2017catcharger}  & 2017 & entropy minimization clustering method &\checkmark &   & \\ \hline
Yan et al.~\cite{yan2018employing}  & 2018 & multi-objective optimization   & & \checkmark & \checkmark\\ \hline
Li et al.~\cite{li2019wireless}  & 2019 & genetic algorithm and particle swarm optimization &\checkmark & &\checkmark \\  \hline
Yan et al.~\cite{yan2021catcharger}  & 2021 & multiobjective optimization & & &\checkmark \\  \hline
Bai et al.~\cite{bai2022robust}  & 2022 & mixed integer programming &\checkmark & & \\  \hline
Elmeligy et al.~\cite{elmeligy2024optimal}  & 2024 & combination  optimization &\checkmark & & \\  \hline
\end{tabular}
}
\end{table*}

\section{Charging Resource Optimization}\label{Sec4-Charging Schedule}

\begin{figure}[htbp]
\centering
    \includegraphics[width = 0.9\linewidth]{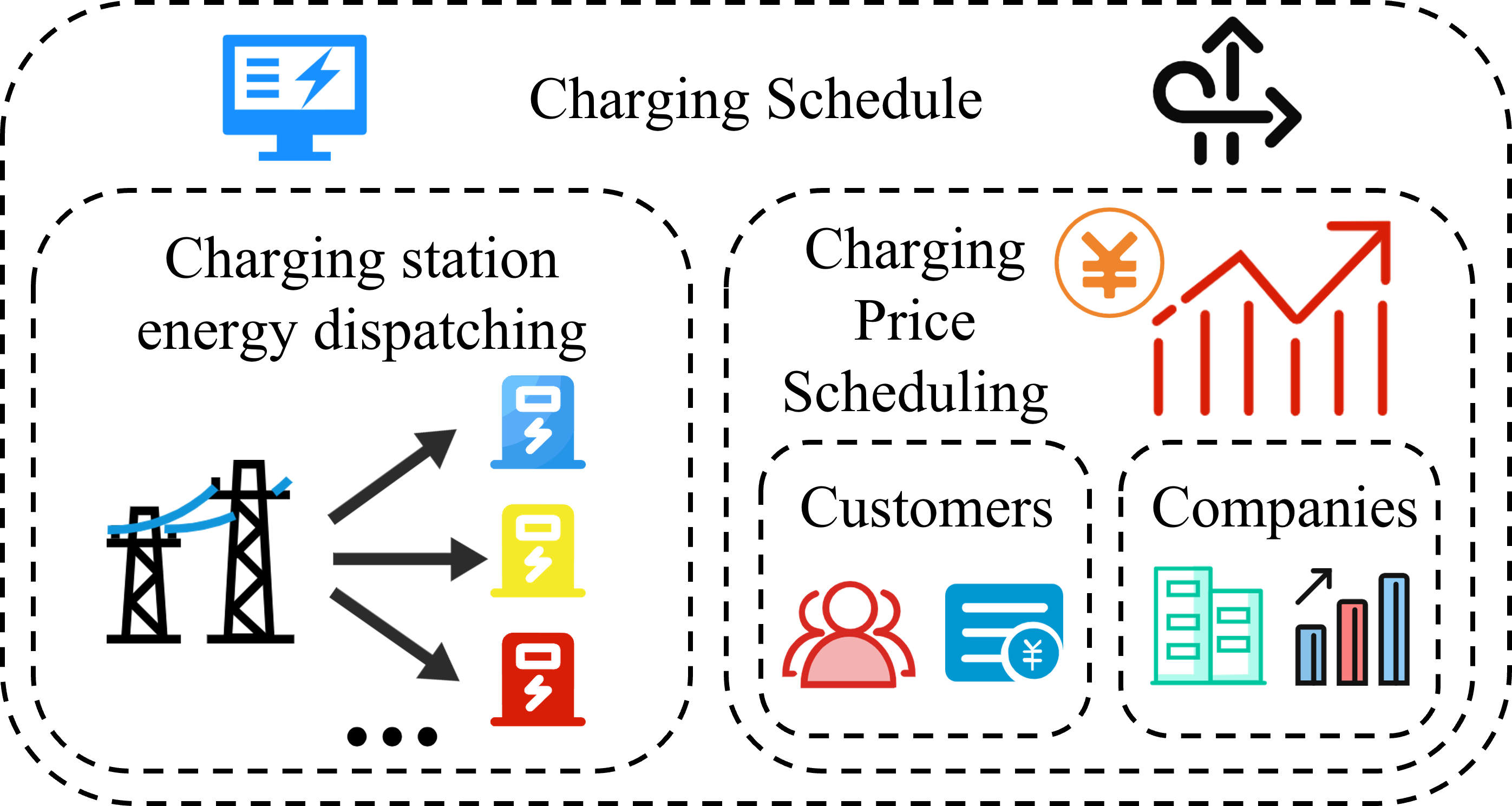}
    \caption{Classification of EV charging schedules}
    \label{Fig.Charging Schedule}   
\end{figure}

In recent years, the surge in EVs has brought significant challenges to charging infrastructure and scheduling. Researchers actively explore the latest developments in energy scheduling and scheduling algorithms for charging stations, with a focus on optimizing models, demand response mechanisms, and real-time charging scheduling systems for EVs. At the same time, they are also committed to researching various charging price scheduling strategies to optimize grid usage and reduce energy costs for consumers, with a particular focus on pricing strategies for customers and companies. Fig.~\ref{Fig.Charging Schedule} illustrates a broad classification of EV charging schedules.

\subsection{Charging Station-level Energy Dispatching} 
This section reviews recent advances in charging station energy dispatching for EVs, highlighting key research directions including (i) network-level optimization, (ii) parking-based charging strategies, (iii) demand response mechanisms, (iv) real-time scheduling systems, and (v) battery management under environmental constraints. These aspects collectively address the operational and systemic challenges of efficient EV energy replenishment.

\subsubsection{\textbf{Network-level optimization}}
To optimize the operation of the charging network, 
Yan et al. \cite{yan2022mobicharger} proposed a pioneering mobile wireless charger guidance system known as MobiCharger. By determining the optimal route and number of mobile energy devices provided, this innovative approach effectively satisfies the continuous EV operation requirements on extensive road networks. In contrast, Gusrialdi et al. \cite{gusrialdi2014scheduling} developed a higher-level distributed scheduling algorithm that focuses on adjusting the percentage of EVs charged at each station. This adjustment enhances the utilization of charging resources while minimizing the total waiting time for EV owners.

\subsubsection{\textbf{Parking-based charging strategies}} 
Parking-based charging strategies~\cite{kong2016line,iversen2014optimal} aim to optimize the charging resources with parking behaviors~\cite{zhu2020sparking} and decisions.
Building upon the optimization of charging networks, Kong et al. \cite{kong2016line} and Iversen et al. \cite{iversen2014optimal} direct their efforts towards optimizing the charging process for parked EVs. 
Kong et al. \cite{kong2016line} introduced the concept of parking charging, enabling EV owners to simultaneously park and charge their vehicles in designated parking lots. This study emphasizes the importance of characterizing charging loads using a metric model to enhance the management of charging demand. In contrast, Iversen et al. \cite{iversen2014optimal} propose a stochastic dynamic programming model that optimizes EV charging while accounting for the inherent uncertainty in their usage patterns. This approach underscores the necessity of considering the randomness of driving modes when planning charging schedules.

\subsubsection{\textbf{Demand response mechanisms}}
Demand response strategies serve as a means to effectively manage the high demand for charging EVs while leveraging the flexibility of demand management. 
To address the challenges associated with peak consumption reduction and the optimization of EV charging within intelligent distribution networks. 
Zhao et al. \cite{zhao2013peak} propose the use of an aggregator to oversee a large number of EV charging tasks, with the ultimate objective of minimizing peak consumption through the sequential planning of charging schedules. 
Alternatively, Shaaban et al. \cite{shaaban2014real} present an online coordination method for charging plug-in electric vehicles (PEVs). The primary goal of this method is to optimize the charging process while ensuring compliance with power system constraints. 
Duan et al.~\cite{duan2025study} propose an integrated energy system with mobile charging stations that uses mobile charging vehicles to enhance energy management flexibility and efficiency. The model optimizes energy scheduling for multiple events, such as supplying and absorbing energy within a park and serving EV customers, resulting in significant cost reductions and improved utilization of photovoltaic energy.
Bokopane et al.~\cite{bokopane2024optimal} propose an optimized model for a solar-grid-electric vehicle charging station with battery storage and peer-to-peer energy sharing, aiming to enhance reliability, profitability, and cost efficiency.

\begin{table*}[h]
\label{private}
\caption{Charging station energy dispatching}
\fontsize{13}{24}
\selectfont
\centering
\resizebox{\textwidth}{!}{
\begin{tabular}{cccccc}
\hline
\multirow{2}{*}{Paper} & \multirow{2}{*}{Year} & \multirow{2}{*}{Methodology} & \multicolumn{3}{c}{Optimization Objective}                                               \\ \cline{4-6} 
    & & & Charging Wait Time & Charging Dispatching & Rechargeable Battery \\ \hline
Zhao et al. \cite{zhao2013peak} & 2013 &  On-line Scheduling Algorithms & \checkmark & \checkmark &  \\ \hline
Shaaban et al. \cite{shaaban2014real} & 2014 &  On-line Scheduling Algorithms & \checkmark & \checkmark &  \\ \hline
Iversen et al. \cite{iversen2014optimal} & 2014 &  Stochastic Dynamic Programming Model & \checkmark & \checkmark &\checkmark \\ \hline
Mou et al. \cite{mou2014decentralized} & 2014 &   Decentralized Water-filling-based Algorithm & \checkmark &  & \\ \hline
Kim et al. \cite{kim2014real} & 2014 &  Adaptive Algorithm & \checkmark &  &\checkmark \\ \hline
Gusrialdi et al. \cite{gusrialdi2014scheduling} & 2014 & Distributed Scheduling Algorithm & \checkmark & \checkmark &  \\ \hline
Kong et al. \cite{kong2015distributed} & 2015 &  Distributed Scheduling Algorithm & \checkmark & \checkmark & \\ \hline
Kim et al. \cite{kim2015modeling} & 2015 &  Real-time Scheduling Algorithm & \checkmark & \checkmark & \\ \hline
Kong et al. \cite{kong2016smart} & 2016 & Approximation Algorithms & \checkmark & \checkmark & \\ \hline
Kong et al. \cite{kong2016line} & 2016 & On-line Scheduling Algorithms & \checkmark &  & \checkmark\\ \hline
Dong et al. \cite{dong2017rec} & 2017 &  Real time EV charging scheduling & \checkmark & \checkmark & \\ \hline
Yang et al. \cite{yang2017charging} & 2017 &  Optimal Charging Scheduling Algorithm & \checkmark & \checkmark & \\ \hline
Du et al. \cite{du2018demand} & 2018 &  An effective greedy approximation algorithm & \checkmark & \checkmark & \\ \hline
Wang et al. \cite{wang2019tcharge} & 2019 &  Online Scheduling & \checkmark & \checkmark & \\ \hline
Li et al. \cite{li2019data} & 2019 &  Battery-lifetime-aware Bus Scheduling Strategy & \checkmark & \checkmark &  \\ \hline
Wang et al. \cite{wang2019sharedcharging} & 2019 &  Heterogeneous EV fleet charging & \checkmark & \checkmark & \\ \hline
Can et al. \cite{cao2021online} & 2021 & Bi-objective Optimization Models & \checkmark & \checkmark & \\ \hline
Yan et al. \cite{yan2022mobicharger} & 2022 & On-line Scheduling Algorithms & \checkmark &  & \checkmark \\ \hline
Bokopane et al.~\cite{bokopane2024optimal} & 2024 & Mathematical Model &   &  & \checkmark \\ \hline
Duan et al.~\cite{duan2025study} & 2025 & Mixed-Integer Linear Programming &  & \checkmark &  \\ \hline
\end{tabular}
}
\end{table*}

\subsubsection{\textbf{Real-time charging scheduling}}
In response to the aforementioned challenges in EV charging, researchers have developed real-time charging scheduling systems. Du et al. \cite{du2018demand}, Wang et al. \cite{wang2019tcharge, wang2019sharedcharging}, and Dong et al. \cite{dong2017rec} independently devise real-time charging scheduling systems with the aim of maximizing efficiency while minimizing wait times. Du et al. \cite{du2018demand} specifically focus on the EV charger planning (EVCP) problem in the context of EV sharing, demonstrating its NP-hardness and proposing an approximation algorithm with a theoretical bound of 1 - 1/e. Wang et al. \cite{wang2019tcharge} introduce tCharge, a fleet-oriented real-time charging scheduling system that utilizes historical GPS data to estimate travel times to various charging stations, and real-time GPS data to infer waiting times at these stations. Subsequently, their study \cite{wang2019sharedcharging} concentrates on a universal real-time shared charging scheduling system, named sharedcharging, which aims to enhance the charging efficiency of heterogeneous EV fleets. This system considers real-world constraints such as station space, charging point availability, and real-time schedule guarantees. Meanwhile, Dong et al. \cite{dong2017rec} develop REC, a real-time EVs charging scheduling framework that caters specifically to EV taxi fleets. REC informs drivers about optimal times and locations for charging their batteries during operations, ensuring predictable waiting times and balanced utilization of charging stations.   

\subsubsection{\textbf{Battery management operations}}
Another critical aspect of EVs research involves battery management systems (BMS). The challenges in EVs' charging infrastructure include demand-supply imbalances, limited battery life, and the impact of environmental factors on charging efficiency.
Kim et al. \cite{kim2014real, kim2015modeling} devise an effective and robust BMS that ensures the protection and efficient power provision for EV batteries. Furthermore, they develop a real-time integrated system for managing battery discharge rates and temperatures.
Li et al. \cite{li2019data} conduct a comprehensive analysis of passenger behavior and electric bus travel, designing an accurate passenger load prediction model that considers external factors such as weather and special holidays. Yang et al. \cite{yang2017charging} explore the optimal charging scheduling for wireless charging electric buses (WCEBs), considering dynamic electricity demand and price fluctuations to minimize the power cost of system operation. To further enhance the coordination of the charging process, Cao et al. \cite{cao2021online} model the online coordinated charging plan for EVs as a dual-objective optimization problem, proposing two heuristic online charging coordination algorithms that balance charging costs and future maintenance costs.

\subsection{Charging Price Scheduling} 

In recent years, the growing popularity of EVs has prompted researchers to explore various charging price scheduling strategies to optimize grid usage and minimize energy costs for consumers. These studies have focused on pricing strategies for both customers and companies, aiming to develop efficient and effective solutions for handling the increasing demand for EV charging.

\subsubsection{Pricing Strategies for Customers}
Real-time electricity price models offer economic and environmental advantages compared to the current universal unified electricity price. Monhsenian et al. \cite{mohsenian2010optimal} propose an automatic optimization control scheme for residential load in the retail electricity market based on a combination of real-time pricing (RTP) and inclining block rates (IBR). This approach encourages consumers to use electricity more efficiently, while reducing their overall energy costs.

Building on this idea, the vehicle-to-grid (V2G) system enables energy to flow from EVs to the grid, providing additional opportunities for customers to benefit from real-time pricing. Shi et al. \cite{shi2011real} study the real-time V2G control problem under price uncertainty and propose a new V2G control algorithm that learns from past experience, automatically adapts to unknown pricing information, and makes optimal hourly control decisions. This solution allows EV owners to capitalize on fluctuations in electricity prices while ensuring grid stability.

In an attempt to balance the total power generation cost with local costs related to overload and battery degradation, Ma et al. \cite{ma2016efficient} establish a plug-in electric vehicle (PEV) charging coordination framework. By addressing the trade-offs between individual consumer costs and overall grid stability, this framework aims to optimize the charging process for both the consumer and the power grid. The decentralized approach involves minimizing the charging cost of each PEV relative to the predicted price curve, while considering the impact of local power grids and batteries.

\subsubsection{Pricing Strategies for Companies}
The escalating proliferation of EVs results in a substantial surge in charging demand, posing a considerable challenge in terms of meeting this demand. In order to address this issue, Zou et al. \cite{zou2014cost} embark on a comprehensive evaluation of two distinct pricing schemes. One of these pricing schemes is predicated on real-time energy prices, while the other relies on basic prices provided by utility companies in conjunction with functional prices that are contingent upon each household's energy consumption/supply. The underlying objective of both strategies is to incentivize more efficient utilization of electricity, while concurrently ensuring that charging stations are capable of generating revenue.

Recognizing the crucial need for optimizing the performance of charging stations, Fan et al. \cite{fan2020enabling} proffer a dynamic charging pricing mechanism that actively governs the length of demand queues for multiple charging stations. This mechanism is designed to maximize the long-term profitability of charging services provided by charging platforms. By taking into account the elasticity of EV charging demand in response to price fluctuations, this approach endeavors to strike a fine balance between the requirements of EV owners and charging service providers.

Building upon this notion, Zhang et al. \cite{zhang2018optimal} expound upon the concept by devising an optimal pricing scheme that aims to minimize the decline rate of service provision at charging stations. Through a meticulous analysis of the correlation between the service decline rate of charging stations and the selection of EVs, the researchers employ this insight to formulate a customer churn minimization problem. Their proposed optimal pricing technique effectively guides and orchestrates the charging process for EVs at charging stations, thereby guaranteeing a more streamlined and efficient charging experience for customers.

In their study, Sarker et al. \cite{sarker2017opportunistic} propose a game theory-based distributed power dispatching framework that integrates game theory into pricing strategies to optimize the dispatching between online electric vehicles (OLEVs) and the smart grid. By establishing a competitive environment, this framework facilitates mutually beneficial decision-making concerning charging and discharging activities. Within this framework, OLEVs receive charging from the smart grid based on the best response strategy and an updated power payment function. As a result, resources are allocated more efficiently, contributing to grid stability. 

Addressing the challenges associated with spatiotemporal charging load, Wang et al. \cite{wang2020pricing} put forward an innovative price adjustment system. This system takes into account the charging and consumption behavior of drivers and offers location information on charging stations, real-time queuing updates, and customized charging fees while considering the revenue of the charging station company. By incorporating both spatial and temporal aspects of charging demand, this approach ensures enhanced utilization of charging stations and a superior charging experience for EV owners. 

Kazemtarghi et al.~\cite{kazemtarghi2024dynamic} propose a dynamic pricing strategy for electric vehicle charging stations to reduce congestion and improve both operator revenue and user satisfaction. Ge et al.~\cite{ge2024distributed} propose a two-stage, distributed charging pricing model for electric vehicles based on game theory and the TAP-UE model, aiming to optimize energy, transportation, and information network interactions.

To summarize, the existing literature on charging price scheduling for EVs has examined diverse strategies to optimize energy costs for consumers, minimize service decline rates for charging stations, and maintain grid stability. By employing real-time pricing, vehicle-to-grid systems, game theory, and spatiotemporal considerations, researchers have developed innovative solutions to meet the emerging requirements in EV charging. However, further research is necessary to refine these strategies and establish even more efficient and effective approaches as the adoption of EVs continues to grow.

\begin{table*}[h]
\label{private}
\caption{Charging Price Scheduling}
\fontsize{10}{16}
\selectfont
\centering
\resizebox{\textwidth}{!}{
\begin{tabular}{ccccc}
\hline
\multirow{2}{*}{Paper} & \multirow{2}{*}{Year} & \multirow{2}{*}{Methodology} & \multicolumn{2}{c}{Optimization Objective}                                               \\ \cline{4-5} 
    & & & Customer's perspective & Company perspective \\ \hline
Mohsenian et al. \cite{mohsenian2010optimal} & 2010 &  Realtime Pricing  &  \checkmark &   \\ \hline
Shi et al. \cite{shi2011real} & 2011 & Markov Decision Process   &   \checkmark &   \\ \hline
Zou et al. \cite{zou2014cost} & 2014 &  Nonlinear Programming  &  & \checkmark    \\ \hline
Ma et al. \cite{ma2016efficient} & 2016 &  PEV Charging Coordination Framework  &      \checkmark  &   \\ \hline
Sarker et al. \cite{sarker2017opportunistic} & 2017 &     Optimal scheduling between OLEV and smart grid   &  & \checkmark\\ \hline
Zhang et al. \cite{zhang2018optimal} & 2018 &  Based on queuing theory &  & \checkmark   \\ \hline
Fan et al. \cite{fan2020enabling} & 2020 & Lyapunov stochastic optimization &  & \checkmark   \\ \hline
Wang et al. \cite{wang2020pricing} & 2020 & Handling spatiotemporal load   &  & \checkmark   \\ \hline
Ren et al.~\cite{ren2023electric}  & 2023 & Reinforcement Learning   & \checkmark  & \checkmark   \\ \hline
Kazemtarghi et al.~\cite{kazemtarghi2024dynamic}  & 2024 & Scenario-based Stochastic Optimization   &   & \checkmark   \\ \hline
Ge et al.~\cite{ge2024distributed}  & 2024 &   Coupled Network &   & \checkmark   \\ \hline
\end{tabular}
}
\end{table*}

\section{Vehicle-Level Charging Scheduling}
\begin{figure}[htbp]
\centering
    \includegraphics[width = 0.9\linewidth]{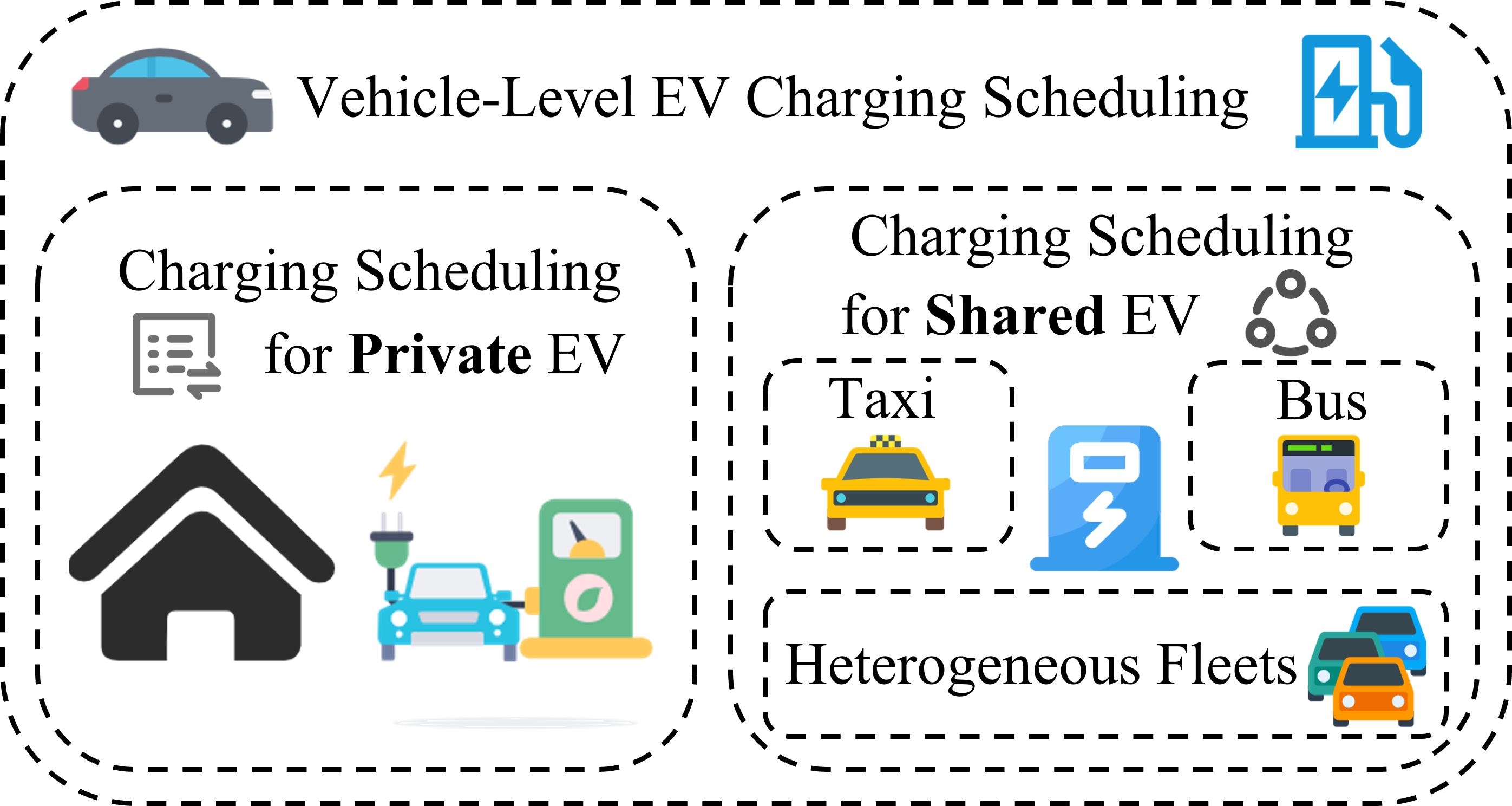}
    \caption{Classification of vehicle-level EV charging scheduling strategies}
    \label{Fig.Vehicle-Level-Charging-Scheduling}   
\end{figure}

In this section, we delve into the intricate realm of scheduling EV charging, wherein the judicious assignment or suggestion of charging stations, charging posts, and charging durations assumes paramount importance. Fig.~\ref{Fig.Vehicle-Level-Charging-Scheduling} illustrates a high-level classification of vehicle-level EV charging scheduling strategies. Our investigation, specifically, revolves around the charging scheduling paradigm for EVs, an endeavour that entails the systematic classification of vehicles into two discrete categories: private EVs and shared EVs.

\subsection{Charging Scheduling for Private EV}
When considering the issue of charging scheduling for private EVs, it is crucial to recognize that these vehicles do not need to take into account operating income. Consequently, the scheduling problem for private EVs primarily revolves around variables such as the cost of charging time, driving time, and charging expenses. 
The primary objective is to minimize the overall time and cost associated with traveling to and waiting at charging stations. 
Some approaches also incorporate the charging cost of EVs and revenue generation for charging stations. The aim is to optimize revenue generation for charging stations while ensuring that EVs have access to charging services when required. 
By integrating these factors, the charging schedule for private EVs can be optimized for both vehicle owners and charging station operators.

\subsubsection{\textbf{User-Centric Charging Scheduling}}
In the context of private EV charging schedules, a critical focus is placed on reducing both driving time to charging stations and waiting time at these stations. 
To tackle this challenge, Gusrialdi et al. \cite{gusrialdi2017distributed} proposed a dynamic model that accounts for average highway EV traffic and queues at charging stations, which takes the form of an aggregated averaging model in the discrete-time domain, and incorporates distributed scheduling of EV charging to minimize waiting times at charging stations. 
Another approach to private EV charging scheduling is CoordiQ \cite{blum_coordiq_2021}, which formulates the problem as a Markov decision process and employs a collaborative Q-learning reinforcement learning approach to optimize the charging process. 
Lin et al. \cite{lin2021toward} put forth a novel approach to address the charging scheduling problem by introducing a Markov Decision Process (MDP) framework. 
The authors develop an online learning-based algorithm that proves to be highly effective in solving the multi-phase MDP model, surpassing the performance of game-theoretic-based and Q-learning-based methods. 
Li et al. \cite{li_coupling_2022} propose a Regularized Actor-Critic (RAC) reinforcement learning method. This approach aims to strike a balance between user preferences, such as historical charging patterns, and external rewards, such as the trade-off between driving time and waiting time, for each private EV driver. 
Furthermore, Liu et al. \cite{liu2021reservation} center their research on devising emergency priority scheduling policies that incorporate calculations of charging urgency based on both EVs' charging demand and parking time. This charging urgency serves as an indicator for prioritizing scheduling, allowing EVs with higher urgency to charge preemptively.

It is noteworthy that previous studies on EV charging scheduling have primarily focused on optimizing the time cost of charging while neglecting the potential variations in charging prices across different charging stations. 
In order to address this limitation, recent approaches have emerged that aim to jointly optimize charging time and expenditure cost as primary objectives. Valogianni et al. \cite{valogianni_effective_2014} propose an Adaptive Management of EV Storage (AMEVS) algorithm, which utilizes a reinforcement learning framework to incorporate charging time and expenditure cost as rewards for an intelligent agent's decision-making process. This approach leads to the development of a charging schedule that maximizes the personal benefits for EV drivers.

\subsubsection{\textbf{System-Level Charging Coordination}}
In the context of large-scale charging scheduling problems involving numerous EVs and charging stations, 
Zhang et al. \cite{zhang_intelligent_2021} propose the Multi-Agent Spatio-Temporal Reinforcement Learning (Master) framework, which formulates the charging scheduling problem as a multi-objective multi-agent reinforcement learning task. 
Suanpang et al. \cite{suanpang2022intelligent} also adopt a multi-agent spatio-temporal reinforcement learning methodology to address the charging scheduling problem. Zhang et al.~\cite{zhang2022rlcharge} introduce RLCharge , which incorporates a spatio-temporal heterogeneous graph convolution module. Additionally, an adaptive imitation learning approach is proposed to facilitate the multi-objective optimization process.

Similarly, Xing et al. \cite{xing2022graph} employ a graph reinforcement learning approach to address the charging scheduling problem. Their method utilizes a graph convolutional network to extract relevant environmental information for charging scheduling, followed by the application of deep reinforcement learning to solve the problem effectively.
In addition to research focused on maximizing personal benefits for individual EV drivers, 
Wei et al.~\cite{wei2015charging} consider the uncertainties associated with renewable energy generation and aggregator capacity when formulating a generalized Nash game to address the charging problem with shared global constraints. 
Sun et al. \cite{sun_orc_2020} examine the charging scheduling problem from the perspective of charging station operators, formulating it as an online decision problem. They introduce the online recommendation and charging scheduling algorithm (ORC), which incorporates a parameterized value function for customized designs that effectively enhance the total revenue of charging stations. Additionally, Xu et al. \cite{xu2022real} conceptualize the recommendation problem as a sequential decision problem and propose a multi-objective system-level approach to fast charging station recommendations. 
Klein et al. \cite{klein2023electric} study the integrated scheduling of charging and service operations for EV fleets, considering battery degradation, nonlinear charging, and time-of-use tariffs. 
Tan et al. \cite{tan2023fair} propose a bi-objective charging and discharging scheduling approach that balances time-aware fairness and overall waiting time for EVs. The method reduces total and individual waiting times while also minimizing station operating costs, demonstrating improved fairness and efficiency in EV charging scheduling.
Yang et al. \cite{yang2023ev} propose a block model predictive control approach to optimize real-time charging scheduling for EVs under demand charge constraints, aiming to minimize peak power costs while efficiently managing deferrable charging demands. 
Li et al. \cite{li2025charging} introduce a two-stage framework for EV task assignment and charging decision-making in urban logistics, aiming to maximize platform revenue by optimizing real-time task allocation and charging scheduling. The approach combines a hybrid weight model with multi-agent reinforcement learning and hierarchical communication to improve task completion and charging efficiency.

Zhang et al. \cite{zhang2024learning} propose a model-free, online learning approach based on deep reinforcement learning to estimate and manage the aggregate flexibility of large-scale EVs in power systems, effectively handling uncertainties in regulation signals and EV behaviors. 
Qureshi et al. \cite{qureshi2024multiobjective} propose a real-time Pareto-optimal scheduling method for bidirectional EV charging in a commercial station with renewable energy and storage, using mixture density networks to model uncertain charging demands and deadlines.

\begin{table*}[htp]
\label{private}
\caption{Charging Scheduling Works for Private EV}
\centering
\fontsize{10}{18}
\selectfont
\resizebox{\textwidth}{!}{
\begin{tabular}{cccccc}
\hline
\multirow{2}{*}{Paper} & \multirow{2}{*}{Year} & \multirow{2}{*}{Model} & \multicolumn{3}{c}{Optimization Objective}                                               \\ \cline{4-6} 
    & & & Charging Wait Time & Charging Price & Charging Station Income \\ \hline
Valogianni et al. \cite{valogianni_effective_2014} & 2014 &  Adaptive Management Algorithm & \checkmark & \checkmark & \\ \hline
Wei et al. \cite{wei2015charging} & 2016 &  Normalized Nash Equilibrium & \checkmark &  & \checkmark \\ \hline
Gusrialdi et al. \cite{gusrialdi2017distributed} & 2017 &  Distributed Scheduling & \checkmark &  &  \\ \hline
Sun et al. \cite{sun_orc_2020} & 2020 & Online Competitive Algorithm & \checkmark &  & \checkmark \\ \hline
Blum et al. \cite{blum_coordiq_2021} & 2021 & Collaborative Q-Learning & \checkmark &  & \\ \hline
Liu et al. \cite{liu2021reservation} & 2021 &  Emergency Priority Scheduling & \checkmark &  & \\ \hline
Zhang et al. \cite{zhang_intelligent_2021} & 2021 &  Multi-Agent Reinforcement Learning & \checkmark & \checkmark & \\ \hline
Lin et al. \cite{lin2021toward} & 2022 &  Multiple-Phase Markov Decision Process Model & \checkmark &  & \\ \hline
Li et al. \cite{li_coupling_2022} & 2022 & Regularized Actor-Critic & \checkmark &  & \\ \hline
Zhang et al. \cite{zhang2022rlcharge} & 2022 &  Multi-Agent Reinforcement Learning & \checkmark & \checkmark & \\ \hline
Xu et al. \cite{xu2022real} & 2022 & Graph Reinforcement Learning & \checkmark &  & \checkmark \\ \hline
Suanpang et al. \cite{suanpang2022intelligent} & 2022 &  Multi-Agent Reinforcement Learning & \checkmark & \checkmark & \\ \hline
Xing et al. \cite{xing2022graph} & 2022 & Graph Reinforcement Learning & \checkmark & \checkmark & \\ \hline
Hossain et al. \cite{hossain2023efficient} & 2023 & Reinforcement Learning &  & \checkmark & \\ \hline
Klein et al. \cite{klein2023electric} & 2023 & Branch-and-price & \checkmark &  & \checkmark \\ \hline
Tan et al. \cite{tan2023fair} & 2023 & Highest Response Ratio Next strategy with a Logistic Function & \checkmark &  & \checkmark \\ \hline
Yang et al. \cite{yang2023ev} & 2023 & Block Model Predictive Control &  &  & \checkmark \\ \hline
Zhang et al. \cite{zhang2024learning} & 2024 & Reinforcement Learning &  & \checkmark &  \\ \hline
Qureshi et al. \cite{qureshi2024multiobjective} & 2024 & Stochastic Convex Optimization &  & \checkmark & \checkmark \\ \hline
Li et al. \cite{li2025charging} & 2025 & Multi-Agent Reinforcement Learning  &  &  & \checkmark \\ \hline
\end{tabular}
}
\end{table*}

\subsection{Charging Scheduling for Shared EV}

EVs are not solely confined to personal transportation; they are extensively utilized in communal mobility contexts, inclusive of bus networks, taxi services, and vehicle rental facilities. Nevertheless, the distinct charging scheduling predicaments encountered by shared EVs deviate from those faced by private EV users, predominantly due to the presence of an operator. Shared EVs can be delineated into two principal categories according to their operational modality: namely, electric taxis and electric buses.

\subsubsection{Charging Scheduling for Taxi}

\begin{figure*}[htbp]
\centering
    \includegraphics[width = 0.75\linewidth]{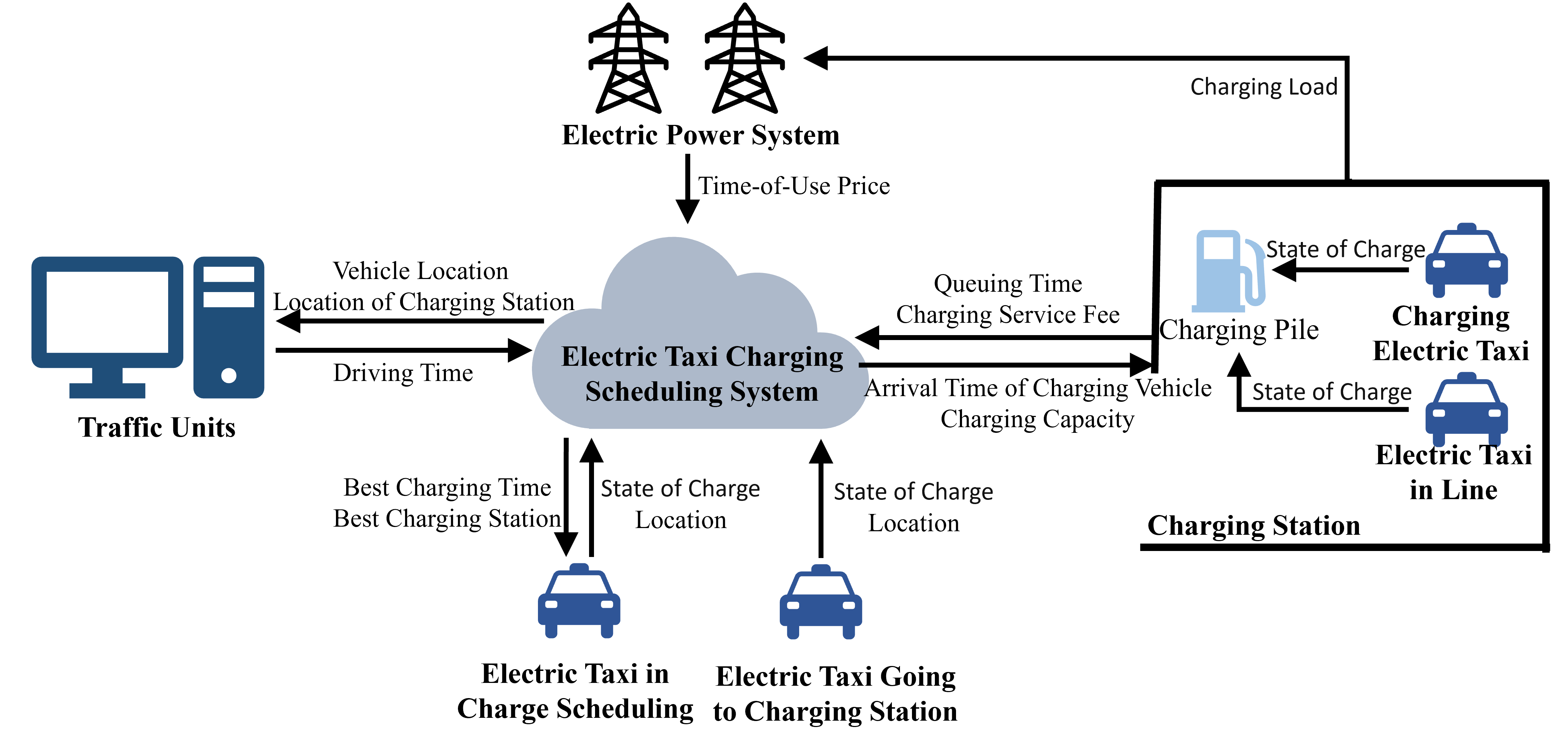}
    \caption{Electric Taxi Charging Scheduling Flow}
    \label{Fig.taxi}   
\end{figure*}

Fig.~\ref{Fig.taxi} shows the flow of electric taxis charing scheduling. The consideration of operational status is crucial in the charging scheduling for electric taxis. In their study, Tian et al. \cite{tian_real-time_2016} have developed a charging station recommendation system specifically for electric taxis. This system predicts the current operational status of electric taxis by integrating historical charging data with real-time GPS trajectory information. However, it is important to note that this system only takes into account individual taxis, without considering the fact that taxis often operate in fleets.

To address this limitation, REC \cite{dong_rec_2017} proposes a real-time EV charging scheduling framework that is specifically designed for electric taxi fleets. This framework incorporates real-time scheduling theory for multiprocessor systems. It effectively informs each driver in the fleet when and where to charge their vehicle's battery, resulting in predictable waiting times for all EVs in the fleet. Additionally, it balances the utilization of charging stations both spatially and temporally.

Another approach, tCharge \cite{wang_tcharge_2019}, employs various sensing and communication devices on the vehicle to estimate travel and wait times for electric cabs at different charging stations. By utilizing this data, tCharge enables more efficient scheduling of electric cab operations and charging. However, it should be noted that tCharge, similar to the previous approaches, does not take into account individual fairness among drivers.

To address this issue, FairCharge \cite{wang_faircharge_2020} introduces fairness as a constraint and formulates the charging scheduling problem as a Pareto optimization problem, as shown in Eq.~\ref{fair}. By doing so, it aims to design a fairness-aware and Pareto-efficient charging schedule. This approach ensures that the schedule not only maximizes efficiency but also considers individual fairness among drivers.
\begin{equation}
\label{fair}
\begin{aligned}
\underset{Q=\left[Q_{i j}\right]_{m \times n}}{\operatorname{minimize}} \mathrm{T}(\mathrm{Q}) & =\left[T_{1}\left(\mathrm{Q}_{1}\right), T_{2}\left(\mathrm{Q}_{2}\right), \cdots, T_{m}\left(\mathrm{Q}_{m}\right)\right]^{\top} \\ \text { s.t. } Q_{i} & =\mathrm{u}_{i},\left\|\mathrm{Q}_{i}\right\|=1, \forall i=1,2, \cdots, m \\ Q^{f} & =\underset{0}{\operatorname{argmin}}\left\{\max _{i}\left\{T_{i}\left(\mathrm{Q}_{\mathrm{i}}\right)-T_{i}\left(\mathrm{Q}_{\mathrm{i}}{ }^{\circ}\right)\right\}\right\}
\end{aligned}
\end{equation}

Where \(Q_i\) is the allocation vector of event \(e_i\) , so we have \(Q = \left[ Q_1, Q_2, \ldots, Q_m \right]^{\top}\) for all charging events. Since a charging event can only be recommended to one charging station, the allocation of \(Q_i\) should be the unit vector  \(u_i\). \(T_i(Q_i)\) represents the idle time spent by the charging station allocated by the charging event ei, which is the sum of travel time and queue time. \(Q_i^{\circ}\) is a utopian recommendation for charging event \(e_i\). Therefore, the charging recommendation for \(m\)  charging events in a short period of time can be formulated as a Pareto optimization problem

ForETaxi is a novel approach presented by Wang et al. \cite{wang_foretaxi_2023} that leverages multiple sensor data to optimize fleet charging from a fleet-wide perspective. By doing so, it successfully minimizes the overall charging overhead, eliminating the need for additional charging stations. In the context of taxi drivers, a key objective is to maximize operating revenue while minimizing the time spent on charging. Consequently, previous research has focused on optimizing both charging overhead and operating revenue as primary goals.

In a related study, R2E \cite{wang_r2e_2018} formulates the electric taxi charging problem as a Markov decision process (MDP), employing MDP to calculate future rewards and propose an optimized service strategy. Similarly, Tseng et al. \cite{tseng_improving_2019} utilize MDP modeling to derive an optimal electric taxi service strategy.

Nonetheless, Wang et al. \cite{wang_joint_2022} take a distinct approach by considering the perspective of electric taxi drivers in the problem formulation. They tackle the issue as a multi-agent reinforcement learning problem, proposing a multi-agent mean-field hierarchical reinforcement learning framework. This framework incorporates the interdependence among electric taxis in decision-making, facilitating joint scheduling of charging operations for drivers. As a result, not only is their charging overhead reduced but their operating revenue is also increased significantly.
Tan et al.~\cite{tan2023joint} introduce a reinforcement learning-based framework for managing shared electric micromobility systems, which incorporates energy-informed demand predictions to optimize vehicle rebalancing and charging.
Yuan et al.~\cite{yuan2025stochastic} propose a coordination framework for electric taxi fleets that integrates renewable energy forecasting to optimize charging schedules and passenger service under spatial-temporal dynamics, aiming to enhance solar power utilization and reduce power system reliability issues.

\begin{table*}[htp]
\label{private}
\caption{Charging Scheduling Works for Taxi}
\fontsize{10}{16}
\selectfont
\centering
\resizebox{\textwidth}{!}{
\begin{tabular}{cccccc}
\hline
\multirow{2}{*}{Paper} & \multirow{2}{*}{Year} & \multirow{2}{*}{Model} & \multirow{2}{*}{Fleet-Oriented} & \multicolumn{2}{c}{Optimization Objective}                                               \\ \cline{5-6} 
    & & & & Charging Overhead & Operating Revenue \\ \hline
Tian et al. \cite{tian_real-time_2016} & 2016 &  Real-time Recommendation &  & \checkmark & \\ \hline
Dong et al. \cite{dong_rec_2017} & 2017 &  Real-time Scheduling & \checkmark & \checkmark & \\ \hline
Wang et al. \cite{wang_r2e_2018} & 2018 &  Markov Decision Process Model &  & \checkmark & \checkmark\\ \hline
Wang et al. \cite{wang_tcharge_2019} & 2019 &  Real-time Scheduling & \checkmark & \checkmark &\\ \hline
Tseng et al. \cite{tseng_improving_2019} & 2019 &  Markov Decision Process Model &  & \checkmark &\checkmark \\ \hline
Wang et al. \cite{wang_faircharge_2020} & 2020 & Pareto Optimality \& Fairness Constraints & \checkmark & \checkmark & \\ \hline
Wang et al. \cite{wang_joint_2022} & 2022 & Multi-Agent Reinforcement Learning & \checkmark & \checkmark & \checkmark\\ \hline
Wang et al. \cite{wang_foretaxi_2023} & 2023 &  Data-Driven Charing Recommendation & \checkmark & \checkmark & \\ \hline
Tan et al.~\cite{tan2023joint} & 2023 & Reinforcement Learning &  & \checkmark & \\ \hline
Yuan et al. \cite{yuan2025stochastic} & 2025 &  Multi-criterion Mixed-integer Linear Programming & \checkmark & \checkmark & \checkmark \\ \hline
\end{tabular}
}
\end{table*}

\subsubsection{Charging Scheduling for Bus}

Unlike electric taxis, electric buses typically adhere to fixed schedules, with the bus company managing their profits and costs holistically. Therefore, the primary optimization objectives for the charging scheduling of electric buses involve charging time, battery aging cost, electricity load cost, and the overall bus system cost. 

bCharge \cite{wang_bcharge_2018} is a real-time charging scheduling system specifically designed for city-scale electric bus fleets. It employs the Markov decision process and takes into account fluctuating electricity prices to reduce the overall charging costs \cite{wang_bcharge_2018}. Additionally, Ma et al. \cite{ma2021optimal} approach the EV charging problem from a two-level optimization perspective and introduce an agent-assisted optimization method to minimize charging time. They demonstrate the effectiveness of their method using bus data. The advancement in charging technology has made fast charging possible during passenger boarding and alighting at bus stops, allowing the use of smaller batteries and consequent cost reduction. Hu et al. \cite{hu2022joint} propose a robust model that incorporates uncertainty in travel time and passenger demand to optimize intermediate station charging scheduling for electric buses. This model reduces additional charging time and enables the use of smaller batteries to fulfill the typical operational requirements of electric buses. 

For electric bus operators, managing battery aging costs and electricity loads resulting from charging is of utmost importance. Houbbady et al. \cite{houbbadi2019optimal} present a method to minimize battery aging costs by managing the nighttime charging of an electric bus fleet. Jahic et al. \cite{jahic2019charging,jahic_preemptive_2019} employ preemptive and quasi-preemptive greedy algorithms to minimize peak charging load. Rodrigues et al. \cite{rodrigues_optimized_2020} propose a charging scheduling approach for urban electric bus fleets that combines solar charging periods to minimize peak charging load. Duan et al. \cite{duan2022bidding} propose a trip-chain-based boundary model to accurately represent the flexible region of electric buses with multiple sequential trips. To handle large-scale problems, this model is aggregated using a state-based approach, enabling electric bus charging scheduling that accounts for energy consumption costs.

To integrate charging costs and battery aging energy load costs, some research considers total transit system costs as an optimization objective. Kang et al. \cite{kang2015centralized} propose a new centralized charging strategy for EVs in a battery swapping scenario, using population-based heuristics to minimize charging costs while accounting for power losses and voltage deviations in the electric network. Li et al. \cite{li2020joint} develop a joint optimization model that considers partial charging policies and time-of-use tariffs to minimize total transit system costs, including capital and maintenance costs of electric buses and chargers, power consumption costs, and time-related costs. A joint optimization model for scheduling and fixed charger deployment in a regularly charged electric bus transit network minimizes the total cost of the transit system. Wang et al. \cite{wang2020pricing} propose a pricing-aware, real-time charging scheduling system that utilizes a Markov Decision Process to minimize overall charging and operating costs for city-scale electric bus fleets. By taking time-variant electricity pricing into account, the model is able to effectively reduce costs. He et al. \cite{he2022integrated} propose a two-stage optimization framework for planning charging stations and scheduling charging for electric bus systems to achieve total cost reduction of the transit system.

Zeng et al. \cite{zeng2024route} propose a robust en-route charging scheduling model for electric buses that accounts for uncertain energy consumption and aims to ensure charging accessibility while minimizing costs under time-of-use electricity pricing. The model combines a deterministic mixed-integer linear program with a robust counterpart formulation, and numerical results demonstrate its effectiveness in maintaining system feasibility and cost-efficiency under uncertainty.

Zhou et al.~\cite{zhou2024electric} address the electric bus charging scheduling problem (EBCSP) by formulating it as a mixed-integer linear program and proposing both exact and heuristic solution methods, including a branch-and-price algorithm and an optimization-based adaptive large neighborhood search, to minimize total operational costs while considering partial charging, battery degradation, and real-world constraints. The methods are evaluated on real-life instances to demonstrate their effectiveness in handling both small- and large-scale EBCSPs. Qi et al.~\cite{qi2025optimizing} propose a hierarchical deep reinforcement learning approach to optimize electric bus charging schedules under uncertainties in travel time, energy consumption, and electricity prices. The method uses a multi-agent framework with attention mechanisms and decentralized decision-making to improve scalability and performance for large electric bus fleets, demonstrating superior results in real-world experiments.

\begin{table*}[htp]
\label{private}
\fontsize{10}{16}
\selectfont
\caption{Charging Scheduling Works for Bus}
\centering
\resizebox{\textwidth}{!}{
\begin{tabular}{cccccc}
\hline
\multirow{2}{*}{Paper} & \multirow{2}{*}{Year} & \multirow{2}{*}{Model}  & \multicolumn{3}{c}{Optimization Objective}                                               \\ \cline{4-6} 
    & & & Charging Time & Energy Cost & Total Transit System Cost \\ \hline
Kang et al. \cite{kang2015centralized} & 2016 & Centralized Charging Strategy  &  &  & \checkmark\\ \hline
Wang et al. \cite{wang_bcharge_2018} & 2018 & Markov Decision Process Model & \checkmark &  & \\ \hline
Houbbadi et al. \cite{houbbadi2019optimal} & 2019 & Nonlinear Programming  &  & \checkmark &\\ \hline
Jahic et al. \cite{jahic2019charging} & 2019 & Greedy Algorithm &  & \checkmark & \\ \hline
Jahic et al. \cite{jahic_preemptive_2019} & 2019 & Greedy Algorithm &  & \checkmark & \\ \hline
Rodrigues et al. \cite{rodrigues_optimized_2020} & 2020 & Linear Programming  &  & \checkmark & \\ \hline
Li et al. \cite{li2020joint} & 2020 & Improved Adaptive Genetic Algorithm  &  &  & \checkmark \\ \hline
Wang et al. \cite{wang2020pricing} & 2020 & Markov Decision Process Model  &  &  & \checkmark \\ \hline
Ma et al. \cite{ma2021optimal} & 2021 & Surrogate-assisted Optimization Model & \checkmark &  & \\ \hline
Hu et al. \cite{hu2022joint} & 2022 & Robust Model  & \checkmark &  & \\ \hline
He et al. \cite{he2022integrated} & 2022 &  Integrated Optimization Model &  &  & \checkmark \\ \hline
Duan et al. \cite{duan2022bidding} & 2023 & Hierarchical Optimization Model &  & \checkmark & \\ \hline
Zeng et al. \cite{zeng2024route} & 2024 &  Deterministic Mixed-integer Linear Programming Model &  &  & \checkmark \\ \hline
Zhou et al. \cite{zhou2024electric} & 2024 & Mixed-integer Linear Programming Model & \checkmark &  & \checkmark \\ \hline
Qi et al. \cite{qi2025optimizing} & 2025 &  Hierarchical Deep Reinforcement Learning &  &  & \checkmark \\ \hline
\end{tabular}
}
\end{table*}

\subsubsection{Charging Scheduling for Heterogeneous EV Fleets}

However, Wang et al. \cite{wang2019sharedcharging} argue against the sole emphasis on specific fleet types, such as electric taxis or electric buses, as it may lead to suboptimal outcomes on a local scale rather than a broader/global scale. To tackle this concern, the authors propose a novel approach called shared charging, which is a versatile real-time shared charging scheduling system designed for EV fleets of heterogeneous nature. This system is backed by data-driven comparative analyses and serves the purpose of enhancing the overall charging efficiency of a city's charging network by accommodating EV fleets of varying sizes and operation modes.

In conclusion, this section provides an introduction to the charging scheduling of private EVs and shared EVs. The operational and charging methods of EVs vary depending on the usage scenario, necessitating a customized approach for designing the scheduling model and system that caters to specific circumstances. When considering private EV charging scheduling, there exist both challenges and opportunities in optimizing the overall benefits for the system, charging stations, and EV users while taking into account individual user preferences. On the other hand, heterogeneous shared EV fleets have diverse mobility and charging patterns, which pose a significant challenge to develop an optimal charging scheduling strategy for shared EVs at a global scale.

\section{EVs Fleet Management}\label{Sec5-EV Fleet Management}

Given the growing prominence of electric vehicles in the passenger vehicle industry, propelled by advancements in vehicle and battery technology, scholarly reports imply a nearly fourfold surge in global electric vehicle sales since 2014 \cite{wang2021data}. Predictions indicate that by 2027, EVs will represent half of all vehicle sales, highlighting the necessity for operators to enhance fleet management practices, as EVs present augmented requirements. Unlike the scheduling approaches employed for traditional gasoline vehicles, scheduling EVs necessitates consideration of factors such as vehicle power, charging station locations, and charging prices, as these factors directly influence vehicle service time. Consequently, these emerging considerations associated with EVs compel operators to strategize EV fleet management with the aim of meeting operational targets.

\subsection{EVs Dispatching}

\begin{figure*}[htbp]
\centering
    \includegraphics[width = 0.8\linewidth]{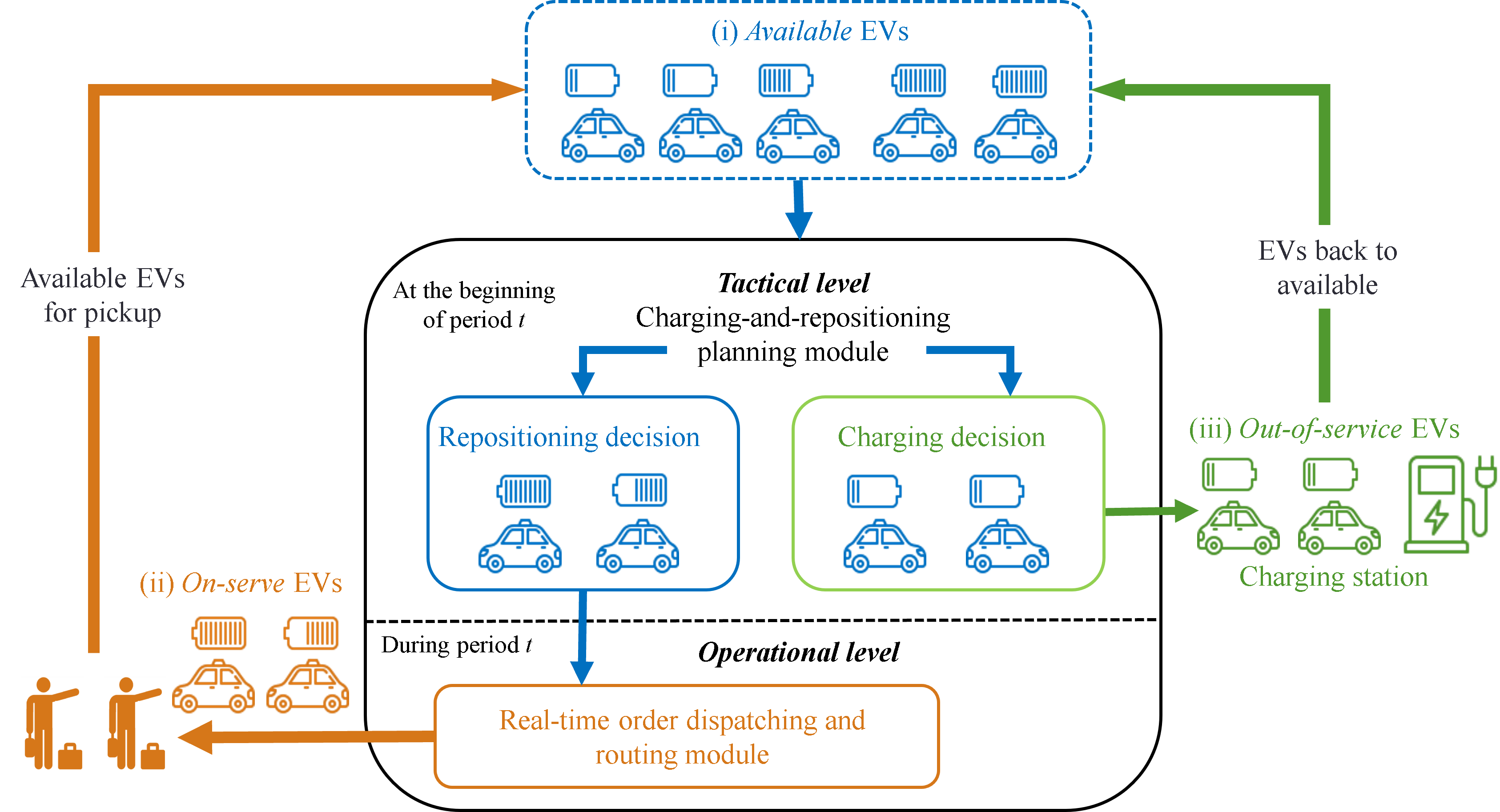}
    \caption{The EVs Dispatching System}
    \label{Fig.EVs_dispatching}   
\end{figure*}

Typically, EV fleet management operators confront a two-module challenge, compromising order dispatching and vehicle repositioning. Order dispatching entails a process in which the fleet system takes into account elements such as driver location, order type, and income potential to allocate drivers to orders. Contrarily, vehicle repositioning refers to the dynamic deployment of idle vehicles from one location to another based on predictive outcomes such as anticipated passenger demand. From a contemporary standpoint, these tasks, considering EVs' characteristics, have only recently become evident. In certain studies, researchers have posited order dispatching and vehicle repositioning as an amalgamated, dynamic scheduling problem. A typical EVs dispatching system like Fig.~\ref{Fig.EVs_dispatching}.

\subsubsection{Vehicle Repositioning}

In the context of vehicle repositioning in urban electric taxi fleets, the consideration of charging station location and charging price has emerged as a crucial factor influencing expected revenue. Tu et al.~\cite{tu2024deep} propose a deep reinforcement learning framework with a spatiotemporal double deep Q-network to optimize the real-time management of electric taxi fleets, improving daily profits by coordinating repositioning, recharging, and destination recommendations. In their pioneering work, Wang et al. \cite{wang2021data} present FairMove, the first data-driven fairness-aware dispatching system to address the problem of taxi repositioning through the application of multi-agent deep reinforcement learning, formulating it as a centralized multi-agent actor-critic framework. 
In a separate study, 
Kullman et al. \cite{kullman2022dynamic} employ deep reinforcement learning in the context of electric ridehail services with centralized control. Their objective is to predict future demand for ridehail services, considering the uncertainty and constraints associated with fleet electrification. The authors introduce two strategies, namely Dart and Drafter, to improve the efficiency and user satisfaction of online car-hailing systems. 
Wang et al.~\cite{wang2023cross} and Hu et al.~\cite{hu2021effective} study the cross-region repositioning of electric vehicles or bikes in on-demand delivery scenarios. They formulate the cross-region repositioning as a multi-agent sequential decision-making problem and propose a collaborative multi-agent reinforcement learning framework to optimize the long-term revenues of platforms.
Iacobucci et al.~\cite{iacobucci2019optimization} propose a mixed-integer linear programming model for optimizing shared autonomous electric vehicles under dynamic time-varying electricity price conditions. The model aims to maximize vehicle utilization and charging efficiency, facilitating vehicle-to-grid charging, routing, and relocation. Lastly, Yi et al. \cite{yi2021framework} investigate the interplay between fleet management strategies and different infrastructural and resource configurations (e.g., fleet sizes, charging infrastructure settings). Their framework, designed to reduce operating costs and increase fleet revenue, integrates relocation and charging decision-making processes. Task priority and distance are used to determine suitable vehicles for relocation, taking into account the vehicles' charging status and real-time road conditions.
Ma et al.~\cite{ma2025crowdsourced} address the shared electric vehicle relocation problem in a one-way vehicle-sharing system, proposing a hybrid optimization algorithm to improve task dispatching and reduce costs. Results show that the hybrid algorithm outperforms the benchmark, with greater efficiency in multitask crowdsourcing under medium to high supply-to-demand scenarios.

\subsubsection{Order Dispatching}

From a perspective centered on order dispatching, 
Yan et al. \cite{yan2023online} present a Markov decision process, which aims to optimize charging and order dispatching strategies concurrently, to maximize the overall fleet income. The vehicles are categorized into three states: available, on-service, and out-of-service. 
Wang et al.~\cite{wang2025electric} propose an EV dispatch model for virtual power plants that enhances participation willingness by addressing range and battery aging anxieties through optimized charging station siting, incentive pricing, and distributed scheduling.
Chen et al. \cite{chen2023electric} propose a centralized matching method that incorporates power management concerns into the order dispatching framework. The authors introduce a new metric, termed the marginal value of charge, to quantify the predictive advantage gained from additional vehicle state of charge within a local market around an electric vehicle. Based on the projected benefits, electric vehicles are chosen to match with either passengers or charging stations. Shi et al. \cite{shi2022memory} propose a dispatching method that leverages the ant colony algorithm and memory mechanism to achieve dynamic dispatch optimization by means of multi-source data association analysis. This article esteems time and space information as crucial factors and employs the ant colony algorithm to identify the most suitable dispatching solution. Additionally, this approach incorporates a memory mechanism to retain the historical best solutions, thereby expediting the solution process. 
In on-demand delivery~\cite{guo2021concurrent,lu2024deco} and last-mile logistics scenarios~\cite{guo2023towards}, the platform will perform order matching and task assignment to workers with electric vehicles or bikes. 
Wang et al.~\cite{wang2023gcrl} propose a Group-based Cooperative Reinforcement Learning (GCRL) framework to optimize the task assignment process in last-mile delivery.
Additionally, Guo et al.~\cite{guo2021concurrent} design a time-constrained actor-critic reinforcement learning to optimize the real-time order dispatching process. Wang et al.~\cite{wang2023time} and Jiang et al.~\cite{jiang2023faircod} further extend the reinforcement learning framework by considering multi-objective optimization and fairness-aware dispatching.

\subsubsection{Joint Optimization}

Several papers have investigated the joint optimization of order dispatching and vehicle repositioning in the context of ride-hailing services. Al-Kanj et al. \cite{al2020approximate} present a novel approach that treats order dispatching and vehicle repositioning as a joint optimization problem with the goal of maximizing order satisfaction, vehicle utilization, passenger satisfaction, and reducing empty mileage and carbon emissions. The authors propose a dynamic programming framework that utilizes state value functions and action value functions to obtain the optimal strategy through continuous iteration. Shi et al. \cite{shi2019operating} introduce a reinforcement learning (RL) based algorithm that incorporates assignment actions into a Markov decision process for dispatching EV fleets in ride-hailing services. The algorithm is built on a decentralized learning component, allowing the EV fleet to share their experiences and approximate state value function parameters during training, thus enhancing scalability. The centralized decision-making process, based on a linear assignment problem formulation, maximizes the action value function of the EV fleet, enabling coordinated actions among the electric vehicles. Yu et al. \cite{yu2021optimal} propose the use of asynchronous learning algorithms to optimize the joint optimization of order dispatching and vehicle repositioning in ride-hailing systems. The author presents an asynchronous RL-based algorithm that evaluates decision quality based on various indicators such as passenger waiting time, driver income, and vehicle charging needs. Liang \cite{liang2020mobility} investigates the joint charging scheduling, order dispatching, and vehicle rebalancing problems faced by large-scale shared EV fleet operators. To maximize fleet operator profit, the paper models joint decision-making as a partially observable Markov decision process and combines deep reinforcement learning with binary linear programming to develop approximate optimal solutions.

\begin{table*}[htp]
\label{private}
\caption{EVs Dispatching}
\fontsize{10}{16}
\selectfont
\centering
\resizebox{\textwidth}{!}{
\begin{tabular}{ccccccc}
\hline
\multirow{2}{*}{Paper} & \multirow{2}{*}{Year} & \multirow{2}{*}{Model} & \multicolumn{4}{c}{Optimization Objective} \\ \cline{4-7} 
& & & Profit Fairness & Charging Efficiency & Passenger Satisfaction & Fleet Revenue \\ \hline
Iacobucci et al. \cite{iacobucci2019optimization} & 2019 &  Mixed-Integer Linear Programming &  & \checkmark & &\checkmark\\ \hline
Shi et al. \cite{shi2022memory} & 2019 &  Reinforcement Learning Algorithm &  &  & \checkmark &\checkmark\\ \hline
Liang et al. \cite{liang2020mobility} & 2020 &  Markov Decision Process and DRL &  & \checkmark & & \checkmark \\ \hline
Al-Kanj et al. \cite{al2020approximate} & 2020 & Dynamic Programming Framework & & & \checkmark & \checkmark  \\ \hline
Yu et al. \cite{yu2021optimal} & 2021 & Reinforcement Learning Algorithm & &\checkmark & \checkmark & \checkmark \\ \hline
Wang et al. \cite{wang2021data} & 2021 &  Multi-Agent Deep Reinforcement Learning & \checkmark &  & & \checkmark \\ \hline
Yi et al. \cite{yi2021framework} & 2021 &  Repositioning and Charging Framerwork & & \checkmark & & \checkmark  \\ \hline
Kullman et al. \cite{kullman2022dynamic} & 2022 & Multi-Agent Deep Reinforcement Learning & & & \checkmark & \checkmark \\ \hline Shi et al. \cite{shi2022memory} & 2022 &  Ant Colony Algorithm  & & \checkmark & \checkmark & \\ \hline
Yan et al. \cite{yan2023online} & 2023 &   Markov Decision Process & & \checkmark & & \checkmark \\ \hline
Chen et al. \cite{chen2023electric} & 2023 &  Probabilistic Matching Method  & & \checkmark & & \checkmark \\ \hline
Wang et al.~\cite{wang2025electric}  & 2025 &    Distributed Scheduling Strategy & & \checkmark & & \checkmark \\ \hline
\end{tabular}
}
\end{table*}

In conclusion, recent research has focused on improving order dispatch and vehicle repositioning to enhance the efficiency and profitability of electric vehicle fleets in urban mobility, with an emphasis on joint optimization. Proposed methods include Markov decision processes, deep reinforcement learning, and centralized matching strategies to address complex fleet management challenges.

\subsection{Route Recommendation}

The limited range of electric vehicles poses a persistent challenge to their widespread adoption, given their reliance on battery energy as a power source. Moreover, the availability of nearby charging stations and their impact on the power grid system further restricts the charging of electric vehicles \cite{abid2022routing}. Hence, it is imperative to devise a more rational approach towards planning the driving routes of electric vehicles. Such an approach must minimize the driving distance while ensuring sufficient endurance, optimizing the power grid load, and improving the charging efficiency of each charging station. 

Existing methods for addressing route recommendation problems can be classified into three main categories: heuristic algorithms, optimization algorithms, and machine learning algorithms. One study by Nolz et al. \cite{nolz2022consistent} explored the domain of heuristic algorithms by proposing an algorithm for the route recommendation problem associated with delivery tasks involving electric vehicles. Their algorithm tackles constraints such as delivery intervals, electric vehicle power, and charging station locations, while aiming to jointly optimize delivery time and operational costs. 
Li et al.~\cite{li2023route} proposed an improved remove–reinsert genetic algorithm to solve the electric vehicle routing problem with time windows, enhancing energy efficiency and reducing transportation costs through optimized route initialization, neighborhood search, and reinsertion strategies.

Similarly, another research effort by Erdelic et al. \cite{erdelic2022goods} devised a self-adaptive large neighborhood search (ALNS) heuristic algorithm based on the destruction-reconstruction strategy for the Vehicle Routing Problem with Time Windows (EVRPTW). This algorithm considers two charging strategies, namely full charging and partial charging, in its optimization of route recommendations.

From an investigation into optimization algorithms, several studies have been conducted to tackle this issue. Stodola et al.~\cite{stodola2020hybrid} explored the utilization of a mixed ant colony algorithm to optimize the Multi-Depot Vehicle Routing Problem (MDVRP). This algorithm integrates probabilistic and precise techniques. Specifically, it leverages probabilistic techniques to emulate ants' foraging behavior while employing complementary precise techniques. In this experiment, the Cordreau benchmark instance was employed, demonstrating superior results when compared to alternative approaches. Another ant colony optimization algorithm based on improved brainstorming optimization, known as IBSO-ACO, was proposed by Wu et al. \cite{wu2019brainstorming}. In contrast to the traditional ant colony algorithm, this methodology applies an enhanced brainstorm optimization algorithm to update the solutions acquired by the ant colony algorithm, effectively tackling the challenge of local optimal solutions. With the ability to achieve more favorable path recommendations while rapidly converging, this algorithm represents a promising advancement in the field.

In the realm of EV path planning, the advent of deep learning and machine learning algorithms has paved the way for numerous methodologies. A recent study conducted by Basso et al. \cite{basso2021electric} focuses on the utilization of probabilistic Bayesian techniques to predict EV trajectories while considering traffic conditions and partial charging. The proposed routing framework can be categorized into two stages: the primary stage entails path selection, while the secondary stage involves routing optimization. 
Zhou et al.~\cite{zhou2021multi,zhou2024multi} perform route prediction and recommendations for electric vehicles or bikes for on-demand delivery.
Reinforcement learning strategies have also gained popularity in addressing EV path planning problems. Notably, Basso et al. \cite{basso2022dynamic} employed Monte Carlo simulation for offline evaluation of customer requests and EV energy consumption, allowing for online resolution of EV route planning challenges. This approach facilitates the design of a secure reinforcement learning algorithm that optimizes energy consumption during route optimization while ensuring adequate charge levels throughout the path-finding process. Conversely, Lin et al. \cite{lin2021deep} devised a composite strategy that incorporates pointer networks, graph embedding layer attention models, and deep reinforcement learning methods to tackle EV path planning issues in the presence of time windows.

\subsection{Car Sharing}
In response to increasing concerns over climate change driven by carbon emissions, low-carbon transportation options have gained widespread recognition. Among these, car-sharing has emerged as a promising and sustainable mobility solution. Car-sharing allows multiple users to share the use of a single vehicle, meeting their transportation needs without the necessity of personal vehicle ownership. This not only reduces transportation costs but also contributes to lower carbon emissions.

Notably, within the realm of electric vehicles, the multifaceted nature of car-sharing has spurred a surge in research endeavors aimed at shared electric vehicles. Focusing on the challenge of vehicle relocation, Folkestad et al. \cite{folkestad2020optimal} delve into the problem of transferring low-battery electric vehicles to charging stations by operators, while considering forthcoming customer demands and relocation expenses. To address this quandary, the authors propose a mixed integer programming heuristic mathematical model and resolve it via a hybrid genetic search algorithm featuring adaptive diversity control. In a parallel investigation, Gambella et al. \cite{gambella2018optimizing} explore the profit generated by shared user travel with the objective of maximizing such gains. The scholars introduce two heuristic mathematical optimization algorithms to tackle the problem, extending their methodology to encompass nocturnal operations through the execution of repositioning tasks while the sharing system is inactive.

Several studies employ reinforcement learning to address the aforementioned problem. Wang et al. \cite{wang2021record} undertake a study primarily focused on the challenges related to vehicle repositioning in the presence of highly dynamic user demands, as well as charging strategies under time-varying electricity prices. Their work introduces the Joint Real-Time Repositioning and Charging for Electric Carsharing with Dynamic Deadlines (Record) algorithm, which aims to simultaneously fulfill user demands while minimizing vehicle charging costs and laborious relocation costs. To achieve this, the authors develop a distributed deep reinforcement learning algorithm based on dynamic deadlines, which is incorporated into the Record algorithm.
The reward function is designed as follows:

\begin{equation}
\label{reward function}
    R_{u}-C_{c}-C_{M} = \sum_{i=1}^m R_{u}^{(i)} - \sum_{j=1}^{n}(\lambda * T_{c}^{(j)}) - \phi \times z
\end{equation}

where $R_{u}$ is the total revenue from serving users;
$C_{c}$ is the total charging cost of the electric carsharing fleet. 
$C_{M}$ is the total labor cost for moving shared EVS. 
$R_{u}^{(i)}$ is the revenue from serving the $i-$th electric carsharing order, and $M$ is the total number of served orders;
$T_{c}^{(j)}$ describes the charging time of $j-$th charging event and $\lambda$ denotes the charging prices during different periods;
$\phi$ is the payment to workers for each movement of shared EVs, and $z$ is the total number of vehicle movements by workers.
This algorithm generates dynamic deadlines using a prediction error compensation mechanism and adapts to search and learn the optimal location in real-time to meet highly dynamic and imbalanced user demands. The ultimate goal of the algorithm is to minimize operational costs while maintaining the quality of service for the electric car-sharing system.

In the study conducted by Luo et al. \cite{luo2021rebalancing}, the authors investigate the issue of repositioning electric vehicles (EVs) within the context of the ever-expanding network of EV sharing systems. This study aims to overcome the challenge posed by the prolonged charging time of EVs and the continuously expanding site network by introducing a rebalancing task modeling approach grounded in Multi-Agent Reinforcement Learning (MARL). The authors propose an optimization strategy that incorporates action cascading along with the utilization of two interconnected networks to effectively address the objective function of the reinforcement learning framework.

\subsection{Multi-task Optimization}
With the increasing demand for enhanced performance and efficiency in EVs, simultaneous optimization of multiple interdependent objectives—such as energy consumption, battery health, thermal management, and driving comfort—has become a critical challenge. Conventional single-task optimization approaches often fail to capture the complex couplings and trade-offs among these objectives, leading to suboptimal overall performance. To address this limitation, multi-task optimization has emerged as a promising paradigm, enabling the coordinated optimization of multiple tasks within a unified framework. By leveraging shared information and exploiting synergies across tasks, multi-task optimization offers the potential to significantly improve the system-level efficiency and robustness of EVs, thereby supporting the development of more intelligent and sustainable electrified transportation systems.

Li et al.~\cite{li2025multi} propose a multi-agent reinforcement learning energy management strategy for Hybrid electric vehicles, where the engine and battery act as cooperative agents to optimize fuel economy. Li et al.~\cite{li2024coordinated} propose a coordinated multitask electric bus scheduling method for post-disaster transportation-power distribution networks, formulating the problem as a mixed-integer linear programming model that integrates evacuation and power supply tasks, and solves it efficiently via a division-based approach. Huang et al.~\cite{huang2024prediction} propose a novel TCGC-ResNet method combined with a knowledge graph to predict and optimize tire/road structure-borne noise in pure electric vehicles, fusing mechanism-driven and data-driven approaches to enhance interpretability, computational efficiency, and robustness by modeling entity correlations and leveraging a task-conditional gate control mechanism for effective multi-task learning. Che et al.~\cite{che2023battery} propose an online end-to-end battery state/temperature monitoring method using transferred multi-task learning, which directly processes measurement data via a convolutional neural network and task-specific layers. Bao et al.~\cite{bao2024dual} propose a dual-task learning framework for joint state-of-charge and state-of-energy estimation of lithium-ion battery packs, featuring a feature attention mechanism with GRU and enhanced data preprocessing.
\section{Impact of EVs}\label{Sec6-Impact of EVs}

\begin{table*}[htp]
\label{private}
\fontsize{12}{18}
\selectfont
\caption{Impact of EVs on Smart grid}
\centering
\resizebox{\textwidth}{!}{
\begin{tabular}{ccccccc}
\hline
\multirow{2}{*}{Paper} & \multirow{2}{*}{Year} & \multirow{2}{*}{Model} & \multicolumn{4}{c}{Optimization Objective} \\ \cline{4-7} 
& & & Grid Charging & Reduce Cost & Grid Load Management & Grid Stability \\ \hline
Saber et al. \cite{saber2010plug} & 2010 &   Intelligent optimization method  &  & \checkmark & \checkmark & \\ \hline
Chatzivasileiadis et al. \cite{chatzivasileiadis2011q} & 2011 &  Q-learning algroithm &  \checkmark&  & & \\ \hline
Galus et al. \cite{galus2011balancing} & 2011 &    Model predictive control (MPC) &  &  &\checkmark &  \\ \hline
Wehinger et al. \cite{wehinger2011assessing} & 2011 & Predictive model & & &  & \checkmark  \\ \hline
Xue et al.~\cite{xue2021impact} & 2021 & Sequential Monte Carlo & & &  & \checkmark  \\ \hline
\end{tabular}
}
\end{table*}

\subsection{Impact of EVs on Human Mobility}

The potential impact of EVs on human mobility is extensively examined in various studies. The International Energy Agency (IEA) conducts research that establishes the possibility of a substantial decrease in dependence on fossil fuels for transportation with the widespread adoption of EVs \cite{global2017outlook}. In a similar vein, the National Renewable Energy Laboratory (NREL) proposes that EVs could be instrumental in reducing petroleum consumption \cite{davis2021transportation}. Moreover, the adoption of EVs has the potential to enhance the daily travel experience by providing greater convenience, courtesy of features such as instant torque and seamless acceleration. Despite these advantages, the limited range of EVs presents obstacles to their practicality for daily travel, particularly for longer journeys. As investigated by Rauh et al. \cite{rauh2015understanding}, the phenomenon known as "EV range anxiety" is a significant deterrent to the widespread adoption of EVs. This underscores the centrality of charging infrastructure, which not only extends the range of EVs but also plays a crucial role in alleviating range anxiety among EV owners. Therefore, the availability of public charging infrastructure assumes critical importance in accommodating the widespread adoption of EVs and enhancing their usability for day-to-day travel.

Furthermore, the implementation of charging infrastructure within urban areas possesses the potential to enhance accessibility and facilitate the acceptance of electric vehicles (EVs) in the context of urban mobility. According to the investigation conducted by Siragusa et al. \cite{siragusa2022electric}, the establishment of charging infrastructure in urban areas drives an increase in the adoption of EVs, particularly for short-distance trips and last-mile delivery endeavors. As a consequence, the concurrent decrease in traffic congestion and emissions emanating from transportation activities delivers an enhanced overall mobility experience for individuals inhabiting urban environments. 

These studies collectively reveal that while EVs offer considerable prospects to revolutionize human mobility, their widespread incorporation and successful assimilation into existing transportation systems stand contingent upon the availability of sufficient charging infrastructure to guarantee their operational viability for everyday travel. Moreover, the installation of charging infrastructure alongside EV-sharing systems \cite{mounce2019potential} within urban settings can furnish an avenue for fortifying accessibility and providing further support for advancing the adoption of EVs in urban mobility domains, consequently ushering in a more expedient and sustainable transportation alternative for urban residents.

\subsection{Impact of EVs on Smart Grid}

 EVs present a notable complication to conventional grid infrastructure due to their dependence on grid connectivity for charging and energy recovery. To address the adverse effects of EVs on the grid, the concept of smart grid technology emerges as a viable solution. Smart grid technology aims to enhance grid system operations and optimize the charging of EVs, making it a promising approach to address the challenges imposed by widespread EV adoption. The potential advantages of smart grid technology for EVs garner recognition from both the academic and governmental communities. In the subsequent subsection, an exhaustive overview of smart grid technologies and their interaction with EVs is presented to further elucidate the subject matter.

 \subsubsection{Smart Grid}
 A smart grid that integrates both conventional and clean energy resources aims to deliver electrical energy with enhanced reliability, reduced consumption, and lower costs. Moreover, the role of EVs in smart grids enables two-way energy exchange, which is a key feature distinguishing EVs from other energy units due to their charging and discharging capabilities. However, the integration of EVs, along with their electronic components into the grid,  introduces various challenges and impacts on the network \cite{sivaraman2021power}. Therefore, this section will focus on the potential power quality issues that EVs may introduce in smart grid systems.

 Xue et al. \cite{xue2021impact} highlight the role of EVs in promoting green transportation and examine the potential of vehicle-to-grid (V2G) technology, which enables bidirectional energy flow between EVs and the grid. It emphasizes the importance of the "5Ds" (decentralization, decarbonization, digitalization, deregulation, and democratization) in overcoming grid challenges and supporting sustainable EV integration.

The notion of the smart grid denotes an advanced grid system that leverages technological developments to address modern requirements. In contrast to the conventional grid, the smart grid facilitates bidirectional transmission of electricity and data \cite{bhatt2014instrumentation}. Consequently, the smart grid incorporates a higher number of distributed sensors, enabling self-monitoring and self-assessment of diverse grid components \cite{yong2015review,qiao2011design}. Moreover, the smart grid possesses the ability to dynamically optimize power supply modes based on varying regional and temporal power demands, enabling real-time pricing by enterprises to cater to distinct power consumption needs \cite{massaoudi2021deep}. Elucidating the essential characteristics of the smart grid, the NIST report \cite{greer2014nist} identifies the subsequent key features:
\begin{itemize}
    \item Enhancing the reliability, security, and efficiency of the electric grid through the heightened utilization of digital information and control technology. 
    \item Implementation and assimilation of distributed resources and generation, encompassing renewable resources, within the grid infrastructure. 
    \item Implementation of cutting-edge technologies, such as metering, communication systems pertaining to grid operations and status, and distribution automation, to significantly augment grid capabilities. 
    \item Adoption and assimilation of advanced electricity storage and peak-shaving technologies, inclusive of plug-in electric and hybrid electric vehicles, as well as thermal-storage air conditioning. 
    \item Effective provision of timely information and control options to consumers.

\end{itemize}

Based on the aforementioned characteristics, the smart grid technology demonstrates unparalleled aptitude in meeting tangible operational necessities with enhanced dependability and effectiveness, thereby establishing itself as the prospective trajectory for grid advancement. As the quantity of EVs proliferates, the repercussions of their integration within the smart grid have captured mounting scrutiny.

\subsubsection{Vehicle-to-Grid (V2G) Technology}

Vehicle-to-Grid (V2G) technology facilitates bidirectional energy exchange between electric vehicle (EV) battery systems and the power grid, effectively treating EVs as distributed energy resources. This has the potential to enhance the reliability, stability, and efficiency of the grid by enabling the provision of clean energy. Moreover, V2G technology presents an opportunity for vehicle owners to sell excess energy, offering mutual benefits to both the grid and EV owners. Numerous researchers dedicate their efforts to exploring the potential applications of this technology \cite{yilmaz2012review}.  Chatzivasileiadis et al. \cite{chatzivasileiadis2011q} pursue the development of optimal deployment strategies for frequency controllers by leveraging the Q-learning algorithm. The proposed approach encompasses the utilization of data obtained from the grid, vehicles, and energy storage systems, allowing for the training of frequency controllers to make real-time decisions related to the charging or discharging of plug-in hybrid electric vehicle (PHEV) batteries. Through such decisions, the frequency controllers effectively contribute to fulfilling the required frequency response of the grid.  In an effort to reduce the cost and emissions associated with smart grids, Saber et al. \cite{saber2010plug} propose a model that incorporates renewable clean energy sources and V2G technology. The authors emphasize the necessity of dynamically modeling energy changes based on the distinctive characteristics of various energy sources within modern smart grids. Correspondingly, the appropriate optimization methods should be employed to optimize resource consumption.  Notably, V2G technology is capable of addressing prevailing challenges existing within the power grid. Galus et al. \cite{galus2011balancing} conduct an analysis of the mismatch between renewable energy sources and the power grid. By leveraging the model predictive control (MPC) algorithm, they propose a methodology to regulate the behavior of a fleet of PHEVs using V2G technology. This necessitates employing driving and charging data of the fleet to predict future battery statuses and power grid demands, thus enabling the adjustment of charging and discharging parameters of the vehicles to effectively manage the power grid load. Turning to the commercial implementation of V2G technology, Wehinger et al. \cite{wehinger2011assessing} evaluate the prevailing electricity market conditions characterized by the growing adoption of renewable energy sources and grid stability concerns. Within this context, the authors propose the utilization of storage devices and a PHEV cluster to mitigate these challenges. Their work introduces a method for modeling storage agents using predictive models, enabling the adjustment of charging and discharging of storage devices in accordance with predicted future spot prices.

\subsection{Impact of EVs on Environment}

Although electric EVs boast lower or no exhaust emissions compared with traditional vehicles during their usage, it is crucial to consider the consumption of fossil fuel energy and mineral resources in the production of electricity and batteries when assessing their overall environmental impact \cite{vidhi2018review}. 

While EVs have been proven effective in enhancing air quality and exhibiting minimal pollutants during the exhaust phase \cite{liu2022exhaust}, their emissions of gases and levels of air pollution vary depending on the technical solutions and power sources employed by different EV types. Notably, the emissions of air pollutants from EVs predominantly occur during the raw material production and new energy utilization stages. To comprehensively evaluate EV emissions, the commonly adopted approach is a wells-to-wheels analysis \cite{ke2017well}, which encompasses two stages: wells-to-tank and tank-to-wheels. Wells-to-tank denotes the conversion of primary energy into vehicle fuel, while tank-to-wheels refers to the process of utilizing the fuel to power the vehicle. The net impact of EVs on air quality hinges on a country's or EVs' industrial chain's technological level and energy structure.

The production of electric vehicle batteries brings about considerable environmental burdens due to energy consumption and emissions from positive material processing and electrode drying. The environmental impact of battery production is influenced by regional factors, particularly the energy mix utilized. A higher proportion of renewable energy in electricity generation can ameliorate the environmental performance of batteries. Additionally, the environmental performance of batteries varies based on their size. Upon reaching the end of their life cycle, EV batteries retain 70-80\% of their capacity, rendering their disposal wasteful and environmentally impactful. Repurposing retired batteries for residential and public utility energy storage presents a more environmentally friendly alternative to producing new batteries of equivalent capacity. Battery recycling and remanufacturing following secondary use further reduce the demand for raw materials in battery production, thereby enabling a closed-loop battery production system.

The impact of EVs on the environment is contingent upon the source of electricity used to power them. If electricity production predominantly relies on fossil fuels, EVs may give rise to higher environmental pollution. However, in cases where the proportion of renewable energy in the electricity mix is substantial, the limitations imposed by environmental and seasonal factors necessitate backup power generation from fossil energy sources. The surge in electric vehicle adoption has added to the power burden, thereby driving the demand for clean energy in the power system. In this regard, governmental efforts should focus on optimizing the energy structure and facilitating the development of smart grid charging technologies.

\section{Future Direction}\label{sec7-Future Direction}
The future research direction for EVs promises significant advancements through the integration of large language models (LLMs). These sophisticated models offer transformative potential in multiple facets of EV technology, including but not limited to energy management, user interaction, and predictive maintenance. By applying LLMs, researchers aim to develop advanced energy management systems that can dynamically optimize the powertrain efficiency based on real-time driving conditions and historical data. Moreover, LLM-driven natural language interfaces could revolutionize human-machine interaction, providing more intuitive and accessible controls for drivers and passengers. 
For instance, Chen et al.~\cite{chen2024integration} propose a LLM-guided hybrid framework that integrates natural language instructions from EV drivers with multi-agent deep reinforcement learning and whale optimization to generate and refine dispatching strategies, enabling commonsense-aware and efficient coordination of connected electric vehicles in complex space-air-ground integrated vehicular networks. 

Beyond interaction and control, LLMs are also being leveraged to enhance charging infrastructure intelligence and user-centric service delivery.
Teimoori et al.~\cite{teimoori2025llm} propose an LLM-powered prompt-based recommender system that leverages natural language reasoning and fine-tuned EV-specific data to dynamically suggest optimal charging stations by integrating real-time heterogeneous information. 
Complementing this, Fan et al.~\cite{fan2025ev} propose a novel spatio-temporal LLM framework for electric vehicle charging demand and station occupancy forecasting, which combines advanced data preprocessing techniques with a partially frozen graph attention module to effectively capture complex spatio-temporal dependencies and domain-specific knowledge, achieving superior prediction accuracy and robustness on real-world data.

Further advancing personalization and grid coordination, Niu et al.~\cite{niu2025ev}  propose a multi-agent LLM framework for EV charging management, where a User Agent and an EV Charging Station Agent interact via a negotiation platform, leveraging fine-tuned LLMs and CGAN-generated behavioral data to enable adaptive, personalized recommendations and dynamic pricing, effectively balancing user preferences with grid demands in a scalable and intelligent manner.
In vehicle-to-grid (V2G) applications, Sun et al.~\cite{sun2025dynamic} propose an enhanced electric vehicle demand response system using a LLM-driven multi-agent framework to create user digital twins for accurate prediction of charging/discharging behavior, combined with a data- and knowledge-driven dynamic incentive mechanism and distributed optimization, significantly improving load regulation, user satisfaction, and grid stability in vehicle-to-grid applications.

Additionally, predictive analytics powered by LLMs can anticipate component failures and recommend preemptive maintenance schedules, thereby enhancing vehicle reliability and longevity. 
The exploration of these areas, alongside the continuous refinement of LLM architectures tailored for specific EV applications, will be pivotal in shaping a new generation of intelligent, efficient, and user-centric electric vehicles. Through interdisciplinary collaboration and innovation, the synergy between LLMs and EV technology is set to unlock unprecedented opportunities for sustainable transportation solutions.

\section{Conclusion}\label{Sec7-Conclusion}

In this paper, we present a comprehensive overview of data-driven applications of electric vehicles (EVs) in the field of intelligent transportation. Our analysis encompasses various perspectives, including the deployment of charging stations, optimization of charging schedules, effective management of EV fleets, and examination of the impacts of these applications. As the EV industry continues to evolve and the scale of EV adoption expands, numerous unexplored avenues present themselves. Of particular interest is the integration of EVs with autonomous driving technologies. In terms of individual vehicles, a multitude of sensors will generate high-quality data, while powerful onboard computers will function as robust edge devices, providing real-time decision-making information. From a multi-vehicle cooperation standpoint, the utilization of multiple EVs, equipped with real-time computing and advanced communication capabilities, has the potential to revolutionize traditional traffic management systems. By employing data-driven approaches and algorithms, greater operational efficiency can be achieved. Equally intriguing is the interaction between EVs and users. This area of study encompasses fair treatment of EV drivers and users, sustainable management of EVs within specific areas, analysis of EV activity and its impact on human mobility patterns in certain regions, as well as innovative trip patterns that incorporate shared EVs. As the number of EVs on the road continues to rise, a wide array of innovative data-driven applications will invariably emerge, providing ample opportunities for further investigation.




%





\ifCLASSOPTIONcaptionsoff
  \newpage
\fi





\bibliographystyle{IEEEtran}
\bibliography{IEEEabrv,Bibliography}

\begin{thebibliography}{100}
\providecommand{\url}[1]{#1}
\csname url@rmstyle\endcsname
\providecommand{\newblock}{\relax}
\providecommand{\bibinfo}[2]{#2}
\providecommand\BIBentrySTDinterwordspacing{\spaceskip=0pt\relax}
\providecommand\BIBentryALTinterwordstretchfactor{4}
\providecommand\BIBentryALTinterwordspacing{\spaceskip=\fontdimen2\font plus
\BIBentryALTinterwordstretchfactor\fontdimen3\font minus \fontdimen4\font\relax}
\providecommand\BIBforeignlanguage[2]{{%
\expandafter\ifx\csname l@#1\endcsname\relax
\typeout{** WARNING: IEEEtran.bst: No hyphenation pattern has been}%
\typeout{** loaded for the language `#1'. Using the pattern for}%
\typeout{** the default language instead.}%
\else
\language=\csname l@#1\endcsname
\fi
#2}}

\bibitem{li2015growing}
Y.~Li, J.~Luo, C.-Y. Chow, K.-L. Chan, Y.~Ding, and F.~Zhang, ``Growing the charging station network for electric vehicles with trajectory data analytics,'' in \emph{2015 IEEE 31st international conference on data engineering}.\hskip 1em plus 0.5em minus 0.4em\relax IEEE, 2015, pp. 1376--1387.

\bibitem{GlobalEV}
IEA, ``Global ev outlook 2025,'' \url{https://www.iea.org/reports/global-ev-outlook-2025}, 2025.

\bibitem{Tesla}
Tesla, ``Tesla,'' \url{https://www.tesla.com}, 2023.

\bibitem{BYD}
BYD, ``Byd,'' \url{https://www.byd.com/us}, 2023.

\bibitem{wang2021data}
G.~Wang, S.~Zhong, S.~Wang, F.~Miao, Z.~Dong, and D.~Zhang, ``Data-driven fairness-aware vehicle displacement for large-scale electric taxi fleets,'' in \emph{2021 IEEE 37th International Conference on Data Engineering (ICDE)}.\hskip 1em plus 0.5em minus 0.4em\relax IEEE, 2021, pp. 1200--1211.

\bibitem{ullah2022prediction}
I.~Ullah, K.~Liu, T.~Yamamoto, M.~Zahid, and A.~Jamal, ``Prediction of electric vehicle charging duration time using ensemble machine learning algorithm and shapley additive explanations,'' \emph{International Journal of Energy Research}, vol.~46, no.~11, pp. 15\,211--15\,230, 2022.

\bibitem{wang2018bcharge}
G.~Wang, X.~Xie, F.~Zhang, Y.~Liu, and D.~Zhang, ``bcharge: Data-driven real-time charging scheduling for large-scale electric bus fleets,'' in \emph{2018 IEEE Real-Time Systems Symposium (RTSS)}.\hskip 1em plus 0.5em minus 0.4em\relax IEEE, 2018, pp. 45--55.

\bibitem{ji2018plug}
Z.~Ji and X.~Huang, ``Plug-in electric vehicle charging infrastructure deployment of china towards 2020: Policies, methodologies, and challenges,'' \emph{Renewable and Sustainable Energy Reviews}, vol.~90, pp. 710--727, 2018.

\bibitem{lamonaca2022state}
S.~LaMonaca and L.~Ryan, ``The state of play in electric vehicle charging services--a review of infrastructure provision, players, and policies,'' \emph{Renewable and sustainable energy reviews}, vol. 154, p. 111733, 2022.

\bibitem{harrison2017exploratory}
G.~Harrison and C.~Thiel, ``An exploratory policy analysis of electric vehicle sales competition and sensitivity to infrastructure in europe,'' \emph{Technological Forecasting and Social Change}, vol. 114, pp. 165--178, 2017.

\bibitem{akinlabi2020configuration}
A.~H. Akinlabi and D.~Solyali, ``Configuration, design, and optimization of air-cooled battery thermal management system for electric vehicles: A review,'' \emph{Renewable and Sustainable Energy Reviews}, vol. 125, p. 109815, 2020.

\bibitem{berkeley2018analysing}
N.~Berkeley, D.~Jarvis, and A.~Jones, ``Analysing the take up of battery electric vehicles: An investigation of barriers amongst drivers in the uk,'' \emph{Transportation Research Part D: Transport and Environment}, vol.~63, pp. 466--481, 2018.

\bibitem{debnath2021political}
R.~Debnath, R.~Bardhan, D.~M. Reiner, and J.~Miller, ``Political, economic, social, technological, legal and environmental dimensions of electric vehicle adoption in the united states: A social-media interaction analysis,'' \emph{Renewable and Sustainable Energy Reviews}, vol. 152, p. 111707, 2021.

\bibitem{lopez2021societal}
N.~S. Lopez, L.~A. Tria, L.~A. Tayo, R.~J. Cruzate, C.~Oppus, P.~Cabacungan, I.~Isla~Jr, A.~Ansay, T.~Garcia, K.~C.-D. Cruz, \emph{et~al.}, ``Societal cost-benefit analysis of electric vehicles in the philippines with the inclusion of impacts to balance of payments,'' \emph{Renewable and Sustainable Energy Reviews}, vol. 150, p. 111492, 2021.

\bibitem{singh2020review}
V.~Singh, V.~Singh, and S.~Vaibhav, ``A review and simple meta-analysis of factors influencing adoption of electric vehicles,'' \emph{Transportation Research Part D: Transport and Environment}, vol.~86, p. 102436, 2020.

\bibitem{Number}
statista, ``Number of electric vehicles in use by type 2016-2022,'' \url{https://www.statista.com/statistics/1101415/number-of-electric-vehicles-by-type/#:~:text=There%20were%20about%2025.9%20million,of%20plug%2Din%20electric%20vehicles.}, 2023.

\bibitem{wang2020understanding}
G.~Wang, F.~Zhang, H.~Sun, Y.~Wang, and D.~Zhang, ``Understanding the long-term evolution of electric taxi networks: A longitudinal measurement study on mobility and charging patterns,'' \emph{ACM Transactions on Intelligent Systems and Technology (TIST)}, vol.~11, no.~4, pp. 1--27, 2020.

\bibitem{hu2014new}
Y.~Hu, X.~Song, W.~Cao, and B.~Ji, ``New sr drive with integrated charging capacity for plug-in hybrid electric vehicles (phevs),'' \emph{IEEE Transactions on Industrial Electronics}, vol.~61, no.~10, pp. 5722--5731, 2014.

\bibitem{sbordone2015ev}
D.~Sbordone, I.~Bertini, B.~Di~Pietra, M.~C. Falvo, A.~Genovese, and L.~Martirano, ``Ev fast charging stations and energy storage technologies: A real implementation in the smart micro grid paradigm,'' \emph{Electric Power Systems Research}, vol. 120, pp. 96--108, 2015.

\bibitem{chen2020enabling}
X.~Chen, Z.~Li, H.~Dong, Z.~Hu, and C.~C. Mi, ``Enabling extreme fast charging technology for electric vehicles,'' \emph{IEEE Transactions on Intelligent Transportation Systems}, vol.~22, no.~1, pp. 466--470, 2020.

\bibitem{mohammed2021comprehensive}
S.~A.~Q. Mohammed and J.-W. Jung, ``A comprehensive state-of-the-art review of wired/wireless charging technologies for battery electric vehicles: Classification/common topologies/future research issues,'' \emph{IEEE Access}, vol.~9, pp. 19\,572--19\,585, 2021.

\bibitem{mohamed2022comprehensive}
N.~Mohamed, F.~Aymen, T.~E. Alharbi, C.~Z. El-Bayeh, S.~Lassaad, S.~S. Ghoneim, and U.~Eicker, ``A comprehensive analysis of wireless charging systems for electric vehicles,'' \emph{IEEE Access}, vol.~10, pp. 43\,865--43\,881, 2022.

\bibitem{machura2019critical}
P.~Machura and Q.~Li, ``A critical review on wireless charging for electric vehicles,'' \emph{Renewable and Sustainable Energy Reviews}, vol. 104, pp. 209--234, 2019.

\bibitem{lam2014electric}
A.~Y. Lam, Y.-W. Leung, and X.~Chu, ``Electric vehicle charging station placement: Formulation, complexity, and solutions,'' \emph{IEEE Transactions on Smart Grid}, vol.~5, no.~6, pp. 2846--2856, 2014.

\bibitem{bae2020game}
S.~Bae, I.~Jang, S.~Gros, B.~Kulcs{\'a}r, and J.~Hellgren, ``A game approach for charging station placement based on user preferences and crowdedness,'' \emph{IEEE Transactions on Intelligent Transportation Systems}, vol.~23, no.~4, pp. 3654--3669, 2020.

\bibitem{zhao2020deployment}
Y.~Zhao, Y.~Guo, Q.~Guo, H.~Zhang, and H.~Sun, ``Deployment of the electric vehicle charging station considering existing competitors,'' \emph{IEEE Transactions on Smart Grid}, vol.~11, no.~5, pp. 4236--4248, 2020.

\bibitem{mirzaei2015probabilistic}
M.~J. Mirzaei, A.~Kazemi, and O.~Homaee, ``A probabilistic approach to determine optimal capacity and location of electric vehicles parking lots in distribution networks,'' \emph{IEEE Transactions on industrial informatics}, vol.~12, no.~5, pp. 1963--1972, 2015.

\bibitem{zhang2016pev}
H.~Zhang, S.~J. Moura, Z.~Hu, and Y.~Song, ``Pev fast-charging station siting and sizing on coupled transportation and power networks,'' \emph{IEEE Transactions on Smart Grid}, vol.~9, no.~4, pp. 2595--2605, 2016.

\bibitem{zhang2017second}
H.~Zhang, S.~J. Moura, Z.~Hu, W.~Qi, and Y.~Song, ``A second-order cone programming model for planning pev fast-charging stations,'' \emph{IEEE Transactions on power Systems}, vol.~33, no.~3, pp. 2763--2777, 2017.

\bibitem{wang2018coordinated}
X.~Wang, M.~Shahidehpour, C.~Jiang, and Z.~Li, ``Coordinated planning strategy for electric vehicle charging stations and coupled traffic-electric networks,'' \emph{IEEE Transactions on Power Systems}, vol.~34, no.~1, pp. 268--279, 2018.

\bibitem{ahmad2023placement}
F.~Ahmad, A.~Iqbal, I.~Asharf, M.~Marzband, and I.~Khan, ``Placement and capacity of ev charging stations by considering uncertainties with energy management strategies,'' \emph{IEEE Transactions on Industry Applications}, vol.~59, no.~3, pp. 3865--3874, 2023.

\bibitem{he2016incorporating}
S.~Y. He, Y.-H. Kuo, and D.~Wu, ``Incorporating institutional and spatial factors in the selection of the optimal locations of public electric vehicle charging facilities: A case study of beijing, china,'' \emph{Transportation Research Part C: Emerging Technologies}, vol.~67, pp. 131--148, 2016.

\bibitem{liu2016optimal}
C.~Liu, K.~Deng, C.~Li, J.~Li, Y.~Li, and J.~Luo, ``The optimal distribution of electric-vehicle chargers across a city,'' in \emph{2016 IEEE 16th International Conference on Data Mining (ICDM)}.\hskip 1em plus 0.5em minus 0.4em\relax IEEE, 2016, pp. 261--270.

\bibitem{liu2017locating}
H.~Liu and D.~Z. Wang, ``Locating multiple types of charging facilities for battery electric vehicles,'' \emph{Transportation Research Part B: Methodological}, vol. 103, pp. 30--55, 2017.

\bibitem{guo2024approach}
D.~Guo, R.~Liu, M.~Li, X.~Tan, P.~Ma, and H.~Zhang, ``An approach to optimizing the layout of charging stations considering differences in user range anxiety,'' \emph{Sustainable Energy, Grids and Networks}, vol.~38, p. 101292, 2024.

\bibitem{zheng2024effects}
Y.~Zheng, D.~R. Keith, S.~Wang, M.~Diao, and J.~Zhao, ``Effects of electric vehicle charging stations on the economic vitality of local businesses,'' \emph{Nature communications}, vol.~15, no.~1, p. 7437, 2024.

\bibitem{liu2023placement}
L.~Liu, S.~Liu, J.~Wu, and J.~Xu, ``A placement strategy for idle mobile charging stations in ioev: From the view of charging demand force,'' \emph{IEEE Transactions on Intelligent Transportation Systems}, vol.~25, no.~5, pp. 3870--3884, 2023.

\bibitem{liu2024random}
Y.~Liu, C.~Tang, and J.~Lu, ``Random scenario-based dynamic location optimization for ev fast-charging station,'' \emph{IEEE Transactions on Transportation Electrification}, 2024.

\bibitem{tang2025stochastic}
Y.~Tang, W.~Liu, K.~T. Chau, Y.~Hou, and J.~Guo, ``Stochastic behavior modeling and optimal bidirectional charging station deployment in ev energy network,'' \emph{IEEE Transactions on Intelligent Transportation Systems}, 2025.

\bibitem{zhao2023optimal}
Z.~Zhao, C.~K. Lee, J.~Ren, and Y.~P. Tsang, ``Optimal ev fast charging station deployment based on a reinforcement learning framework,'' \emph{IEEE transactions on intelligent transportation systems}, vol.~24, no.~8, pp. 8053--8065, 2023.

\bibitem{yan2018employing}
L.~Yan, H.~Shen, Z.~Li, A.~Sarker, J.~A. Stankovic, C.~Qiu, J.~Zhao, and C.~Xu, ``Employing opportunistic charging for electric taxicabs to reduce idle time,'' \emph{Proceedings of the ACM on Interactive, Mobile, Wearable and Ubiquitous Technologies}, vol.~2, no.~1, pp. 1--25, 2018.

\bibitem{riemann2015optimal}
R.~Riemann, D.~Z. Wang, and F.~Busch, ``Optimal location of wireless charging facilities for electric vehicles: flow-capturing location model with stochastic user equilibrium,'' \emph{Transportation Research Part C: Emerging Technologies}, vol.~58, pp. 1--12, 2015.

\bibitem{yan2017catcharger}
L.~Yan, H.~Shen, J.~Zhao, C.~Xu, F.~Luo, and C.~Qiu, ``Catcharger: Deploying wireless charging lanes in a metropolitan road network through categorization and clustering of vehicle traffic,'' in \emph{IEEE INFOCOM 2017-IEEE conference on computer communications}.\hskip 1em plus 0.5em minus 0.4em\relax IEEE, 2017, pp. 1--9.

\bibitem{li2019wireless}
M.~Li, X.~Wu, Z.~Zhang, G.~Yu, Y.~Wang, and W.~Ma, ``A wireless charging facilities deployment problem considering optimal traffic delay and energy consumption on signalized arterial,'' \emph{IEEE Transactions on Intelligent Transportation Systems}, vol.~20, no.~12, pp. 4427--4438, 2019.

\bibitem{elmeligy2024optimal}
A.~O. Elmeligy, E.~Elghanam, M.~S. Hassan, A.~H. Osman, A.~A. Shalaby, and M.~Shaaban, ``Optimal planning of dynamic wireless charging infrastructure for electric vehicles,'' \emph{IEEE Access}, vol.~12, pp. 30\,661--30\,673, 2024.

\bibitem{bai2022robust}
Z.~Bai, L.~Yang, C.~Fu, Z.~Liu, Z.~He, and N.~Zhu, ``A robust approach to integrated wireless charging infrastructure design and bus fleet size optimization,'' \emph{Computers \& Industrial Engineering}, vol. 168, p. 108046, 2022.

\bibitem{yan2022mobicharger}
L.~Yan, H.~Shen, L.~Kang, J.~Zhao, Z.~Zhang, and C.~Xu, ``Mobicharger: Optimal scheduling for cooperative ev-to-ev dynamic wireless charging,'' \emph{IEEE Transactions on Mobile Computing}, 2022.

\bibitem{gusrialdi2014scheduling}
A.~Gusrialdi, Z.~Qu, and M.~A. Simaan, ``Scheduling and cooperative control of electric vehicles' charging at highway service stations,'' in \emph{53rd IEEE conference on decision and control}.\hskip 1em plus 0.5em minus 0.4em\relax IEEE, 2014, pp. 6465--6471.

\bibitem{kong2016line}
F.~Kong, Q.~Xiang, L.~Kong, and X.~Liu, ``On-line event-driven scheduling for electric vehicle charging via park-and-charge,'' in \emph{2016 IEEE Real-Time Systems Symposium (RTSS)}.\hskip 1em plus 0.5em minus 0.4em\relax IEEE, 2016, pp. 69--78.

\bibitem{iversen2014optimal}
E.~B. Iversen, J.~M. Morales, and H.~Madsen, ``Optimal charging of an electric vehicle using a markov decision process,'' \emph{Applied Energy}, vol. 123, pp. 1--12, 2014.

\bibitem{zhao2013peak}
S.~Zhao, X.~Lin, and M.~Chen, ``Peak-minimizing online ev charging,'' in \emph{2013 51st Annual Allerton Conference on Communication, Control, and Computing (Allerton)}.\hskip 1em plus 0.5em minus 0.4em\relax IEEE, 2013, pp. 46--53.

\bibitem{shaaban2014real}
M.~F. Shaaban, M.~Ismail, E.~F. El-Saadany, and W.~Zhuang, ``Real-time pev charging/discharging coordination in smart distribution systems,'' \emph{IEEE Transactions on Smart Grid}, vol.~5, no.~4, pp. 1797--1807, 2014.

\bibitem{duan2025study}
S.~Duan, Z.~Zhang, Z.~Wang, X.~Xiong, X.~Chen, and X.~Que, ``A study on mobile charging station combined with integrated energy system: Emphasis on energy dispatch strategy and multi-scenario analysis,'' \emph{Renewable Energy}, vol. 239, p. 122111, 2025.

\bibitem{bokopane2024optimal}
L.~Bokopane, K.~Kusakana, H.~Vermaak, and A.~Hohne, ``Optimal power dispatching for a grid-connected electric vehicle charging station microgrid with renewable energy, battery storage and peer-to-peer energy sharing,'' \emph{Journal of Energy Storage}, vol.~96, p. 112435, 2024.

\bibitem{du2018demand}
B.~Du, Y.~Tong, Z.~Zhou, Q.~Tao, and W.~Zhou, ``Demand-aware charger planning for electric vehicle sharing,'' in \emph{Proceedings of the 24th ACM SIGKDD International Conference on Knowledge Discovery \& Data Mining}, 2018, pp. 1330--1338.

\bibitem{wang2019tcharge}
G.~Wang, F.~Zhang, and D.~Zhang, ``tcharge-a fleet-oriented real-time charging scheduling system for electric taxi fleets,'' in \emph{Proceedings of the 17th Conference on Embedded Networked Sensor Systems}, 2019, pp. 440--441.

\bibitem{wang2019sharedcharging}
G.~Wang, W.~Li, J.~Zhang, Y.~Ge, Z.~Fu, F.~Zhang, Y.~Wang, and D.~Zhang, ``sharedcharging: Data-driven shared charging for large-scale heterogeneous electric vehicle fleets,'' \emph{Proceedings of the ACM on Interactive, Mobile, Wearable and Ubiquitous Technologies}, vol.~3, no.~3, pp. 1--25, 2019.

\bibitem{dong2017rec}
Z.~Dong, C.~Liu, Y.~Li, J.~Bao, Y.~Gu, and T.~He, ``Rec: Predictable charging scheduling for electric taxi fleets,'' in \emph{2017 IEEE Real-Time Systems Symposium (RTSS)}.\hskip 1em plus 0.5em minus 0.4em\relax IEEE, 2017, pp. 287--296.

\bibitem{kim2014real}
E.~Kim, K.~G. Shin, and J.~Lee, ``Real-time discharge/charge rate management for hybrid energy storage in electric vehicles,'' in \emph{2014 IEEE Real-Time Systems Symposium}.\hskip 1em plus 0.5em minus 0.4em\relax IEEE, 2014, pp. 228--237.

\bibitem{kim2015modeling}
E.~Kim, J.~Lee, and K.~G. Shin, ``Modeling and real-time scheduling of large-scale batteries for maximizing performance,'' in \emph{2015 IEEE Real-Time Systems Symposium}.\hskip 1em plus 0.5em minus 0.4em\relax IEEE, 2015, pp. 33--42.

\bibitem{li2019data}
S.~Li, S.~He, S.~Wang, T.~He, and J.~Chen, ``Data-driven battery-lifetime-aware scheduling for electric bus fleets,'' \emph{Proceedings of the ACM on Interactive, Mobile, Wearable and Ubiquitous Technologies}, vol.~3, no.~4, pp. 1--22, 2019.

\bibitem{yang2017charging}
C.~Yang, W.~Lou, J.~Yao, and S.~Xie, ``On charging scheduling optimization for a wirelessly charged electric bus system,'' \emph{IEEE Transactions on Intelligent Transportation Systems}, vol.~19, no.~6, pp. 1814--1826, 2017.

\bibitem{cao2021online}
Y.~Cao, S.~Qian, W.~Zhu, J.~Cao, G.~Xue, Y.~Zhu, and M.~Li, ``Online charging coordination of electric vehicles to optimize cost and smoothness,'' \emph{Pervasive and Mobile Computing}, vol.~73, p. 101391, 2021.

\bibitem{mohsenian2010optimal}
A.-H. Mohsenian-Rad and A.~Leon-Garcia, ``Optimal residential load control with price prediction in real-time electricity pricing environments,'' \emph{IEEE transactions on Smart Grid}, vol.~1, no.~2, pp. 120--133, 2010.

\bibitem{shi2011real}
W.~Shi and V.~W. Wong, ``Real-time vehicle-to-grid control algorithm under price uncertainty,'' in \emph{2011 IEEE international conference on smart grid communications (SmartGridComm)}.\hskip 1em plus 0.5em minus 0.4em\relax IEEE, 2011, pp. 261--266.

\bibitem{ma2016efficient}
Z.~Ma, S.~Zou, L.~Ran, X.~Shi, and I.~A. Hiskens, ``Efficient decentralized coordination of large-scale plug-in electric vehicle charging,'' \emph{Automatica}, vol.~69, pp. 35--47, 2016.

\bibitem{zou2014cost}
N.~Zou, L.~Qian, J.~Attia, and C.~Ai, ``Cost and peak-to-average ratio reduction of electricity usage via intelligent ev charging,'' in \emph{2014 9th IEEE Conference on Industrial Electronics and Applications}.\hskip 1em plus 0.5em minus 0.4em\relax IEEE, 2014, pp. 584--589.

\bibitem{fan2020enabling}
G.~Fan, Z.~Yang, H.~Jin, X.~Gan, and X.~Wang, ``Enabling optimal control under demand elasticity for electric vehicle charging systems,'' \emph{IEEE Transactions on Mobile Computing}, vol.~21, no.~3, pp. 955--970, 2020.

\bibitem{zhang2018optimal}
Y.~Zhang, P.~You, and L.~Cai, ``Optimal charging scheduling by pricing for ev charging station with dual charging modes,'' \emph{IEEE Transactions on Intelligent Transportation Systems}, vol.~20, no.~9, pp. 3386--3396, 2018.

\bibitem{sarker2017opportunistic}
A.~Sarker, Z.~Li, W.~Kolodzey, and H.~Shen, ``Opportunistic energy sharing between power grid and electric vehicles: A game theory-based pricing policy,'' in \emph{2017 IEEE 37th International Conference on Distributed Computing Systems (ICDCS)}.\hskip 1em plus 0.5em minus 0.4em\relax IEEE, 2017, pp. 1197--1207.

\bibitem{wang2020pricing}
G.~Wang, Z.~Fang, X.~Xie, S.~Wang, H.~Sun, F.~Zhang, Y.~Liu, and D.~Zhang, ``Pricing-aware real-time charging scheduling and charging station expansion for large-scale electric buses,'' \emph{ACM Transactions on Intelligent Systems and Technology (TIST)}, vol.~12, no.~1, pp. 1--26, 2020.

\bibitem{gusrialdi2017distributed}
A.~Gusrialdi, Z.~Qu, and M.~A. Simaan, ``Distributed scheduling and cooperative control for charging of electric vehicles at highway service stations,'' \emph{IEEE Transactions on Intelligent Transportation Systems}, vol.~18, no.~10, pp. 2713--2727, 2017.

\bibitem{lin2021toward}
H.~Lin, X.~Lin, H.~Labiod, and L.~Chen, ``Toward multiple-phase mdp model for charging station recommendation,'' \emph{IEEE Transactions on Intelligent Transportation Systems}, vol.~23, no.~8, pp. 10\,583--10\,595, 2021.

\bibitem{blum_coordiq_2021}
C.~Blum, H.~Liu, and H.~Xiong, ``{CoordiQ} : {Coordinated} {Q}-learning for {Electric} {Vehicle} {Charging} {Recommendation},'' Jan. 2021.

\bibitem{li_coupling_2022}
C.~Li, Z.~Dong, N.~Fisher, and D.~Zhu, ``Coupling {User} {Preference} with {External} {Rewards} to {Enable} {Driver}-centered and {Resource}-aware {EV} {Charging} {Recommendation},'' Oct. 2022.

\bibitem{liu2021reservation}
S.~Liu, X.~Xia, Y.~Cao, Q.~Ni, X.~Zhang, and L.~Xu, ``Reservation-based ev charging recommendation concerning charging urgency policy,'' \emph{Sustainable Cities and Society}, vol.~74, p. 103150, 2021.

\bibitem{valogianni_effective_2014}
K.~Valogianni, W.~Ketter, J.~Collins, and D.~Zhdanov, ``\BIBforeignlanguage{en}{Effective {Management} of {Electric} {Vehicle} {Storage} {Using} {Smart} {Charging}},'' \emph{\BIBforeignlanguage{en}{Proceedings of the AAAI Conference on Artificial Intelligence}}, vol.~28, no.~1, June 2014.

\bibitem{zhang_intelligent_2021}
W.~Zhang, H.~Liu, F.~Wang, T.~Xu, H.~Xin, D.~Dou, and H.~Xiong, ``Intelligent {Electric} {Vehicle} {Charging} {Recommendation} {Based} on {Multi}-{Agent} {Reinforcement} {Learning},'' in \emph{Proceedings of the {Web} {Conference} 2021}, ser. {WWW} '21.\hskip 1em plus 0.5em minus 0.4em\relax New York, NY, USA: Association for Computing Machinery, June 2021, pp. 1856--1867.

\bibitem{suanpang2022intelligent}
P.~Suanpang, P.~Jamjuntr, P.~Kaewyong, C.~Niamsorn, and K.~Jermsittiparsert, ``An intelligent recommendation for intelligently accessible charging stations: Electronic vehicle charging to support a sustainable smart tourism city,'' \emph{Sustainability}, vol.~15, no.~1, p. 455, 2022.

\bibitem{zhang2022rlcharge}
W.~Zhang, H.~Liu, H.~Xiong, T.~Xu, F.~Wang, H.~Xin, and H.~Wu, ``Rlcharge: Imitative multi-agent spatiotemporal reinforcement learning for electric vehicle charging station recommendation,'' \emph{IEEE Transactions on Knowledge and Data Engineering}, 2022.

\bibitem{xing2022graph}
Q.~Xing, Y.~Xu, Z.~Chen, Z.~Zhang, and Z.~Shi, ``A graph reinforcement learning-based decision-making platform for real-time charging navigation of urban electric vehicles,'' \emph{IEEE Transactions on Industrial Informatics}, vol.~19, no.~3, pp. 3284--3295, 2022.

\bibitem{wei2015charging}
W.~Wei, F.~Liu, and S.~Mei, ``Charging strategies of ev aggregator under renewable generation and congestion: A normalized nash equilibrium approach,'' \emph{IEEE Transactions on Smart Grid}, vol.~7, no.~3, pp. 1630--1641, 2015.

\bibitem{xu2022real}
P.~Xu, J.~Zhang, T.~Gao, S.~Chen, X.~Wang, H.~Jiang, and W.~Gao, ``Real-time fast charging station recommendation for electric vehicles in coupled power-transportation networks: A graph reinforcement learning method,'' \emph{International Journal of Electrical Power \& Energy Systems}, vol. 141, p. 108030, 2022.

\bibitem{sun_orc_2020}
B.~Sun, T.~Li, S.~H. Low, and D.~H.~K. Tsang, ``{ORC}: {An} {Online} {Competitive} {Algorithm} for {Recommendation} and {Charging} {Schedule} in {Electric} {Vehicle} {Charging} {Network},'' in \emph{Proceedings of the {Eleventh} {ACM} {International} {Conference} on {Future} {Energy} {Systems}}, ser. e-{Energy} '20.\hskip 1em plus 0.5em minus 0.4em\relax New York, NY, USA: Association for Computing Machinery, June 2020, pp. 144--155.

\bibitem{klein2023electric}
P.~S. Klein and M.~Schiffer, ``Electric vehicle charge scheduling with flexible service operations,'' \emph{Transportation Science}, vol.~57, no.~6, pp. 1605--1626, 2023.

\bibitem{tan2023fair}
M.~Tan, Y.~Ren, R.~Pan, L.~Wang, and J.~Chen, ``Fair and efficient electric vehicle charging scheduling optimization considering the maximum individual waiting time and operating cost,'' \emph{IEEE Transactions on Vehicular Technology}, vol.~72, no.~8, pp. 9808--9820, 2023.

\bibitem{li2025charging}
Y.~Li, Y.~Pan, G.~Zhu, S.~He, M.~Xu, and J.~Xu, ``Charging-aware task assignment for urban logistics with electric vehicles,'' \emph{IEEE Transactions on Knowledge and Data Engineering}, 2025.

\bibitem{tian_real-time_2016}
Z.~Tian, T.~Jung, Y.~Wang, F.~Zhang, L.~Tu, C.~Xu, C.~Tian, and X.-Y. Li, ``Real-{Time} {Charging} {Station} {Recommendation} {System} for {Electric}-{Vehicle} {Taxis},'' \emph{IEEE Transactions on Intelligent Transportation Systems}, vol.~17, no.~11, pp. 3098--3109, Nov. 2016.

\bibitem{dong_rec_2017}
Z.~Dong, C.~Liu, Y.~Li, J.~Bao, Y.~Gu, and T.~He, ``{REC}: {Predictable} {Charging} {Scheduling} for {Electric} {Taxi} {Fleets},'' in \emph{2017 {IEEE} {Real}-{Time} {Systems} {Symposium} ({RTSS})}, Dec. 2017, pp. 287--296.

\bibitem{wang_tcharge_2019}
G.~Wang, F.~Zhang, and D.~Zhang, ``{tCharge} - {A} fleet-oriented real-time charging scheduling system for electric taxi fleets: poster abstract,'' in \emph{Proceedings of the 17th {Conference} on {Embedded} {Networked} {Sensor} {Systems}}, ser. {SenSys} '19.\hskip 1em plus 0.5em minus 0.4em\relax New York, NY, USA: Association for Computing Machinery, Nov. 2019, pp. 440--441.

\bibitem{wang_foretaxi_2023}
G.~Wang, Y.~Chen, S.~Wang, F.~Zhang, and D.~Zhang, ``{ForETaxi}: {Data}-{Driven} {Fleet}-{Oriented} {Charging} {Resource} {Allocation} in {Large}-{Scale} {Electric} {Taxi} {Networks},'' \emph{ACM Transactions on Sensor Networks}, vol.~19, no.~3, pp. 63:1--63:25, Mar. 2023.

\bibitem{wang_r2e_2018}
E.~Wang, C.~Qiao, L.~Su, W.~Dong, and {University at Buffalo. Department of Computer Science and Engineering.}, ``\BIBforeignlanguage{English}{{R2E}: {A} {Real}-{Time} {Routing} and {Recharging} {Recommendation} {System} for {Electric} {Taxi} {Drivers}},'' Tech. Rep., Nov. 2018.

\bibitem{wang_joint_2022}
E.~Wang, R.~Ding, Z.~Yang, H.~Jin, C.~Miao, L.~Su, F.~Zhang, C.~Qiao, and X.~Wang, ``Joint {Charging} and {Relocation} {Recommendation} for {E}-{Taxi} {Drivers} via {Multi}-{Agent} {Mean} {Field} {Hierarchical} {Reinforcement} {Learning},'' \emph{IEEE Transactions on Mobile Computing}, vol.~21, no.~4, pp. 1274--1290, Apr. 2022.

\bibitem{yuan2025stochastic}
Y.~Yuan, M.~Jia, Y.~Zhao, and S.~Lin, ``Stochastic model predictive control-based electric taxi fleet coordination under solar power uncertainty,'' \emph{ACM Transactions on Cyber-Physical Systems}, 2025.

\bibitem{wang_faircharge_2020}
G.~Wang, Y.~Zhang, Z.~Fang, S.~Wang, F.~Zhang, and D.~Zhang, ``{FairCharge}: {A} {Data}-{Driven} {Fairness}-{Aware} {Charging} {Recommendation} {System} for {Large}-{Scale} {Electric} {Taxi} {Fleets},'' \emph{Proceedings of the ACM on Interactive, Mobile, Wearable and Ubiquitous Technologies}, vol.~4, no.~1, pp. 28:1--28:25, Mar. 2020.

\bibitem{wang_bcharge_2018}
G.~Wang, X.~Xie, F.~Zhang, Y.~Liu, and D.~Zhang, ``{bCharge}: {Data}-{Driven} {Real}-{Time} {Charging} {Scheduling} for {Large}-{Scale} {Electric} {Bus} {Fleets},'' in \emph{2018 {IEEE} {Real}-{Time} {Systems} {Symposium} ({RTSS})}, Dec. 2018, pp. 45--55.

\bibitem{ma2021optimal}
T.-Y. Ma and S.~Xie, ``Optimal fast charging station locations for electric ridesharing with vehicle-charging station assignment,'' \emph{Transportation Research Part D: Transport and Environment}, vol.~90, p. 102682, 2021.

\bibitem{hu2022joint}
H.~Hu, B.~Du, W.~Liu, and P.~Perez, ``A joint optimisation model for charger locating and electric bus charging scheduling considering opportunity fast charging and uncertainties,'' \emph{Transportation Research Part C: Emerging Technologies}, vol. 141, p. 103732, 2022.

\bibitem{houbbadi2019optimal}
A.~Houbbadi, R.~Trigui, S.~Pelissier, E.~Redondo-Iglesias, and T.~Bouton, ``Optimal scheduling to manage an electric bus fleet overnight charging,'' \emph{Energies}, vol.~12, no.~14, p. 2727, 2019.

\bibitem{jahic2019charging}
A.~Jahic, M.~Eskander, and D.~Schulz, ``Charging schedule for load peak minimization on large-scale electric bus depots,'' \emph{Applied Sciences}, vol.~9, no.~9, p. 1748, 2019.

\bibitem{rodrigues_optimized_2020}
N.~Rodrigues, S.~Thakare, S.~Vyas, and R.~Kumar, ``Optimized {Charge} {Scheduling} of {Electric} {Buses} in a {City} {Bus} {Network},'' in \emph{2020 {IEEE} {International} {Conference} on {Power} {Electronics}, {Drives} and {Energy} {Systems} ({PEDES})}, Dec. 2020, pp. 1--6.

\bibitem{duan2022bidding}
X.~Duan, Z.~Hu, Y.~Song, Y.~Cui, and Y.~Wen, ``Bidding and charging scheduling optimization for the urban electric bus operator,'' \emph{IEEE Transactions on Smart Grid}, vol.~14, no.~1, pp. 489--501, 2022.

\bibitem{kang2015centralized}
Q.~Kang, J.~Wang, M.~Zhou, and A.~C. Ammari, ``Centralized charging strategy and scheduling algorithm for electric vehicles under a battery swapping scenario,'' \emph{IEEE Transactions on Intelligent Transportation Systems}, vol.~17, no.~3, pp. 659--669, 2015.

\bibitem{li2020joint}
X.~Li, T.~Wang, L.~Li, F.~Feng, W.~Wang, and C.~Cheng, ``Joint optimization of regular charging electric bus transit network schedule and stationary charger deployment considering partial charging policy and time-of-use electricity prices,'' \emph{Journal of Advanced Transportation}, vol. 2020, pp. 1--16, 2020.

\bibitem{qi2025optimizing}
J.~Qi, L.~Lei, T.~Jonsson, and D.~Niyato, ``Optimizing electric bus charging scheduling with uncertainties using hierarchical deep reinforcement learning,'' \emph{IEEE Internet of Things Journal}, 2025.

\bibitem{zhou2024electric}
Y.~Zhou, Q.~Meng, G.~P. Ong, and H.~Wang, ``Electric bus charging scheduling on a bus network,'' \emph{Transportation Research Part C: Emerging Technologies}, vol. 161, p. 104553, 2024.

\bibitem{tu2024deep}
W.~Tu, H.~Ye, K.~Mai, M.~Zhou, J.~Jiang, T.~Zhao, S.~Yi, and Q.~Li, ``Deep online recommendations for connected e-taxis by coupling trajectory mining and reinforcement learning,'' \emph{International Journal of Geographical Information Science}, vol.~38, no.~2, pp. 216--242, 2024.

\bibitem{kullman2022dynamic}
N.~D. Kullman, M.~Cousineau, J.~C. Goodson, and J.~E. Mendoza, ``Dynamic ride-hailing with electric vehicles,'' \emph{Transportation Science}, vol.~56, no.~3, pp. 775--794, 2022.

\bibitem{hu2021effective}
S.~Hu, B.~Guo, S.~Wang, and X.~Zhou, ``Effective cross-region courier-displacement for instant delivery via reinforcement learning,'' in \emph{International Conference on Wireless Algorithms, Systems, and Applications}.\hskip 1em plus 0.5em minus 0.4em\relax Springer, 2021, pp. 288--300.

\bibitem{ma2025crowdsourced}
G.~Ma, W.~Wang, B.~Sun, W.~Wu, and Y.~Zhou, ``Crowdsourced task dispatching for the shared electric vehicle relocation problem: a hybrid variable neighbourhood search and genetic algorithm,'' \emph{Transportmetrica B: Transport Dynamics}, vol.~13, no.~1, p. 2490511, 2025.

\bibitem{yi2021framework}
Z.~Yi and J.~Smart, ``A framework for integrated dispatching and charging management of an autonomous electric vehicle ride-hailing fleet,'' \emph{Transportation Research Part D: Transport and Environment}, vol.~95, p. 102822, 2021.

\bibitem{yan2023online}
P.~Yan, K.~Yu, X.~Chao, and Z.~Chen, ``An online reinforcement learning approach to charging and order-dispatching optimization for an e-hailing electric vehicle fleet,'' \emph{European Journal of Operational Research}, 2023.

\bibitem{chen2023electric}
L.~Chen, H.~Hamedmoghadam, M.~Jalili, and M.~Ramezani, ``Electric vehicle e-hailing fleet dispatching and charge scheduling,'' \emph{arXiv preprint arXiv:2302.12650}, 2023.

\bibitem{guo2021concurrent}
B.~Guo, S.~Wang, Y.~Ding, G.~Wang, S.~He, D.~Zhang, and T.~He, ``Concurrent order dispatch for instant delivery with time-constrained actor-critic reinforcement learning,'' in \emph{2021 IEEE Real-Time Systems Symposium (RTSS)}.\hskip 1em plus 0.5em minus 0.4em\relax IEEE, 2021, pp. 176--187.

\bibitem{wang2023gcrl}
H.~Wang, S.~Wang, Y.~Yang, and D.~Zhang, ``Gcrl: Efficient delivery area assignment for last-mile logistics with group-based cooperative reinforcement learning,'' in \emph{2023 IEEE 39th International Conference on Data Engineering (ICDE)}.\hskip 1em plus 0.5em minus 0.4em\relax IEEE, 2023, pp. 3522--3534.

\bibitem{shi2022memory}
L.~Shi, Z.-H. Zhan, D.~Liang, and J.~Zhang, ``Memory-based ant colony system approach for multi-source data associated dynamic electric vehicle dispatch optimization,'' \emph{IEEE Transactions on Intelligent Transportation Systems}, vol.~23, no.~10, pp. 17\,491--17\,505, 2022.

\bibitem{nolz2022consistent}
P.~C. Nolz, N.~Absi, D.~Feillet, and C.~Seragiotto, ``The consistent electric-vehicle routing problem with backhauls and charging management,'' \emph{European Journal of Operational Research}, vol. 302, no.~2, pp. 700--716, 2022.

\bibitem{li2023route}
C.~Li, Y.~Zhu, and K.~Y. Lee, ``Route optimization of electric vehicles based on reinsertion genetic algorithm,'' \emph{IEEE Transactions on Transportation Electrification}, vol.~9, no.~3, pp. 3753--3768, 2023.

\bibitem{erdelic2022goods}
T.~Erdeli{\'c} and T.~Cari{\'c}, ``Goods delivery with electric vehicles: Electric vehicle routing optimization with time windows and partial or full recharge,'' \emph{Energies}, vol.~15, no.~1, p. 285, 2022.

\bibitem{stodola2020hybrid}
P.~Stodola, ``Hybrid ant colony optimization algorithm applied to the multi-depot vehicle routing problem,'' \emph{Natural Computing}, vol.~19, no.~2, pp. 463--475, 2020.

\bibitem{wu2019brainstorming}
L.~Wu, Z.~He, Y.~Chen, D.~Wu, and J.~Cui, ``Brainstorming-based ant colony optimization for vehicle routing with soft time windows,'' \emph{IEEE Access}, vol.~7, pp. 19\,643--19\,652, 2019.

\bibitem{basso2021electric}
R.~Basso, B.~Kulcs{\'a}r, and I.~Sanchez-Diaz, ``Electric vehicle routing problem with machine learning for energy prediction,'' \emph{Transportation Research Part B: Methodological}, vol. 145, pp. 24--55, 2021.

\bibitem{basso2022dynamic}
R.~Basso, B.~Kulcs{\'a}r, I.~Sanchez-Diaz, and X.~Qu, ``Dynamic stochastic electric vehicle routing with safe reinforcement learning,'' \emph{Transportation research part E: logistics and transportation review}, vol. 157, p. 102496, 2022.

\bibitem{lin2021deep}
B.~Lin, B.~Ghaddar, and J.~Nathwani, ``Deep reinforcement learning for the electric vehicle routing problem with time windows,'' \emph{IEEE Transactions on Intelligent Transportation Systems}, vol.~23, no.~8, pp. 11\,528--11\,538, 2021.

\bibitem{folkestad2020optimal}
C.~A. Folkestad, N.~Hansen, K.~Fagerholt, H.~Andersson, and G.~Pantuso, ``Optimal charging and repositioning of electric vehicles in a free-floating carsharing system,'' \emph{Computers \& Operations Research}, vol. 113, p. 104771, 2020.

\bibitem{gambella2018optimizing}
C.~Gambella, E.~Malaguti, F.~Masini, and D.~Vigo, ``Optimizing relocation operations in electric car-sharing,'' \emph{Omega}, vol.~81, pp. 234--245, 2018.

\bibitem{wang2021record}
G.~Wang, Z.~Qin, S.~Wang, H.~Sun, Z.~Dong, and D.~Zhang, ``Record: Joint real-time repositioning and charging for electric carsharing with dynamic deadlines,'' in \emph{Proceedings of the 27th ACM SIGKDD Conference on Knowledge Discovery \& Data Mining}, 2021, pp. 3660--3669.

\bibitem{luo2021rebalancing}
M.~Luo, W.~Zhang, T.~Song, K.~Li, H.~Zhu, B.~Du, and H.~Wen, ``Rebalancing expanding ev sharing systems with deep reinforcement learning,'' in \emph{Proceedings of the Twenty-Ninth International Conference on International Joint Conferences on Artificial Intelligence}, 2021, pp. 1338--1344.

\bibitem{li2025multi}
X.~Li, Z.~Zhou, C.~Wei, X.~Gao, and Y.~Zhang, ``Multi-objective optimization of hybrid electric vehicles energy management using multi-agent deep reinforcement learning framework,'' \emph{Energy and AI}, vol.~20, p. 100491, 2025.

\bibitem{huang2024prediction}
H.~Huang, Y.~Wang, J.~Wu, W.~Ding, and J.~Pang, ``Prediction and optimization of pure electric vehicle tire/road structure-borne noise based on knowledge graph and multi-task resnet,'' \emph{Expert Systems with Applications}, vol. 255, p. 124536, 2024.

\bibitem{che2023battery}
Y.~Che, Y.~Zheng, Y.~Wu, X.~Lin, J.~Li, X.~Hu, and R.~Teodorescu, ``Battery states monitoring for electric vehicles based on transferred multi-task learning,'' \emph{IEEE Transactions on Vehicular Technology}, vol.~72, no.~8, pp. 10\,037--10\,047, 2023.

\bibitem{bao2024dual}
Z.~Bao, J.~Nie, H.~Lin, Z.~Li, K.~Gao, Z.~He, and M.~Gao, ``Dual-task learning for joint state-of-charge and state-of-energy estimation of lithium-ion battery in electric vehicle,'' \emph{IEEE Transactions on Transportation Electrification}, vol.~11, no.~1, pp. 558--569, 2024.

\bibitem{madaram2024advancement}
V.~G. Madaram, P.~K. Biswas, C.~Sain, S.~B. Thanikanti, and P.~K. Balachandran, ``Advancement of electric vehicle technologies, classification of charging methodologies, and optimization strategies for sustainable development-a comprehensive review,'' \emph{Heliyon}, vol.~10, no.~20, 2024.

\bibitem{singh2023comprehensive}
S.~Singh, R.~K. Saket, and B.~Khan, ``A comprehensive state-of-the-art review on reliability assessment and charging methodologies of grid-integrated electric vehicles,'' \emph{IET Electrical systems in Transportation}, vol.~13, no.~1, p. e12073, 2023.

\bibitem{saraswathi2024comprehensive}
V.~Saraswathi and V.~P. Ramachandran, ``A comprehensive review on charger technologies, types, and charging stations models for electric vehicles,'' \emph{Heliyon}, vol.~10, no.~20, 2024.

\bibitem{garofalaki2022electric}
Z.~Garofalaki, D.~Kosmanos, S.~Moschoyiannis, D.~Kallergis, and C.~Douligeris, ``Electric vehicle charging: A survey on the security issues and challenges of the open charge point protocol (ocpp),'' \emph{IEEE Communications Surveys \& Tutorials}, vol.~24, no.~3, pp. 1504--1533, 2022.

\bibitem{suhail2023objective}
M.~Suhail, I.~Akhtar, and S.~Kirmani, ``Objective functions and infrastructure for optimal placement of electrical vehicle charging station: a comprehensive survey,'' \emph{IETE Journal of Research}, vol.~69, no.~8, pp. 5250--5260, 2023.

\bibitem{ren2024understanding}
M.~Ren, H.~Dai, T.~Liu, X.~Deng, W.~Dou, Y.~Yang, and G.~Chen, ``Understanding wireless charger networks: Concepts, current research, and future directions,'' \emph{IEEE Communications Surveys \& Tutorials}, 2024.

\bibitem{zhao2024reinforcement}
Z.~Zhao, C.~K. Lee, X.~Yan, and H.~Wang, ``Reinforcement learning for electric vehicle charging scheduling: A systematic review,'' \emph{Transportation Research Part E: Logistics and Transportation Review}, vol. 190, p. 103698, 2024.

\bibitem{salam2024charge}
S.~S.~A. Salam, V.~Raj, M.~I. Petra, A.~K. Azad, S.~Mathew, and S.~M. Sulthan, ``Charge scheduling optimization of electric vehicles: A comprehensive review of essentiality, perspectives, techniques and security,'' \emph{IEEE Access}, 2024.

\bibitem{al2024optimization}
H.~M. Al-Alwash, E.~Borcoci, M.-C. Vochin, I.~A. Balapuwaduge, and F.~Y. Li, ``Optimization schedule schemes for charging electric vehicles: Overview, challenges, and solutions,'' \emph{IEEE access}, vol.~12, pp. 32\,801--32\,818, 2024.

\bibitem{elghanam2024optimization}
E.~Elghanam, A.~Abdelfatah, M.~S. Hassan, and A.~H. Osman, ``Optimization techniques in electric vehicle charging scheduling, routing and spatio-temporal demand coordination: A systematic review,'' \emph{IEEE Open Journal of Vehicular Technology}, vol.~5, pp. 1294--1313, 2024.

\bibitem{alaee2023review}
P.~Alaee, J.~Bems, and A.~Anvari-Moghaddam, ``A review of the latest trends in technical and economic aspects of ev charging management,'' \emph{Energies}, vol.~16, no.~9, p. 3669, 2023.

\bibitem{zhang2023vehicle}
L.~Zhang, Y.~Han, J.~Peng, and Y.~Wang, ``Vehicle and charging scheduling of electric bus fleets: A comprehensive review,'' \emph{Journal of Intelligent and Connected Vehicles}, vol.~6, no.~3, pp. 116--124, 2023.

\bibitem{kalakanti2023computational}
A.~K. Kalakanti and S.~Rao, ``Computational challenges and approaches for electric vehicles,'' \emph{ACM Computing Surveys}, vol.~55, no. 14s, pp. 1--35, 2023.

\bibitem{dahiwale2024comprehensive}
P.~V. Dahiwale, Z.~H. Rather, and I.~Mitra, ``A comprehensive review of smart charging strategies for electric vehicles and way forward,'' \emph{IEEE Transactions on Intelligent Transportation Systems}, vol.~25, no.~9, pp. 10\,462--10\,482, 2024.

\bibitem{wen2024survey}
D.~Wen, Y.~Li, and F.~C. Lau, ``A survey of machine learning-based ride-hailing planning,'' \emph{IEEE Transactions on Intelligent Transportation Systems}, vol.~25, no.~6, pp. 4734--4753, 2024.

\bibitem{teusch2023systematic}
J.~Teusch, J.~N. Gremmel, C.~Koetsier, F.~T. Johora, M.~Sester, D.~M. Woisetschl{\"a}ger, and J.~P. M{\"u}ller, ``A systematic literature review on machine learning in shared mobility,'' \emph{IEEE Open Journal of Intelligent Transportation Systems}, vol.~4, pp. 870--899, 2023.

\bibitem{bruglieri2023survey}
M.~Bruglieri and O.~Pisacane, ``A survey on emergent trends in the optimization of car-sharing systems,'' \emph{International Transactions in Operational Research}, vol.~30, no.~6, pp. 2867--2908, 2023.

\bibitem{soto2025vehicle}
R.~Soto-Concha, J.~W. Escobar, D.~Morillo-Torres, and R.~Linfati, ``The vehicle-routing problem with satellites utilization: A systematic review of the literature,'' \emph{Mathematics}, vol.~13, no.~7, pp. 1--29, 2025.

\bibitem{wang2021grid}
L.~Wang, Z.~Qin, T.~Slangen, P.~Bauer, and T.~Van~Wijk, ``Grid impact of electric vehicle fast charging stations: Trends, standards, issues and mitigation measures-an overview,'' \emph{IEEE Open Journal of Power Electronics}, vol.~2, pp. 56--74, 2021.

\bibitem{franzese2023fast}
P.~Franzese, D.~D. Patel, A.~A. Mohamed, D.~Iannuzzi, B.~Fahimi, M.~Risso, and J.~M. Miller, ``Fast dc charging infrastructures for electric vehicles: Overview of technologies, standards, and challenges,'' \emph{IEEE Transactions on Transportation Electrification}, vol.~9, no.~3, pp. 3780--3800, 2023.

\bibitem{inci2024power}
M.~Inci, {\"O}.~{\c{C}}elik, A.~Lashab, K.~{\c{C}}. Bay{\i}nd{\i}r, J.~C. Vasquez, and J.~M. Guerrero, ``Power system integration of electric vehicles: A review on impacts and contributions to the smart grid,'' \emph{Applied Sciences}, vol.~14, no.~6, p. 2246, 2024.

\bibitem{ismail2023impact}
A.~A. Ismail, N.~T. Mbungu, A.~Elnady, R.~C. Bansal, A.-K. Hamid, and M.~AlShabi, ``Impact of electric vehicles on smart grid and future predictions: a survey,'' \emph{International Journal of Modelling and Simulation}, vol.~43, no.~6, pp. 1041--1057, 2023.

\bibitem{rashid2024comprehensive}
H.~Rashid, L.~M. Hua, L.~Guanghua, R.~Hasan, A.~AlKaseem, A.~Ali, S.~H.~H. Shah, S.~Shaikh, A.~M. Soomar, and P.~Musznicki, ``A comprehensive review on economic, environmental impacts and future challenges for photovoltaic-based electric vehicle charging infrastructures,'' \emph{Frontiers in Energy Research}, vol.~12, p. 1411440, 2024.

\bibitem{bai2022charging}
H.~K. Bai, D.~Costinett, L.~M. Tolbert, R.~Qin, L.~Zhu, Z.~Liang, and Y.~Huang, ``Charging electric vehicle batteries: Wired and wireless power transfer: Exploring ev charging technologies,'' \emph{IEEE Power Electronics Magazine}, vol.~9, no.~2, pp. 14--29, 2022.

\bibitem{hemavathi2022study}
S.~Hemavathi and A.~Shinisha, ``A study on trends and developments in electric vehicle charging technologies,'' \emph{Journal of energy storage}, vol.~52, p. 105013, 2022.

\bibitem{chen2020review}
T.~Chen, X.-P. Zhang, J.~Wang, J.~Li, C.~Wu, M.~Hu, and H.~Bian, ``A review on electric vehicle charging infrastructure development in the uk,'' \emph{Journal of Modern Power Systems and Clean Energy}, vol.~8, no.~2, pp. 193--205, 2020.

\bibitem{EVStationTrend}
{U.S. DEPARTMENT OF ENERGY}, ``Electric vehicle charging infrastructure trends,'' \url{https://afdc.energy.gov/fuels/electricity_infrastructure_trends.html}, 2023.

\bibitem{sathik2022comprehensive}
J.~Sathik Mohamed~Ali, D.~Almakhles, \emph{et~al.}, ``A comprehensive review of the on-road wireless charging system for e-mobility applications,'' \emph{Frontiers in Energy Research}, vol.~10, p. 926270, 2022.

\bibitem{rahman2016review}
I.~Rahman, P.~M. Vasant, B.~S.~M. Singh, M.~Abdullah-Al-Wadud, and N.~Adnan, ``Review of recent trends in optimization techniques for plug-in hybrid, and electric vehicle charging infrastructures,'' \emph{Renewable and Sustainable Energy Reviews}, vol.~58, pp. 1039--1047, 2016.

\bibitem{kettles2015electric}
D.~Kettles, ``Electric vehicle charging technology analysis and standards,'' \emph{Florida Solar Energy Center, FSEC Report Number: FSEC-CR-1996-15}, 2015.

\bibitem{wu2011review}
H.~H. Wu, A.~Gilchrist, K.~Sealy, P.~Israelsen, and J.~Muhs, ``A review on inductive charging for electric vehicles,'' in \emph{2011 IEEE international electric machines \& drives conference (IEMDC)}.\hskip 1em plus 0.5em minus 0.4em\relax IEEE, 2011, pp. 143--147.

\bibitem{arias2017robust}
N.~B. Arias, A.~Tabares, J.~F. Franco, M.~Lavorato, and R.~Romero, ``Robust joint expansion planning of electrical distribution systems and ev charging stations,'' \emph{IEEE Transactions on Sustainable Energy}, vol.~9, no.~2, pp. 884--894, 2017.

\bibitem{gan2020fast}
X.~Gan, H.~Zhang, G.~Hang, Z.~Qin, and H.~Jin, ``Fast-charging station deployment considering elastic demand,'' \emph{IEEE Transactions on Transportation Electrification}, vol.~6, no.~1, pp. 158--169, 2020.

\bibitem{sun2021data}
C.~Sun, T.~Li, and X.~Tang, ``Data-driven electric vehicle charging station placement for incentivizing potential demand,'' in \emph{2021 IEEE International Conference on Communications, Control, and Computing Technologies for Smart Grids (SmartGridComm)}.\hskip 1em plus 0.5em minus 0.4em\relax IEEE, 2021, pp. 27--32.

\bibitem{vazifeh2019optimizing}
M.~M. Vazifeh, H.~Zhang, P.~Santi, and C.~Ratti, ``Optimizing the deployment of electric vehicle charging stations using pervasive mobility data,'' \emph{Transportation Research Part A: Policy and Practice}, vol. 121, pp. 75--91, 2019.

\bibitem{liu2019social}
Q.~Liu, Y.~Zeng, L.~Chen, and X.~Zheng, ``Social-aware optimal electric vehicle charger deployment on road network,'' in \emph{Proceedings of the 27th ACM SIGSPATIAL International Conference on Advances in Geographic Information Systems}, 2019, pp. 398--407.

\bibitem{von2022reinforcement}
L.~von Wahl, N.~Tempelmeier, A.~Sao, and E.~Demidova, ``Reinforcement learning-based placement of charging stations in urban road networks,'' in \emph{Proceedings of the 28th ACM SIGKDD Conference on Knowledge Discovery and Data Mining}, 2022, pp. 3992--4000.

\bibitem{sanguesa2021review}
J.~A. Sanguesa, V.~Torres-Sanz, P.~Garrido, F.~J. Martinez, and J.~M. Marquez-Barja, ``A review on electric vehicles: Technologies and challenges,'' \emph{Smart Cities}, vol.~4, no.~1, pp. 372--404, 2021.

\bibitem{li2018electric}
C.~Li, T.~Ding, X.~Liu, and C.~Huang, ``An electric vehicle routing optimization model with hybrid plug-in and wireless charging systems,'' \emph{IEEE Access}, vol.~6, pp. 27\,569--27\,578, 2018.

\bibitem{yan2021catcharger}
L.~Yan, H.~Shen, J.~Zhao, C.~Xu, F.~Luo, C.~Qiu, Z.~Zhang, and S.~Mahmud, ``Catcharger: Deploying in-motion wireless chargers in a metropolitan road network via categorization and clustering of vehicle traffic,'' \emph{IEEE Internet of Things Journal}, vol.~9, no.~12, pp. 9525--9541, 2021.

\bibitem{zhu2020sparking}
X.~Zhu, S.~Wang, B.~Guo, T.~Ling, Z.~Zhou, L.~Tu, and T.~He, ``Sparking: A win-win data-driven contract parking sharing system,'' in \emph{Adjunct Proceedings of the 2020 ACM International Joint Conference on Pervasive and Ubiquitous Computing and Proceedings of the 2020 ACM International Symposium on Wearable Computers}, 2020, pp. 596--604.

\bibitem{mou2014decentralized}
Y.~Mou, H.~Xing, Z.~Lin, and M.~Fu, ``Decentralized optimal demand-side management for phev charging in a smart grid,'' \emph{IEEE Transactions on Smart Grid}, vol.~6, no.~2, pp. 726--736, 2014.

\bibitem{kong2015distributed}
F.~Kong and X.~Liu, ``Distributed deadline and renewable aware electric vehicle demand response in the smart grid,'' in \emph{2015 IEEE Real-Time Systems Symposium}.\hskip 1em plus 0.5em minus 0.4em\relax IEEE, 2015, pp. 23--32.

\bibitem{kong2016smart}
F.~Kong, X.~Liu, Z.~Sun, and Q.~Wang, ``Smart rate control and demand balancing for electric vehicle charging,'' in \emph{2016 ACM/IEEE 7th International Conference on Cyber-Physical Systems (ICCPS)}.\hskip 1em plus 0.5em minus 0.4em\relax IEEE, 2016, pp. 1--10.

\bibitem{kazemtarghi2024dynamic}
A.~Kazemtarghi, A.~Mallik, and Y.~Chen, ``Dynamic pricing strategy for electric vehicle charging stations to distribute the congestion and maximize the revenue,'' \emph{International Journal of Electrical Power \& Energy Systems}, vol. 158, p. 109946, 2024.

\bibitem{ge2024distributed}
X.~Ge, G.~Wang, R.~Sun, and F.~Wang, ``A distributed and game-theoretic based ev charging pricing model under coupled energy-transportation-information networks,'' \emph{IEEE Transactions on Smart Grid}, 2024.

\bibitem{ren2023electric}
L.~Ren, M.~Yuan, and X.~Jiao, ``Electric vehicle charging and discharging scheduling strategy based on dynamic electricity price,'' \emph{Engineering Applications of Artificial Intelligence}, vol. 123, p. 106320, 2023.

\bibitem{yang2023ev}
L.~Yang, X.~Geng, X.~Guan, and L.~Tong, ``Ev charging scheduling under demand charge: A block model predictive control approach,'' \emph{IEEE Transactions on Automation Science and Engineering}, vol.~21, no.~2, pp. 2125--2138, 2023.

\bibitem{zhang2024learning}
M.~Zhang, H.~Yang, Y.~Xu, and H.~Sun, ``Learning-based real-time aggregate flexibility provision and scheduling of electric vehicles,'' \emph{IEEE Transactions on Smart Grid}, vol.~15, no.~6, pp. 5840--5852, 2024.

\bibitem{qureshi2024multiobjective}
U.~Qureshi, A.~Ghosh, and B.~K. Panigrahi, ``Multiobjective pareto-optimal intelligent electric vehicle charging schedule in a commercial charging station: A stochastic convex optimization approach,'' \emph{IEEE Transactions on Industrial Informatics}, 2024.

\bibitem{hossain2023efficient}
M.~B. Hossain, S.~R. Pokhrel, and H.~L. Vu, ``Efficient and private scheduling of wireless electric vehicles charging using reinforcement learning,'' \emph{IEEE Transactions on Intelligent Transportation Systems}, vol.~24, no.~4, pp. 4089--4102, 2023.

\bibitem{tseng_improving_2019}
C.-M. Tseng, S.~C.-K. Chau, and X.~Liu, ``Improving {Viability} of {Electric} {Taxis} by {Taxi} {Service} {Strategy} {Optimization}: {A} {Big} {Data} {Study} of {New} {York} {City},'' \emph{IEEE Transactions on Intelligent Transportation Systems}, vol.~20, no.~3, pp. 817--829, Mar. 2019.

\bibitem{tan2023joint}
H.~Tan, Y.~Yuan, S.~Zhong, and Y.~Yang, ``Joint rebalancing and charging for shared electric micromobility vehicles with energy-informed demand,'' in \emph{Proceedings of the 32nd ACM International Conference on Information and Knowledge Management}, 2023, pp. 2392--2401.

\bibitem{jahic_preemptive_2019}
A.~Jahic, M.~Eskander, and D.~Schulz, ``Preemptive vs. non-preemptive charging schedule for large-scale electric bus depots,'' in \emph{2019 {IEEE} {PES} {Innovative} {Smart} {Grid} {Technologies} {Europe} ({ISGT}-{Europe})}, Sept. 2019, pp. 1--5.

\bibitem{he2022integrated}
Y.~He, Z.~Liu, and Z.~Song, ``Integrated charging infrastructure planning and charging scheduling for battery electric bus systems,'' \emph{Transportation Research Part D: Transport and Environment}, vol. 111, p. 103437, 2022.

\bibitem{zeng2024route}
Z.~Zeng, T.~Wang, and X.~Qu, ``En-route charge scheduling for an electric bus network: Stochasticity and real-world practice,'' \emph{Transportation Research Part E: Logistics and Transportation Review}, vol. 185, p. 103498, 2024.

\bibitem{wang2023cross}
S.~Wang, S.~Hu, B.~Guo, and G.~Wang, ``Cross-region courier displacement for on-demand delivery with multi-agent reinforcement learning,'' \emph{IEEE Transactions on Big Data}, vol.~9, no.~5, pp. 1321--1333, 2023.

\bibitem{iacobucci2019optimization}
R.~Iacobucci, B.~McLellan, and T.~Tezuka, ``Optimization of shared autonomous electric vehicles operations with charge scheduling and vehicle-to-grid,'' \emph{Transportation Research Part C: Emerging Technologies}, vol. 100, pp. 34--52, 2019.

\bibitem{wang2025electric}
Y.~Wang, D.~Xie, C.~Gu, P.~Zhao, and X.~Wang, ``Electric vehicle scheduling model with strategic siting and incentive pricing considering participation willingness,'' \emph{IEEE Transactions on Smart Grid}, 2025.

\bibitem{lu2024deco}
Y.~Lu, S.~Wang, Y.~Yang, H.~Wang, B.~Guo, D.~Zhang, S.~Wang, and T.~He, ``Deco: Cooperative order dispatching for on-demand delivery with real-time encounter detection,'' in \emph{Proceedings of the 33rd ACM International Conference on Information and Knowledge Management}, 2024, pp. 4734--4742.

\bibitem{guo2023towards}
B.~Guo, S.~Wang, H.~Wang, Y.~Liu, F.~Kong, D.~Zhang, and T.~He, ``Towards equitable assignment: Data-driven delivery zone partition at last-mile logistics,'' in \emph{Proceedings of the 29th ACM SIGKDD Conference on Knowledge Discovery and Data Mining}, 2023, pp. 4078--4088.

\bibitem{wang2023time}
S.~Wang, B.~Guo, Y.~Ding, G.~Wang, S.~He, D.~Zhang, and T.~He, ``Time-constrained actor-critic reinforcement learning for concurrent order dispatch in on-demand delivery,'' \emph{IEEE Transactions on Mobile Computing}, vol.~23, no.~8, pp. 8175--8192, 2023.

\bibitem{jiang2023faircod}
L.~Jiang, S.~Wang, B.~Guo, H.~Wang, D.~Zhang, and G.~Wang, ``Faircod: A fairness-aware concurrent dispatch system for large-scale instant delivery services,'' in \emph{Proceedings of the 29th ACM SIGKDD Conference on knowledge discovery and data mining}, 2023, pp. 4229--4238.

\bibitem{al2020approximate}
L.~Al-Kanj, J.~Nascimento, and W.~B. Powell, ``Approximate dynamic programming for planning a ride-hailing system using autonomous fleets of electric vehicles,'' \emph{European Journal of Operational Research}, vol. 284, no.~3, pp. 1088--1106, 2020.

\bibitem{shi2019operating}
J.~Shi, Y.~Gao, W.~Wang, N.~Yu, and P.~A. Ioannou, ``Operating electric vehicle fleet for ride-hailing services with reinforcement learning,'' \emph{IEEE Transactions on Intelligent Transportation Systems}, vol.~21, no.~11, pp. 4822--4834, 2019.

\bibitem{yu2021optimal}
G.~Yu, A.~Liu, J.~Zhang, and H.~Sun, ``Optimal operations planning of electric autonomous vehicles via asynchronous learning in ride-hailing systems,'' \emph{Omega}, vol. 103, p. 102448, 2021.

\bibitem{liang2020mobility}
Y.~Liang, Z.~Ding, T.~Ding, and W.-J. Lee, ``Mobility-aware charging scheduling for shared on-demand electric vehicle fleet using deep reinforcement learning,'' \emph{IEEE Transactions on Smart Grid}, vol.~12, no.~2, pp. 1380--1393, 2020.

\bibitem{abid2022routing}
M.~Abid, M.~Tabaa, A.~Chakir, and H.~Hachimi, ``Routing and charging of electric vehicles: Literature review,'' \emph{Energy Reports}, vol.~8, pp. 556--578, 2022.

\bibitem{zhou2021multi}
Z.~Zhou, X.~Zhou, Y.~Lu, H.~Yan, B.~Guo, and S.~Wang, ``Multi-source data-driven route prediction for instant delivery,'' in \emph{2021 17th International Conference on Mobility, Sensing and Networking (MSN)}.\hskip 1em plus 0.5em minus 0.4em\relax IEEE, 2021, pp. 374--381.

\bibitem{zhou2024multi}
Z.~Zhou, X.~Zhou, B.~Guo, S.~Wang, and T.~He, ``Multi-sensor data-driven route prediction in instant delivery with a 3-conversion network,'' \emph{ACM Transactions on Sensor Networks}, vol.~20, no.~2, pp. 1--21, 2024.

\bibitem{li2024coordinated}
J.~Li, X.~Xu, H.~Wang, Z.~Yan, M.~Shahidehpour, and B.~Yang, ``Coordinated multi-task scheduling of electric buses in post-disaster transportation-power distribution systems,'' \emph{IEEE Transactions on Transportation Electrification}, 2024.

\bibitem{saber2010plug}
A.~Y. Saber and G.~K. Venayagamoorthy, ``Plug-in vehicles and renewable energy sources for cost and emission reductions,'' \emph{IEEE Transactions on Industrial electronics}, vol.~58, no.~4, pp. 1229--1238, 2010.

\bibitem{chatzivasileiadis2011q}
S.~Chatzivasileiadis, M.~D. Galus, Y.~Reckinger, and G.~Andersson, ``Q-learning for optimal deployment strategies of frequency controllers using the aggregated storage of phev fleets,'' in \emph{2011 IEEE Trondheim PowerTech}.\hskip 1em plus 0.5em minus 0.4em\relax IEEE, 2011, pp. 1--8.

\bibitem{galus2011balancing}
M.~D. Galus and G.~Andersson, ``Balancing renewable energy source with vehicle to grid services from a large fleet of plug-in hybrid electric vehicles controlled in a metropolitan area distribution network,'' in \emph{CIGRE International Symposium}.\hskip 1em plus 0.5em minus 0.4em\relax Citeseer, 2011.

\bibitem{wehinger2011assessing}
\BIBentryALTinterwordspacing
L.~A. Wehinger, G.~Hug, M.~D. Galus, and G.~Andersson, ``Assessing the effect of storage devices and a phev cluster on german spot prices by using model predictive and profit maximizing agents,'' 2011. [Online]. Available: \url{https://api.semanticscholar.org/CorpusID:14220219}
\BIBentrySTDinterwordspacing

\bibitem{xue2021impact}
P.~Xue, Y.~Xiang, J.~Gou, W.~Xu, W.~Sun, Z.~Jiang, S.~Jawad, H.~Zhao, and J.~Liu, ``Impact of large-scale mobile electric vehicle charging in smart grids: A reliability perspective,'' \emph{Frontiers in Energy Research}, vol.~9, p. 688034, 2021.

\bibitem{global2017outlook}
E.~Global, ``Outlook 2018, international energy agency,'' 2017.

\bibitem{davis2021transportation}
S.~C. Davis and R.~G. Boundy, ``Transportation energy data book: Edition 39,'' Oak Ridge National Lab.(ORNL), Oak Ridge, TN (United States), Tech. Rep., 2021.

\bibitem{rauh2015understanding}
N.~Rauh, T.~Franke, and J.~F. Krems, ``Understanding the impact of electric vehicle driving experience on range anxiety,'' \emph{Human factors}, vol.~57, no.~1, pp. 177--187, 2015.

\bibitem{siragusa2022electric}
C.~Siragusa, A.~Tumino, R.~Mangiaracina, and A.~Perego, ``Electric vehicles performing last-mile delivery in b2c e-commerce: An economic and environmental assessment,'' \emph{International Journal of Sustainable Transportation}, vol.~16, no.~1, pp. 22--33, 2022.

\bibitem{mounce2019potential}
R.~Mounce and J.~D. Nelson, ``On the potential for one-way electric vehicle car-sharing in future mobility systems,'' \emph{Transportation Research Part A: Policy and Practice}, vol. 120, pp. 17--30, 2019.

\bibitem{sivaraman2021power}
P.~Sivaraman, J.~S.~S. Raj, and P.~A. Kumar, ``Power quality impact of electric vehicle charging station on utility grid,'' in \emph{2021 IEEE Madras Section Conference (MASCON)}.\hskip 1em plus 0.5em minus 0.4em\relax IEEE, 2021, pp. 1--4.

\bibitem{bhatt2014instrumentation}
J.~Bhatt, V.~Shah, and O.~Jani, ``An instrumentation engineer’s review on smart grid: Critical applications and parameters,'' \emph{Renewable and Sustainable Energy Reviews}, vol.~40, pp. 1217--1239, 2014.

\bibitem{yong2015review}
J.~Y. Yong, V.~K. Ramachandaramurthy, K.~M. Tan, and N.~Mithulananthan, ``A review on the state-of-the-art technologies of electric vehicle, its impacts and prospects,'' \emph{Renewable and sustainable energy reviews}, vol.~49, pp. 365--385, 2015.

\bibitem{qiao2011design}
L.~Qiao, X.~Liu, and B.~Jiang, ``Design and implementation of the smart meter in vehicle-to-grid,'' in \emph{2011 4th international conference on electric utility deregulation and restructuring and power technologies (DRPT)}.\hskip 1em plus 0.5em minus 0.4em\relax IEEE, 2011, pp. 618--621.

\bibitem{massaoudi2021deep}
M.~Massaoudi, H.~Abu-Rub, S.~S. Refaat, I.~Chihi, and F.~S. Oueslati, ``Deep learning in smart grid technology: A review of recent advancements and future prospects,'' \emph{IEEE Access}, vol.~9, pp. 54\,558--54\,578, 2021.

\bibitem{greer2014nist}
\BIBentryALTinterwordspacing
C.~Greer, D.~Wollman, D.~Prochaska, P.~Boynton, J.~Mazer, C.~Nguyen, G.~FitzPatrick, T.~Nelson, G.~Koepke, A.~H. Jr., V.~Pillitteri, T.~Brewer, N.~Golmie, D.~Su, A.~Eustis, D.~Holmberg, and S.~Bushby, ``\BIBforeignlanguage{en}{Nist framework and roadmap for smart grid interoperability standards, release 3.0},'' 2014-10-01 00:10:00 2014. [Online]. Available: \url{https://tsapps.nist.gov/publication/get_pdf.cfm?pub_id=916755}
\BIBentrySTDinterwordspacing

\bibitem{yilmaz2012review}
M.~Yilmaz and P.~T. Krein, ``Review of benefits and challenges of vehicle-to-grid technology,'' in \emph{2012 IEEE Energy Conversion Congress and Exposition (ECCE)}.\hskip 1em plus 0.5em minus 0.4em\relax IEEE, 2012, pp. 3082--3089.

\bibitem{vidhi2018review}
R.~Vidhi and P.~Shrivastava, ``A review of electric vehicle lifecycle emissions and policy recommendations to increase ev penetration in india,'' \emph{Energies}, vol.~11, no.~3, p. 483, 2018.

\bibitem{liu2022exhaust}
Y.~Liu, H.~Chen, Y.~Li, J.~Gao, K.~Dave, J.~Chen, T.~Li, and R.~Tu, ``Exhaust and non-exhaust emissions from conventional and electric vehicles: A comparison of monetary impact values,'' \emph{Journal of Cleaner Production}, vol. 331, p. 129965, 2022.

\bibitem{ke2017well}
W.~Ke, S.~Zhang, X.~He, Y.~Wu, and J.~Hao, ``Well-to-wheels energy consumption and emissions of electric vehicles: Mid-term implications from real-world features and air pollution control progress,'' \emph{Applied Energy}, vol. 188, pp. 367--377, 2017.

\bibitem{chen2024integration}
X.~Chen, X.~Lu, Q.~Li, D.~Li, and F.~Zhu, ``Integration of llm and human-ai coordination for power dispatching with connected electric vehicles under sagvns,'' \emph{IEEE Transactions on Vehicular Technology}, 2024.

\bibitem{teimoori2025llm}
Z.~Teimoori, ``Llm-enabled ev charging stations recommendation,'' \emph{arXiv preprint arXiv:2505.01447}, 2025.

\bibitem{fan2025ev}
H.~Fan, Y.~Chai, C.~Liu, W.~Liu, Z.~Zhang, W.~Run, and D.~Liu, ``Ev-stllm: Electric vehicle charging forecasting based on spatio-temporal large language models with multi-frequency and multi-scale information fusion,'' \emph{arXiv preprint arXiv:2507.09527}, 2025.

\bibitem{niu2025ev}
Z.~Niu, J.~Li, Q.~Ai, J.~Jiang, Q.~Yang, and H.~Zhou, ``Ev charging system considering power dispatching based on multi-agent llms and cgan,'' \emph{IEEE Transactions on Intelligent Transportation Systems}, 2025.

\bibitem{sun2025dynamic}
Y.~Sun, C.~Cui, C.~Zhang, and C.~Gong, ``Dynamic incentive strategies for smart ev charging stations: An llm-driven user digital twin approach,'' \emph{arXiv preprint arXiv:2504.01423}, 2025.

\end{thebibliography}

\vfill


\end{document}